# Open Data from LIGO, Virgo, and KAGRA through the First Part of the Fourth Observing Run

A. G. Abac[1], I. Abouelfettouh[2], F. Acernese[3,4], K. Ackley[5], C. Adamcewicz[6], S. Adhicary[7], D. Adhikari[8,9], N. Adhikari[10], R. X. Adhikari[11], V. K. Adkins[12], S. Afroz[13], A. Agapito[14], D. Agarwal[15], M. Agathos[16], N. Aggarwal[17], S. Aggarwal[18], O. D. Aguiar[19], I.-L. Ahrend[20], L. Aiello[21,22], A. Ain[23], P. Ajith[24], T. Akutsu[25,26], S. Albanesi[27,28], W. Ali[29,30], S. Al-Kershi[8,9], C. Alléné[31], A. Allocca[4,32], S. Al-Shammari[33], P. A. Altin[34], S. Alvarez-Lopez[35], W. Amar[31], O. Amarasinghe[33], A. Amato[36,37], F. Amicucci[38,39], C. Amra[40], A. Ananyeva[11], S. B. Anderson[11], W. G. Anderson[11], M. Andia[41], M. Ando[42], M. Andrés-Carcasona[43], T. Andrić[8,9,44,45], J. Anglin[46], S. Ansoldi[47,48], J. M. Antelis[49], S. Antier[41], M. Aoumi[50], E. Z. Appavuravther[51,52], S. Appert[11], S. K. Apple[53], K. Arai[11], A. Araya[42], M. C. Araya[11], M. Arca Sedda[44,45], J. S. Areeda[54], N. Aritomi[2], F. Armato[29,30], S. Armstrong[55], N. Arnaud[56], M. Arogeti[57], S. M. Aronson[12], G. Ashton[58], Y. Aso[25,59], L. Asprea[28], M. Assiduo[60,61], S. Assis de Souza Melo[62], S. M. Aston[63], P. Astone[38], F. Attadio[38,39], F. Aubin[64], K. AultONeal[65], G. Avallone[66], E. A. Avila[49], S. Babak[20], C. Badger[67], S. Bae[68], S. Bagnasco[28], L. Baiotti[69], R. Bajpai[70], T. Baka[37,71], A. M. Baker[6], K. A. Baker[72], T. Baker[73], G. Baldi[74,75], N. Baldicchi[51,76], M. Ball[77], G. Ballardin[62], S. W. Ballmer[78], S. Banagiri[6], B. Banerjee[44], D. Bankar[79], T. M. Baptiste[12], P. Baral[10], M. Baratti[80,81], J. C. Barayoga[11], B. C. Barish[11], D. Barker[2], N. Barman[79], P. Barneo[82,83,84], F. Barone[4,85], B. Barr[86], L. Barsotti[35], M. Barsuglia[20], D. Barta[87], A. M. Bartoletti[88], M. A. Barton[86], I. Bartos[46], A. Basalaev[8,9], R. Bassiri[89], A. Basti[80,81], M. Bawaj[51,76], P. Baxi[90], J. C. Bayley[86], A. C. Baylor[10], P. A. Baynard II[57], M. Bazzan[91,92], V. M. Bedakihale[93], F. Beirnaert[94], M. Bejger[95], D. Belardinelli[22], A. S. Bell[86], D. S. Bellie[96], L. Bellizzi[80,81], W. Benoit[18], I. Bentara[56], J. D. Bentley[97], M. Ben Yaala[55], S. Bera[98,99], F. Bergamin[33], B. K. Berger[89], S. Bernuzzi[27], M. Beroiz[11], C. P. L. Berry[86], D. Bersanetti[29], T. Bertheas[100], A. Bertolini[36,37], J. Betzwieser[63], D. Beveridge[72], G. Bevilacqua[101], N. Bevins[102], R. Bhandare[103], R. Bhatt[11], D. Bhattacharjee[104,105], S. Bhattacharyya[106], S. Bhaumik[46], V. Biancalana[101], A. Bianchi[37,107], I. A. Bilenko[108], G. Billingsley[11], A. Binetti[109], S. Bini[11,74,75], C. Binu[110], S. Biot[111], O. Birnholtz[112], S. Biscoveanu[96], A. Bisht[9], M. Bitossi[62,80], M.-A. Bizouard[113], S. Blaber[114], J. K. Blackburn[11], L. A. Blagg[77], C. D. Blair[63,72], D. G. Blair[72], N. Bode[8,9], N. Boettner[97], G. Boileau[113], M. Boldrini[38], G. N. Bolingbroke[115], A. Bolliand[40,116], L. D. Bonavena[46], R. Bondarescu[82], F. Bondu[117], E. Bonilla[89], M. S. Bonilla[54], A. Bonino[118], R. Bonnand[31,116], A. Borchers[8,9], S. Borhanian[7], V. Boschi[80], S. Bose[119], V. Bossilkov[63], Y. Bothra[37,107], A. Boudon[56], L. Bourg[57], G. Bouyer[120], M. Boyle[121], A. Bozzi[62], C. Bradaschia[80], P. R. Brady[10], A. Branch[63], M. Branchesi[44,45], I. Braun[104], T. Briant[122], A. Brillet[113], M. Brinkmann[8,9], P. Brockill[10], E. Brockmueller[8,9], A. F. Brooks[11], B. C. Brown[46], D. D. Brown[115], M. L. Brozzetti[76,51], S. Brunett[11], G. Bruno[15], R. Bruntz[123], J. Bryant[118], Y. Bu[124], F. Bucci[61], J. Buchanan[123], O. Bulashenko[82,83], T. Bulik[125], H. J. Bulten[37], A. Buonanno[1,126], K. Burtnyk[2], R. Buscicchio[127,128], D. Buskulic[31], C. Buy[100], R. L. Byer[89], G. S. Cabourn Davies[73], R. Cabrita[15], V. Cáceres-Barbosa[7], L. Cadonati[57], G. Cagnoli[129], C. Cahillane[78], A. Calafat[98], T. A. Callister[130], E. Calloni[4,32], S. R. Callos[77], M. Canepa[29,30], G. Caneva Santoro[43], K. C. Cannon[42], H. Cao[35], L. A. Capistran[131], E. Capocasa[20], E. Capote[2,11], G. Capurri[81,80], G. Carapella[66,132], F. Carbognani[62], M. Carlassara[8,9], J. B. Carlin[124], T. K. Carlson[133], M. F. Carney[104], M. Carpinelli[62,127], G. Carrillo[77], J. J. Carter[8,9], G. Carullo[118,134], A. Casallas-Lagos[135], J. Casanueva Diaz[62], C. Casentini[22,136], S. Y. Castro-Lucas[137], S. Caudill[133], M. Cavaglià[105], R. Cavalieri[62], A. Ceja[54], G. Cella[80], P. Cerdá-Durán[138,139], E. Cesarini[22], N. Chabbra[34], W. Chaibi[113], A. Chakraborty[13], P. Chakraborty[8,9], S. Chakraborty[103], S. Chalathadka Subrahmanya[97], J. C. L. Chan[140], M. Chan[114], K. Chang[141], S. Chao[141,142], P. Charlton[143], E. Chassande-Mottin[20], C. Chatterjee[144], Debarati Chatterjee[79], Deep Chatterjee[35], M. Chaturvedi[103], S. Chaty[20], K. Chatziioannou[11], A. Chen[145], A. H.-Y. Chen[146], D. Chen[147], H. Chen[142], H. Y. Chen[120], S. Chen[144], Yanbei Chen[148], Yitian Chen[121], H. P. Cheng[149], P. Chessa[51,76], H. T. Cheung[90], S. Y. Cheung[6], F. Chiadini[132,150], G. Chiarini[8,9,92], A. Chiba[151], A. Chincarini[29], M. L. Chiofalo[80,81], A. Chiummo[4,62], C. Chou[146], S. Choudhary[72], N. Christensen[113,152], S. S. Y. Chua[34], G. Ciani[74,75], P. Ciecielag[95], M. Cieślar[125], M. Cifaldi[22], B. Cirok[153], F. Clara[2], J. A. Clark[11,57], T. A. Clarke[6], P. Clearwater[154], S. Clesse[111], F. Cleva[113,116], E. Coccia[43,44,45], E. Codazzo[155,156], P.-F. Cohadon[122], S. Colace[30], E. Colangeli[73], M. Colleoni[98], C. G. Collette[157], J. Collins[63], S. Colloms[86], A. Colombo[128,158], C. M. Compton[2], G. Connolly[77], L. Conti[92], T. R. Corbitt[12], I. Cordero-Carrión[159], S. Corezzi[51,76], N. J. Cornish[160], I. Coronado[161], A. Corsi[162], R. Cottingham[63], M. W. Coughlin[18], A. Couineaux[38], P. Couvares[11,57], D. M. Coward[72], R. Coyne[163], A. Cozzumbo[44], J. D. E. Creighton[10], T. D. Creighton[164], P. Cremonese[98], S. Crook[63], R. Crouch[2], J. Csizmazia[2], J. R. Cudell[165], T. J. Cullen[11], A. Cumming[86], E. Cuoco[166,167], M. Cusinato[138], L. V. Da Conceição[168], T. Dal Canton[41], S. Dal Pra[169], G. Dálya[100], B. D'Angelo[29], S. Danilishin[36,37], S. D'Antonio[38], K. Danzmann[8,9], K. E. Darroch[123], L. P. Dartez[63], R. Das[106], A. Dasgupta[93], V. Dattilo[62], A. Daumas[20], N. Davari[170,171], I. Dave[103], A. Davenport[137], M. Davier[41], T. F. Davies[72], D. Davis[11], L. Davis[72], M. C. Davis[18], P. Davis[172,173], E. J. Daw[174], M. Dax[1], J. De Bolle[94], M. Deenadayalan[79],

J. Degallaix[175], M. De Laurentis[4,32], F. De Lillo[23], S. Della Torre[128], W. Del Pozzo[80,81], A. Demagny[31], F. De Marco[38,39], G. Demasi[61,176], F. De Matteis[21,22], N. Demos[35], T. Dent[177], A. Depasse[15], N. DePergola[102], R. De Pietri[178,179], R. De Rosa[4,32], C. De Rossi[62], M. Desai[35], R. DeSalvo[180], A. DeSimone[181], R. De Simone[132,150], A. Dhani[1], R. Diab[46], M. C. Díaz[164], M. Di Cesare[4,32], G. Dideron[182], T. Dietrich[1], L. Di Fiore[4], C. Di Fronzo[72], M. Di Giovanni[38,39], T. Di Girolamo[32,4], D. Diksha[36,37], J. Ding[183,20], S. Di Pace[38,39], I. Di Palma[38,39], D. Di Piero[48,184], F. Di Renzo[56], Divyajyoti[33], A. Dmitriev[118], J. P. Docherty[86], Z. Doctor[96], N. Doerksen[168], E. Dohmen[2], A. Doke[133], A. Domiciano de Souza[185], L. D'Onofrio[38], F. Donovan[35], K. L. Dooley[33], T. Dooney[71], S. Doravari[79], O. Dorosh[186], W. J. D. Doyle[123], M. Drago[39,38], J. C. Driggers[2], M. Dubois[100], L. Dunn[124], U. Dupletsa[44], P.-A. Duverne[20], D. D'Urso[155,170], P. Dutta Roy[46], H. Duval[187], S. E. Dwyer[2], C. Eassa[2], M. Ebersold[188,31], T. Eckhardt[97], G. Eddolls[78], A. Effler[63], J. Eichholz[34], H. Einsle[113], M. Eisenmann[25], M. Emma[58], K. Endo[151], R. Enficiaud[1], L. Errico[32,4], R. Espinosa[164], M. C. Espitia[189], M. Esposito[4,32], R. C. Essick[190], H. Estellés[1], T. Etzel[11], M. Evans[35], T. Evstafyeva[182], B. E. Ewing[7], J. M. Ezquiaga[140], F. Fabrizi[60,61], V. Fafone[21,22], S. Fairhurst[33], A. M. Farah[130], B. Farr[77], W. M. Farr[191,192], G. Favaro[91], M. Favata[193], M. Fays[165], M. Fazio[55], J. Feicht[11], M. M. Fejer[89], R. Felicetti[184,48], E. Fenyvesi[87,194], J. Fernandes[195], T. Fernandes[196,138], D. Fernando[110], S. Ferraiuolo[197,39,38], T. A. Ferreira[12], F. Fidecaro[81,80], P. Figura[95], A. Fiori[80,81], I. Fiori[62], M. Fishbach[190], R. P. Fisher[123], R. Fittipaldi[132,198], V. Fiumara[132,199], R. Flaminio[31], S. M. Fleischer[200], L. S. Fleming[201], E. Floden[18], H. Fong[114], J. A. Font[138,139], F. Fontinele-Nunes[18], C. Foo[1], B. Fornal[202], K. Franceschetti[178], F. Frappez[31], S. Frasca[39,38], F. Frasconi[80], J. P. Freed[65], Z. Frei[203], A. Freise[37,107], O. Freitas[196,138], R. Frey[77], W. Frischhertz[63], P. Fritschel[35], V. V. Frolov[63], G. G. Fronzé[28], M. Fuentes-Garcia[11], S. Fujii[204], T. Fujimori[205], P. Fulda[46], M. Fyffe[63], B. Gadre[71], J. R. Gair[1], S. Galaudage[185], V. Galdi[206], R. Gamba[7], A. Gamboa[1], S. Gamoji[180], D. Ganapathy[318], A. Ganguly[79], B. Garaventa[29], J. García-Bellido[207], C. García-Quirós[188], J. W. Gardner[34], K. A. Gardner[114], S. Garg[42], J. Gargiulo[62], X. Garrido[41], A. Garron[98], F. Garufi[32,4], P. A. Garver[89], C. Gasbarra[21,22], B. Gateley[2], F. Gautier[208], V. Gayathri[10], T. Gayer[78], G. Gemme[29], A. Gennai[80], V. Gennari[100], J. George[103], R. George[120], O. Gerberding[97], L. Gergely[153], A. Ghosh[94], Sayantan Ghosh[195], Shaon Ghosh[193], Shrobana Ghosh[8,9], Suprovo Ghosh[209], T. Ghosh[79], J. A. Giaime[12,63], K. D. Giardina[63], D. R. Gibson[201], C. Gier[55], S. Gkaitatzis[81,80], J. Glanzer[11], F. Glotin[41], J. Godfrey[77], R. V. Godley[8,9], P. Godwin[11], A. S. Goettel[33], E. Goetz[114], J. Golomb[11], S. Gomez Lopez[39,38], B. Goncharov[44], G. González[12], P. Goodarzi[210], S. Goode[6], A. W. Goodwin-Jones[15], M. Gosselin[62], R. Gouaty[31], D. W. Gould[34], K. Govorkova[35], A. Grado[76,51], V. Graham[86], A. E. Granados[18], M. Granata[175], V. Granata[132,211], S. Gras[35], P. Grassia[11], J. Graves[57], C. Gray[2], R. Gray[86], G. Greco[51], A. C. Green[37,107], L. Green[212], S. M. Green[73], S. R. Green[213], C. Greenberg[133], A. M. Gretarsson[65], H. K. Griffin[18], D. Griffith[11], H. L. Griggs[57], G. Grignani[51,76], C. Grimaud[31], H. Grote[33], S. Grunewald[1], D. Guerra[138], D. Guetta[214], G. M. Guidi[60,61], A. R. Guimaraes[12], H. K. Gulati[93], F. Gulminelli[172,173], H. Guo[145], W. Guo[72], Y. Guo[37,36], A. Gupta[215], I. Gupta[7], N. C. Gupta[93], S. K. Gupta[46], V. Gupta[18], N. Gupte[1], J. Gurs[97], N. Gutierrez[175], N. Guttman[6], F. Guzman[131], D. Haba[216], M. Haberland[1], S. Haino[217], E. D. Hall[35], E. Z. Hamilton[98], G. Hammond[86], M. Haney[37], J. Hanks[2], C. Hanna[7], M. D. Hannam[33], O. A. Hannuksela[218], A. G. Hanselman[130], H. Hansen[2], J. Hanson[63], S. Hanumasagar[57], R. Harada[42], A. R. Hardison[181], S. Harikumar[186], K. Haris[37,71], I. Harley-Trochimczyk[131], T. Harmark[134], J. Harms[44,45], G. M. Harry[219], I. W. Harry[73], J. Hart[104], B. Haskell[95,220,221], C. J. Haster[212], K. Haughian[86], H. Hayakawa[50], K. Hayama[222], M. C. Heintze[63], J. Heinze[118], J. Heinzel[35], H. Heitmann[113], F. Hellman[318], A. F. Helmling-Cornell[77], G. Hemming[62], O. Henderson-Sapir[115], M. Hendry[86], I. S. Heng[86], M. H. Hennig[86], C. Henshaw[57], M. Heurs[8,9], A. L. Hewitt[223,224], J. Heynen[15], J. Heyns[35], S. Higginbotham[33], S. Hild[36,37], S. Hill[86], Y. Himemoto[225], N. Hirata[25], C. Hirose[226], D. Hofman[175], B. E. Hogan[65], N. A. Holland[37,107], I. J. Hollows[174], D. E. Holz[130], L. Honet[111], D. J. Horton-Bailey[318], J. Hough[86], S. Hourihane[11], N. T. Howard[144], E. J. Howell[72], C. G. Hoy[73], C. A. Hrishikesh[21], P. Hsi[35], H.-F. Hsieh[142], H.-Y. Hsieh[142], C. Hsiung[227], S.-H. Hsu[146], W.-F. Hsu[109], Q. Hu[86], H. Y. Huang[141], Y. Huang[7], Y. T. Huang[78], A. D. Huddart[228], B. Hughey[65], V. Hui[31], S. Husa[98], R. Huxford[7], L. Iampieri[39,38], G. A. Iandolo[36], M. Ianni[21,22], G. Iannone[132], J. Iascau[77], K. Ide[229], R. Iden[216], A. Ierardi[44,45], S. Ikeda[147], H. Imafuku[42], Y. Inoue[141], G. Iorio[91], P. Iosif[184,48], M. H. Iqbal[34], J. Irwin[86], R. Ishikawa[229], M. Isi[191,192], K. S. Isleif[230], Y. Itoh[205,231], M. Iwaya[204], B. R. Iyer[24], C. Jacquet[100], P.-E. Jacquet[122], T. Jacquot[41], S. J. Jadhav[232], S. P. Jadhav[154], M. Jain[133], T. Jain[223], A. L. James[11], K. Jani[144], J. Janquart[15], N. N. Janthalur[232], S. Jaraba[233], P. Jaranowski[234], R. Jaume[98], W. Javed[33], A. Jennings[2], M. Jensen[2], W. Jia[35], J. Jiang[149], H.-B. Jin[235,236], G. R. Johns[123], N. A. Johnson[46], M. C. Johnston[212], R. Johnston[86], N. Johny[8,9], D. H. Jones[34], D. I. Jones[209], R. Jones[86], H. E. Jose[77], P. Joshi[7], S. K. Joshi[79], G. Joubert[56], J. Ju[237], L. Ju[72], K. Jung[238], J. Junker[34], V. Juste[111], H. B. Kabagoz[63,35], T. Kajita[239], I. Kaku[205], V. Kalogera[96], M. Kalomenopoulos[212], M. Kamiizumi[50], N. Kanda[231,205], S. Kandhasamy[79], G. Kang[240], N. C. Kannachel[6], J. B. Kanner[11], S. A. KantiMahanty[18], S. J. Kapadia[79], D. P. Kapasi[54], M. Karthikeyan[133], M. Kasprzack[11], H. Kato[151], T. Kato[204], E. Katsavounidis[35], W. Katzman[63], R. Kaushik[103], K. Kawabe[2], R. Kawamoto[205], D. Keitel[98], L. J. Kemperman[115], J. Kennington[7], F. A. Kerkow[18], R. Kesharwani[79], J. S. Key[241], R. Khadela[8,9], S. Khadka[89], S. S. Khadkikar[7], F. Y. Khalili[108], F. Khan[8,9], T. Khanam[162], M. Khursheed[103], N. M. Khusid[191,192], W. Kiendrebeogo[113,242], N. Kijbunchoo[115], C. Kim[243],

J. C. Kim[244], K. Kim[245], M. H. Kim[237], S. Kim[246], Y.-M. Kim[245], C. Kimball[96], K. Kimes[54], M. Kinnear[33], J. S. Kissel[2], S. Klimenko[46], A. M. Knee[114], E. J. Knox[77], N. Knust[8,9], K. Kobayashi[204], S. M. Koehlenbeck[89], G. Koekoek[37,36], K. Kohri[247,248], K. Kokeyama[33,249], S. Koley[44,165], P. Kolitsidou[118], A. E. Koloniari[250], K. Komori[42], A. K. H. Kong[142], A. Kontos[251], L. M. Koponen[118], M. Korobko[97], X. Kou[18], A. Koushik[23], N. Kouvatsos[67], M. Kovalam[72], T. Koyama[151], D. B. Kozak[11], S. L. Kranzhoff[36,37], V. Kringel[8,9], N. V. Krishnendu[118], S. Kroker[252], A. Królak[186,253], K. Kruska[8,9], J. Kubisz[254], G. Kuehn[8,9], S. Kulkarni[215], A. Kulur Ramamohan[34], Achal Kumar[46], Anil Kumar[232], Praveen Kumar[177], Prayush Kumar[24], Rahul Kumar[2], Rakesh Kumar[93], J. Kume[255,256,42], K. Kuns[35], N. Kuntimaddi[33], S. Kuroyanagi[207,257], S. Kuwahara[42], K. Kwak[238], K. Kwan[34], S. Kwon[42], G. Lacaille[86], D. Laghi[188,100], A. H. Laity[163], E. Lalande[258], M. Lalleman[23], P. C. Lalremruati[259], M. Landry[2], B. B. Lane[35], R. N. Lang[35], J. Lange[120], R. Langgin[212], B. Lantz[89], I. L. Rosa[98], J. Larsen[200], A. Lartaux-Vollard[41], P. D. Lasky[6], J. Lawrence[164], M. Laxen[63], C. Lazarte[138], A. Lazzarini[11], C. Lazzaro[155,156], P. Leaci[38,39], L. Leali[18], Y. K. Lecoeuche[114], H. M. Lee[260], H. W. Lee[261], J. Lee[78], K. Lee[237], R.-K. Lee[142], R. Lee[35], Sungho Lee[245], Sunjae Lee[237], Y. Lee[141], I. N. Legred[11], J. Lehmann[8,9], L. Lehner[182], M. Le Jean[175,116], A. Lemaître[262], M. Lenti[61,176], M. Leonardi[74,75,263], M. Lequime[40], N. Leroy[41], M. Lesovsky[11], N. Letendre[31], M. Lethuillier[56], Y. Levin[6], K. Leyde[73], A. K. Y. Li[11], K. L. Li[264], T. G. F. Li[109], X. Li[148], Y. Li[96], Z. Li[86], A. Lihos[123], E. T. Lin[142], F. Lin[141], L. C.-C. Lin[264], Y.-C. Lin[142], C. Lindsay[201], S. D. Linker[180], A. Liu[218], G. C. Liu[227], J. Liu[72], F. Llamas Villarreal[164], J. Llobera-Querol[98], R. K. L. Lo[140], J.-P. Locquet[109], S. C. G. Loggins[265], M. R. Loizou[133], L. T. London[67], A. Longo[60,61], D. Lopez[165], M. Lopez Portilla[71], A. Lorenzo-Medina[177], V. Loriette[41], M. Lormand[63], G. Losurdo[266,80], E. Lotti[133], T. P. Lott IV[57], J. D. Lough[8,9], H. A. Loughlin[35], C. O. Lousto[110], N. Low[124], N. Lu[34], L. Lucchesi[80], H. Lück[9,8,9], D. Lumaca[22], A. P. Lundgren[267,268], A. W. Lussier[258], R. Macas[73], M. MacInnis[35], D. M. Macleod[33], I. A. O. MacMillan[11], A. Macquet[41], K. Maeda[151], S. Maenaut[109], S. S. Magare[79], R. M. Magee[11], E. Maggio[1], R. Maggiore[37,107], M. Magnozzi[29,30], M. Mahesh[97], M. Maini[163], S. Majhi[79], E. Majorana[39,38], C. N. Makarem[11], D. Malakar[105], J. A. Malaquias-Reis[19], U. Mali[190], S. Maliakal[11], A. Malik[103], L. Mallick[168,190], A.-K. Malz[58], N. Man[113], M. Mancarella[99], V. Mandic[18], V. Mangano[155,170], B. Mannix[77], G. L. Mansell[78], M. Manske[10], M. Mantovani[62], M. Mapelli[91,92,269], C. Marinelli[101], F. Marion[31], A. S. Markosyan[89], A. Markowitz[11], E. Maros[11], S. Marsat[100], F. Martelli[60,61], I. W. Martin[86], R. M. Martin[193], B. B. Martinez[131], D. A. Martinez[54], M. Martinez[43,270], V. Martinez[129], A. Martini[74,75], J. C. Martins[19], D. V. Martynov[118], E. J. Marx[35], L. Massaro[36,37], A. Masserot[31], M. Masso-Reid[86], S. Mastrogiovanni[38], T. Matcovich[51], M. Matiushechkina[8,9], L. Maurin[208], N. Mavalvala[35], N. Maxwell[2], G. McCarrol[63], R. McCarthy[2], D. E. McClelland[34], S. McCormick[63], L. McCuller[11], S. McEachin[123], C. McElhenny[123], G. I. McGhee[86], J. McGinn[86], K. B. M. McGowan[144], J. McIver[114], A. McLeod[72], I. McMahon[188], T. McRae[34], R. McTeague[86], D. Meacher[10], B. N. Meagher[78], R. Mechum[110], Q. Meijer[71], A. Melatos[124], C. S. Menoni[137], F. Mera[2], R. A. Mercer[10], L. Mereni[175], K. Merfeld[162], E. L. Merilh[63], J. R. Mérou[98], J. D. Merritt[77], M. Merzougui[113], C. Messick[10], B. Mestichelli[44], M. Meyer-Conde[271], F. Meylahn[8,9], A. Mhaske[79], A. Miani[74,75], H. Miao[272], C. Michel[175], Y. Michimura[42], H. Middleton[118], D. P. Mihaylov[104], S. J. Miller[11], M. Millhouse[57], E. Milotti[184,48], V. Milotti[91], Y. Minenkov[22], E. M. Minihan[65], Ll. M. Mir[43], L. Mirasola[155,156], M. Miravet-Tenés[138], C.-A. Miritescu[43], A. Mishra[24], C. Mishra[106], T. Mishra[46], A. L. Mitchell[37,107], J. G. Mitchell[65], S. Mitra[79], V. P. Mitrofanov[108], K. Mitsuhashi[25], R. Mittleman[35], O. Miyakawa[50], S. Miyoki[50], A. Miyoko[65], G. Mo[35], L. Mobilia[60,61], S. R. P. Mohapatra[11], S. R. Mohite[7], M. Molina-Ruiz[318], M. Mondin[180], J. K. Monsalve[189], M. Montani[60,61], C. J. Moore[223], D. Moraru[2], A. More[79], S. More[79], C. Moreno[135], E. A. Moreno[35], G. Moreno[2], A. Moreso Serra[82], S. Morisaki[42,204], Y. Moriwaki[151], G. Morras[207], A. Moscatello[91], M. Mould[35], B. Mours[64], C. M. Mow-Lowry[37,107], L. Muccillo[176,61], F. Muciaccia[39,38], D. Mukherjee[118], Samanwaya Mukherjee[24], Soma Mukherjee[164], Subroto Mukherjee[93], Suvodip Mukherjee[13], N. Mukund[35], A. Mullavey[63], H. Mullock[114], J. Mundi[219], C. L. Mungioli[72], M. Murakoshi[229], P. G. Murray[86], D. Nabari[74,75], S. L. Nadji[8,9], A. Nagar[28,273], N. Nagarajan[86], K. Nakagaki[50], K. Nakamura[25], H. Nakano[274], M. Nakano[11], D. Nanadoumgar-Lacroze[43], D. Nandi[12], V. Napolano[62], P. Narayan[215], I. Nardecchia[22], T. Narikawa[204], H. Narola[71], L. Naticchioni[38], R. K. Nayak[259], L. Negri[71], A. Nela[86], C. Nelle[77], A. Nelson[131], T. J. N. Nelson[63], M. Nery[8,9], A. Neunzert[2], S. Ng[54], L. Nguyen Quynh[275], S. A. Nichols[12], A. B. Nielsen[276], Y. Nishino[25,42], A. Nishizawa[277], S. Nissanke[278,37], W. Niu[7], F. Nocera[62], J. Noller[279], M. Norman[33], C. North[33], J. Novak[116,233,280], R. Nowicki[144], J. F. Nuño Siles[207], L. K. Nuttall[73], K. Obayashi[229], J. Oberling[2], J. O'Dell[228], E. Oelker[35], M. Oertel[233,116,281,280], G. Oganesyan[44,45], T. O'Hanlon[63], M. Ohashi[50], F. Ohme[8,9], R. Oliveri[116,281,280], R. Omer[18], B. O'Neal[123], M. Onishi[151], K. Oohara[282], B. O'Reilly[63], M. Orselli[51,76], R. O'Shaughnessy[110], S. O'Shea[86], S. Oshino[50], C. Osthelder[11], I. Ota[12], D. J. Ottaway[115], A. Ouzriat[56], H. Overmier[63], B. J. Owen[283], R. Ozaki[229], A. E. Pace[7], R. Pagano[12], M. A. Page[25], A. Pai[195], L. Paiella[44], A. Pal[284], S. Pal[259], M. A. Palaia[80,81], M. Pálfi[203], P. P. Palma[39,21,22], C. Palomba[38], P. Palud[20], H. Pan[142], J. Pan[72], K. C. Pan[142], P. K. Panda[232], Shiksha Pandey[7], Swadha Pandey[35], P. T. H. Pang[37,71], F. Pannarale[39,38], K. A. Pannone[54], B. C. Pant[103], F. H. Panther[72], M. Panzeri[60,61], F. Paoletti[80], A. Paolone[38,285], A. Papadopoulos[86], E. E. Papalexakis[210], L. Papalini[80,81], G. Papigkiotis[250], A. Paquis[41], A. Parisi[76,51], B.-J. Park[245], J. Park[286], W. Parker[63], G. Pascale[8,9], D. Pascucci[94], A. Pasqualetti[62], R. Passaquieti[81,80], L. Passenger[6], D. Passuello[80], O. Patane[2], A. V. Patel[141],

D. Pathak[79], A. Patra[33], B. Patricelli[81,80], B. G. Patterson[33], K. Paul[106], S. Paul[77], E. Payne[11], T. Pearce[33], M. Pedraza[11], A. Pele[11], F. E. Peña Arellano[287], X. Peng[118], Y. Peng[57], S. Penn[288], M. D. Penuliar[54], A. Perego[74,75], Z. Pereira[133], C. Périgois[289,92,91], G. Perna[91], A. Perreca[74,75,44], J. Perret[20], S. Perriès[56], J. W. Perry[37,107], D. Pesios[250], S. Peters[165], S. Petracca[206], C. Petrillo[76], H. P. Pfeiffer[1], H. Pham[63], K. A. Pham[18], K. S. Phukon[118], H. Phurailatpam[218], M. Piarulli[100], L. Piccari[39,38], O. J. Piccinni[34], M. Pichot[113], M. Piendibene[81,80], F. Piergiovanni[60,61], L. Pierini[38], G. Pierra[38], V. Pierro[290,132], M. Pietrzak[95], M. Pillas[165], F. Pilo[80], L. Pinard[175], I. M. Pinto[290,132,291,32], M. Pinto[62], B. J. Piotrzkowski[10], M. Pirello[2], M. D. Pitkin[223,86], A. Placidi[51], E. Placidi[39,38], M. L. Planas[98], W. Plastino[211,22], C. Plunkett[35], R. Poggiani[81,80], E. Polini[35], J. Pomper[80,81], L. Pompili[1], J. Poon[218], E. Porcelli[37], E. K. Porter[20], C. Posnansky[7], R. Poulton[62], J. Powell[154], G. S. Prabhu[79], M. Pracchia[165], B. K. Pradhan[79], T. Pradier[64], A. K. Prajapati[93], K. Prasai[292], R. Prasanna[232], P. Prasia[79], G. Pratten[118], G. Principe[184,48], G. A. Prodi[74,75], P. Prosperi[80], P. Prosposito[21,22], A. C. Providence[65], A. Puecher[1], J. Pullin[12], P. Puppo[38], M. Pürrer[163], H. Qi[16], J. Qin[34], G. Quéméner[173,116], V. Quetschke[164], L. H. Quiceno[189], P. J. Quinonez[65], N. Qutob[57], R. Rading[230], I. Rainho[138], S. Raja[103], C. Rajan[103], B. Rajbhandari[110], K. E. Ramirez[63], F. A. Ramis Vidal[98], M. Ramos Arevalo[164], A. Ramos-Buades[98,37], S. Ranjan[57], K. Ransom[63], P. Rapagnani[39,38], B. Ratto[65], A. Ravichandran[133], A. Ray[96], V. Raymond[33], M. Razzano[81,80], J. Read[54], T. Regimbau[31], S. Reid[55], C. Reissel[35], D. H. Reitze[11], A. I. Renzini[11,127], B. Revenu[293,41], A. Revilla Pe na[82], R. Reyes[180], L. Ricca[15], F. Ricci[39,38], M. Ricci[38,39], A. Ricciardone[81,80], J. Rice[78], J. W. Richardson[210], M. L. Richardson[115], A. Rijal[65], K. Riles[90], H. K. Riley[33], S. Rinaldi[269], J. Rittmeyer[97], C. Robertson[228], F. Robinet[41], M. Robinson[2], A. Rocchi[22], L. Rolland[31], J. G. Rollins[11], A. E. Romano[189], R. Romano[3,4], A. Romero[31], I. M. Romero-Shaw[223], J. H. Romie[63], S. Ronchini[7], T. J. Roocke[115], L. Rosa[4,32], T. J. Rosauer[210], C. A. Rose[57], D. Rosińska[125], M. P. Ross[53], M. Rossello-Sastre[98], S. Rowan[86], S. K. Roy[191,192], S. Roy[15], D. Rozza[127,128], P. Ruggi[62], N. Ruhama[238], E. Ruiz Morales[294,207], K. Ruiz-Rocha[144], S. Sachdev[57], T. Sadecki[2], P. Saffarieh[37,107], S. Safi-Harb[168], M. R. Sah[13], S. Saha[142], T. Sainrat[64], S. Sajith Menon[214,39,38], K. Sakai[295], Y. Sakai[271], M. Sakellariadou[67], S. Sakon[7], O. S. Salafia[158,128,127], F. Salces-Carcoba[11], L. Salconi[62], M. Saleem[120], F. Salemi[39,38], M. Sallé[37], S. U. Salunkhe[79], S. Salvador[173,172], A. Salvarese[120], A. Samajdar[71,37], A. Sanchez[2], E. J. Sanchez[11], L. E. Sanchez[11], N. Sanchis-Gual[138], J. R. Sanders[181], E. M. Sänger[1], F. Santoliquido[44,45], F. Sarandrea[28], T. R. Saravanan[79], N. Sarin[6], P. Sarkar[8,9], A. Sasli[250], P. Sassi[51,76], B. Sassolas[175], B. S. Sathyaprakash[7,33], R. Sato[226], S. Sato[151], Yukino Sato[151], Yu Sato[151], O. Sauter[46], R. L. Savage[2], T. Sawada[50], H. L. Sawant[79], S. Sayah[175], V. Scacco[21,22], D. Schaetzl[11], M. Scheel[148], A. Schiebelbein[190], M. G. Schiworski[78], P. Schmidt[118], S. Schmidt[71], R. Schnabel[97], M. Schneewind[8,9], R. M. S. Schofield[77], K. Schouteden[109], B. W. Schulte[8,9], B. F. Schutz[33,8,9], E. Schwartz[296], M. Scialpi[297], J. Scott[86], S. M. Scott[34], R. M. Sedas[63], T. C. Seetharamu[86], M. Seglar-Arroyo[43], Y. Sekiguchi[298], D. Sellers[63], N. Sembo[205], A. S. Sengupta[299], E. G. Seo[86], J. W. Seo[109], V. Sequino[32,4], M. Serra[38], A. Sevrin[187], T. Shaffer[2], U. S. Shah[57], M. A. Shaikh[260], L. Shao[300], A. K. Sharma[98], Preeti Sharma[12], Prianka Sharma[103], Ritwik Sharma[18], S. Sharma Chaudhary[105], P. Shawhan[126], N. S. Shcheblanov[301,262], E. Sheridan[144], Z.-H. Shi[142], M. Shikauchi[42], R. Shimomura[302], H. Shinkai[302], S. Shirke[79], D. H. Shoemaker[35], D. M. Shoemaker[120], R. W. Short[2], S. ShyamSundar[103], A. Sider[157], H. Siegel[191,192], D. Sigg[2], L. Silenzi[36,37], L. Silvestri[39,169], M. Simmonds[115], L. P. Singer[303], Amitesh Singh[215], Anika Singh[11], D. Singh[318], N. Singh[98], S. Singh[216,59], A. M. Sintes[98], V. Sipala[155,170], V. Skliris[33], B. J. J. Slagmolen[34], D. A. Slater[200], T. J. Slaven-Blair[72], J. Smetana[118], J. R. Smith[54], L. Smith[86,184,48], R. J. E. Smith[6], W. J. Smith[144], S. Soares de Albuquerque Filho[60], M. Soares-Santos[188], K. Somiya[216], I. Song[142], S. Soni[35], V. Sordini[56], F. Sorrentino[29], H. Sotani[304], F. Spada[80], V. Spagnuolo[37], A. P. Spencer[86], P. Spinicelli[62], A. K. Srivastava[93], F. Stachurski[86], C. J. Stark[123], D. A. Steer[305], N. Steinle[168], J. Steinlechner[36,37], S. Steinlechner[36,37], N. Stergioulas[250], P. Stevens[41], M. St. Pierre[163], M. D. Strong[12], A. Strunk[2], A. L. Stuver[102], M. Suchenek[95], S. Sudhagar[95], Y. Sudo[229], N. Sueltmann[97], L. Suleiman[54], K. D. Sullivan[12], J. Sun[240], L. Sun[34], S. Sunil[93], J. Suresh[113], B. J. Sutton[67], P. J. Sutton[33], K. Suzuki[216], M. Suzuki[204], B. L. Swinkels[37], A. Syx[116], M. J. Szczepańczyk[306], P. Szewczyk[125], M. Tacca[37], H. Tagoshi[204], K. Takada[204], H. Takahashi[271], R. Takahashi[25], A. Takamori[42], S. Takano[307], H. Takeda[308,309], K. Takeshita[216], I. Takimoto Schmiegelow[44,45], M. Takou-Ayaoh[78], C. Talbot[130], M. Tamaki[204], N. Tamanini[100], D. Tanabe[141], K. Tanaka[50], S. J. Tanaka[229], S. Tanioka[33], D. B. Tanner[46], W. Tanner[8,9], L. Tao[210], R. D. Tapia[7], E. N. Tapia San Martín[37], C. Taranto[21,22], A. Taruya[310], J. D. Tasson[152], J. G. Tau[110], D. Tellez[54], R. Tenorio[98], H. Themann[180], A. Theodoropoulos[138], M. P. Thirugnanasambandam[79], L. M. Thomas[11], M. Thomas[63], P. Thomas[2], J. E. Thompson[209], S. R. Thondapu[103], K. A. Thorne[63], E. Thrane[6], J. Tissino[44,45], A. Tiwari[79], Pawan Tiwari[44], Praveer Tiwari[195], S. Tiwari[188], V. Tiwari[118], M. R. Todd[78], M. Toffano[91], A. M. Toivonen[18], K. Toland[86], A. E. Tolley[73], T. Tomaru[25], V. Tommasini[11], T. Tomura[50], H. Tong[6], C. Tong-Yu[141], A. Torres-Forné[138,139], C. I. Torrie[11], I. Tosta e Melo[311], E. Tournefier[31], M. Trad Nery[113], K. Tran[123], A. Trapananti[52,51], R. Travaglini[167], F. Travasso[52,51], G. Traylor[63], M. Trevor[126], M. C. Tringali[62], A. Tripathee[90], G. Troian[184,48], A. Trovato[184,48], L. Trozzo[4], R. J. Trudeau[11], T. Tsang[33], S. Tsuchida[312], L. Tsukada[212], K. Turbang[187,23], M. Turconi[113], C. Turski[94], H. Ubach[82,83], N. Uchikata[204], T. Uchiyama[50], R. P. Udall[11], T. Uehara[313], K. Ueno[42], V. Undheim[276], L. E. Uronen[218], T. Ushiba[50], M. Vacatello[80,81], H. Vahlbruch[8,9], N. Vaidya[11], G. Vajente[11], A. Vajpeyi[6],

J. Valencia[98], M. Valentini[107,37], S. A. Vallejo Peña[189], S. Vallero[28], V. Valsan[10], M. van Dael[37,314], E. Van den Bossche[187], J. F. J. van den Brand[36,107,37], C. Van Den Broeck[71,37], M. van der Sluys[37,71], A. Van de Walle[41], J. van Dongen[37,107], K. Vandra[102], M. VanDyke[119], H. van Haevermaet[23], J. V. van Heijningen[37,107], P. Van Hove[64], J. Vanier[258], M. Van Keuren[104], J. Vanosky[2], N. van Remortel[23], M. Vardaro[36,37], A. F. Vargas[124], V. Varma[133], A. N. Vazquez[89], A. Vecchio[118], G. Vedovato[92], J. Veitch[86], P. J. Veitch[115], S. Venikoudis[15], R. C. Venterea[18], P. Verdier[56], M. Vereecken[15], D. Verkindt[31], B. Verma[133], Y. Verma[103], S. M. Vermeulen[11], F. Vetrano[60], A. Veutro[38,39], A. Viceré[60,61], S. Vidyant[78], A. D. Viets[88], A. Vijaykumar[190], A. Vilkha[110], N. Villanueva Espinosa[138], V. Villa-Ortega[177], E. T. Vincent[57], J.-Y. Vinet[113], S. Viret[56], S. Vitale[35], H. Vocca[76,51], D. Voigt[97], E. R. G. von Reis[2], J. S. A. von Wrangel[8,9], W. E. Vossius[230], L. Vujeva[140], S. P. Vyatchanin[108], J. Wack[11], L. E. Wade[104], M. Wade[104], K. J. Wagner[110], L. Wallace[11], E. J. Wang[89], H. Wang[216], J. Z. Wang[90], W. H. Wang[164], Y. F. Wang[1], G. Waratkar[195], J. Warner[2], M. Was[31], T. Washimi[25], N. Y. Washington[11], D. Watarai[42], B. Weaver[2], S. A. Webster[86], N. L. Weickhardt[97], M. Weinert[8,9], A. J. Weinstein[11], R. Weiss[35], L. Wen[72], K. Wette[34], J. T. Whelan[110], B. F. Whiting[46], C. Whittle[11], E. G. Wickens[73], D. Wilken[8,9], A. T. Wilkin[210], B. M. Williams[119], D. Williams[86], M. J. Williams[73], N. S. Williams[1], J. L. Willis[11], B. Willke[9,8,9], M. Wils[109], L. Wilson[104], C. W. Winborn[105], J. Winterflood[72], C. C. Wipf[11], G. Woan[86], J. Woehler[36,37], N. E. Wolfe[35], H. T. Wong[141], I. C. F. Wong[218,109], K. Wong[190], T. Wouters[71,37], J. L. Wright[2], M. Wright[86,71], B. Wu[78], C. Wu[142], D. S. Wu[8,9], H. Wu[142], K. Wu[119], Q. Wu[53], Y. Wu[96], Z. Wu[100], E. Wuchner[54], D. M. Wysocki[10], V. A. Xu[318], Y. Xu[98], N. Yadav[28], H. Yamamoto[11], K. Yamamoto[151], T. S. Yamamoto[42], T. Yamamoto[50], R. Yamazaki[229], T. Yan[118], K. Z. Yang[18], Y. Yang[146], Z. Yarbrough[12], J. Yebana[98], S.-W. Yeh[142], A. B. Yelikar[144], X. Yin[35], J. Yokoyama[315,42], T. Yokozawa[50], S. Yuan[72], H. Yuzurihara[50], M. Zanolin[65], M. Zeeshan[110], T. Zelenova[62], J.-P. Zendri[92], M. Zeoli[15], M. Zerrad[40], M. Zevin[96], L. Zhang[11], N. Zhang[57], R. Zhang[149], T. Zhang[118], C. Zhao[72], Yue Zhao[161], Yuhang Zhao[20], Z.-C. Zhao[316], Y. Zheng[105], H. Zhong[18], H. Zhou[78], H. O. Zhu[72], Z.-H. Zhu[316,317], A. B. Zimmerman[120], L. Zimmermann[56], M. E. Zucker[35,11], J. Zweizig[11]

(The LIGO Scientific Collaboration, The Virgo Collaboration, and The KAGRA Collaboration)

[1] Max Planck Institute for Gravitational Physics (Albert Einstein Institute), D-14476 Potsdam, Germany
[2] LIGO Hanford Observatory, Richland, WA 99352, USA
[3] Dipartimento di Farmacia, Università di Salerno, I-84084 Fisciano, Salerno, Italy
[4] INFN Sezione di Napoli, I-80126 Napoli, Italy
[5] University of Warwick, Coventry CV4 7AL, UK
[6] OzGrav, School of Physics & Astronomy, Monash University, Clayton 3800, Victoria, Australia
[7] The Pennsylvania State University, University Park, PA 16802, USA
[8] Max Planck Institute for Gravitational Physics (Albert Einstein Institute), D-30167 Hannover, Germany
[9] Leibniz Universität Hannover, D-30167 Hannover, Germany
[10] University of Wisconsin-Milwaukee, Milwaukee, WI 53201, USA
[11] LIGO Laboratory, California Institute of Technology, Pasadena, CA 91125, USA
[12] Louisiana State University, Baton Rouge, LA 70803, USA
[13] Tata Institute of Fundamental Research, Mumbai 400005, India
[14] Centre de Physique Théorique, Aix-Marseille Université, Campus de Luminy, 163 Av. de Luminy, 13009 Marseille, France
[15] Université catholique de Louvain, B-1348 Louvain-la-Neuve, Belgium
[16] Queen Mary University of London, London E1 4NS, UK
[17] University of California, Davis, Davis, CA 95616, USA
[18] University of Minnesota, Minneapolis, MN 55455, USA
[19] Instituto Nacional de Pesquisas Espaciais, 12227-010 São José dos Campos, São Paulo, Brazil
[20] Université Paris Cité, CNRS, Astroparticule et Cosmologie, F-75013 Paris, France
[21] Università di Roma Tor Vergata, I-00133 Roma, Italy
[22] INFN Sezione di Roma, Tor Vergata, I-00133 Roma, Italy
[23] Universiteit Antwerpen, 2000 Antwerpen, Belgium
[24] International Centre for Theoretical Sciences, Tata Institute of Fundamental Research, Bengaluru 560089, India
[25] Gravitational Wave Science Project, National Astronomical Observatory of Japan, 2-21-1 Osawa, Mitaka City, Tokyo 181-8588, Japan
[26] Advanced Technology Center, National Astronomical Observatory of Japan, 2-21-1 Osawa, Mitaka City, Tokyo 181-8588, Japan
[27] Theoretisch-Physikalisches Institut, Friedrich-Schiller-Universität Jena, D-07743 Jena, Germany
[28] INFN Sezione di Torino, I-10125 Torino, Italy
[29] INFN Sezione di Genova, I-16146 Genova, Italy
[30] Dipartimento di Fisica, Università degli Studi di Genova, I-16146 Genova, Italy
[31] Université Savoie Mont Blanc, CNRS, Laboratoire d'Annecy de Physique des Particules - IN2P3, F-74000 Annecy, France
[32] Università di Napoli "Federico II," I-80126 Napoli, Italy
[33] Cardiff University, Cardiff CF24 3AA, UK
[34] OzGrav, Australian National University, Canberra, Australian Capital Territory 0200, Australia
[35] LIGO Laboratory, Massachusetts Institute of Technology, Cambridge, MA 02139, USA
[36] Maastricht University, 6200 MD Maastricht, The Netherlands
[37] Nikhef, 1098 XG Amsterdam, The Netherlands
[38] INFN Sezione di Roma, I-00185 Roma, Italy
[39] Università di Roma "La Sapienza," I-00185 Roma, Italy
[40] Aix Marseille Université, CNRS, Centrale Med, Institut Fresnel, F-13013 Marseille, France
[41] Université Paris-Saclay, CNRS/IN2P3, IJCLab, 91405 Orsay, France
[42] University of Tokyo, Tokyo, 113-0033, Japan
[43] Institut de Física d'Altes Energies (IFAE), The Barcelona Institute of Science and Technology, Campus UAB, E-08193 Bellaterra (Barcelona), Spain

[44] Gran Sasso Science Institute (GSSI), I-67100 L'Aquila, Italy
[45] INFN, Laboratori Nazionali del Gran Sasso, I-67100 Assergi, Italy
[46] University of Florida, Gainesville, FL 32611, USA
[47] Dipartimento di Scienze Matematiche, Informatiche e Fisiche, Università di Udine, I-33100 Udine, Italy
[48] INFN Sezione di Trieste, I-34127 Trieste, Italy
[49] Tecnologico de Monterrey, Escuela de Ingeniería y Ciencias, 64849, Monterrey, Nuevo León, Mexico
[50] Institute for Cosmic Ray Research, KAGRA Observatory, The University of Tokyo, 238 Higashi-Mozumi, Kamioka-cho, Hida City, Gifu 506-1205, Japan
[51] INFN Sezione di Perugia, I-06123 Perugia, Italy
[52] Università di Camerino, I-62032 Camerino, Italy
[53] University of Washington, Seattle, WA 98195, USA
[54] California State University, Fullerton, Fullerton, CA 92831, USA
[55] SUPA, University of Strathclyde, Glasgow G1 1XQ, UK
[56] Université Claude Bernard Lyon 1, CNRS, IP2I Lyon/IN2P3, UMR 5822, F-69622 Villeurbanne, France
[57] Georgia Institute of Technology, Atlanta, GA 30332, USA
[58] Royal Holloway, University of London, London TW20 0EX, UK
[59] The Graduate University for Advanced Studies (SOKENDAI), 2-21-1 Osawa, Mitaka City, Tokyo 181-8588, Japan
[60] Università degli Studi di Urbino "Carlo Bo," I-61029 Urbino, Italy
[61] INFN Sezione di Firenze, I-50019 Sesto Fiorentino, Firenze, Italy
[62] European Gravitational Observatory (EGO), I-56021 Cascina, Pisa, Italy
[63] LIGO Livingston Observatory, Livingston, LA 70754, USA
[64] Université de Strasbourg, CNRS, IPHC UMR 7178, F-67000 Strasbourg, France
[65] Embry-Riddle Aeronautical University, Prescott, AZ 86301, USA
[66] Dipartimento di Fisica "E.R. Caianiello," Università di Salerno, I-84084 Fisciano, Salerno, Italy
[67] King's College London, University of London, London WC2R 2LS, UK
[68] Korea Institute of Science and Technology Information, Daejeon 34141, Republic of Korea
[69] International College, Osaka University, 1-1 Machikaneyama-cho, Toyonaka City, Osaka 560-0043, Japan
[70] Accelerator Laboratory, High Energy Accelerator Research Organization (KEK), 1-1 Oho, Tsukuba City, Ibaraki 305-0801, Japan
[71] Institute for Gravitational and Subatomic Physics (GRASP), Utrecht University, 3584 CC Utrecht, The Netherlands
[72] OzGrav, University of Western Australia, Crawley, Western Australia 6009, Australia
[73] University of Portsmouth, Portsmouth, PO1 3FX, UK
[74] Università di Trento, Dipartimento di Fisica, I-38123 Povo, Trento, Italy
[75] INFN, Trento Institute for Fundamental Physics and Applications, I-38123 Povo, Trento, Italy
[76] Università di Perugia, I-06123 Perugia, Italy
[77] University of Oregon, Eugene, OR 97403, USA
[78] Syracuse University, Syracuse, NY 13244, USA
[79] Inter-University Centre for Astronomy and Astrophysics, Pune 411007, India
[80] INFN, Sezione di Pisa, I-56127 Pisa, Italy
[81] Università di Pisa, I-56127 Pisa, Italy
[82] Institut de Ciències del Cosmos (ICCUB), Universitat de Barcelona (UB), c. Martí i Franquès, 1, 08028 Barcelona, Spain
[83] Departament de Física Quàntica i Astrofísica (FQA), Universitat de Barcelona (UB), c. Martí i Franqués, 1, 08028 Barcelona, Spain
[84] Institut d'Estudis Espacials de Catalunya, c. Gran Capità, 2-4, 08034 Barcelona, Spain
[85] Dipartimento di Medicina, Chirurgia e Odontoiatria "Scuola Medica Salernitana," Università di Salerno, I-84081 Baronissi, Salerno, Italy
[86] IGR, University of Glasgow, Glasgow G12 8QQ, UK
[87] HUN-REN Wigner Research Centre for Physics, H-1121 Budapest, Hungary
[88] Concordia University Wisconsin, Mequon, WI 53097, USA
[89] Stanford University, Stanford, CA 94305, USA
[90] University of Michigan, Ann Arbor, MI 48109, USA
[91] Università di Padova, Dipartimento di Fisica e Astronomia, I-35131 Padova, Italy
[92] INFN Sezione di Padova, I-35131 Padova, Italy
[93] Institute for Plasma Research, Bhat, Gandhinagar 382428, India
[94] Universiteit Gent, B-9000 Gent, Belgium
[95] Nicolaus Copernicus Astronomical Center, Polish Academy of Sciences, 00-716, Warsaw, Poland
[96] Northwestern University, Evanston, IL 60208, USA
[97] Universität Hamburg, D-22761 Hamburg, Germany
[98] IAC3–IEEC, Universitat de les Illes Balears, E-07122 Palma de Mallorca, Spain
[99] Aix-Marseille Université, Université de Toulon, CNRS, CPT, Marseille, France
[100] Laboratoire des 2 Infinis - Toulouse (L2IT-IN2P3), F-31062 Toulouse Cedex 9, France
[101] Università di Siena, Dipartimento di Scienze Fisiche, della Terra e dell'Ambiente, I-53100 Siena, Italy
[102] Villanova University, Villanova, PA 19085, USA
[103] RRCAT, Indore, Madhya Pradesh 452013, India
[104] Kenyon College, Gambier, OH 43022, USA
[105] Missouri University of Science and Technology, Rolla, MO 65409, USA
[106] Indian Institute of Technology Madras, Chennai 600036, India
[107] Department of Physics and Astronomy, Vrije Universiteit Amsterdam, 1081 HV Amsterdam, The Netherlands
[108] Lomonosov Moscow State University, Moscow 119991, Russia
[109] Katholieke Universiteit Leuven, Oude Markt 13, 3000 Leuven, Belgium
[110] Rochester Institute of Technology, Rochester, NY 14623, USA
[111] Université libre de Bruxelles, 1050 Bruxelles, Belgium
[112] Bar-Ilan University, Ramat Gan, 5290002, Israel
[113] Université Côte d'Azur, Observatoire de la Côte d'Azur, CNRS, Artemis, F-06304 Nice, France
[114] University of British Columbia, Vancouver, BC V6T 1Z4, Canada
[115] OzGrav, University of Adelaide, Adelaide, South Australia 5005, Australia
[116] Centre national de la recherche scientifique, 75016 Paris, France
[117] Université Rennes, CNRS, Institut FOTON - UMR 6082, F-35000 Rennes, France
[118] University of Birmingham, Birmingham B15 2TT, UK
[119] Washington State University, Pullman, WA 99164, USA

[120] University of Texas, Austin, TX 78712, USA
[121] Cornell University, Ithaca, NY 14850, USA
[122] Laboratoire Kastler Brossel, Sorbonne Université, CNRS, ENS-Université PSL, Collège de France, F-75005 Paris, France
[123] Christopher Newport University, Newport News, VA 23606, USA
[124] OzGrav, University of Melbourne, Parkville, Victoria 3010, Australia
[125] Astronomical Observatory of the University of Warsaw, 00-478 Warsaw, Poland
[126] University of Maryland, College Park, MD 20742, USA
[127] Università degli Studi di Milano-Bicocca, I-20126 Milano, Italy
[128] INFN Sezione di Milano-Bicocca, I-20126 Milano, Italy
[129] Université de Lyon, Université Claude Bernard Lyon 1, CNRS, Institut Lumière Matière, F-69622 Villeurbanne, France
[130] University of Chicago, Chicago, IL 60637, USA
[131] University of Arizona, Tucson, AZ 85721, USA
[132] INFN Sezione di Napoli, Gruppo Collegato di Salerno, I-80126 Napoli, Italy
[133] University of Massachusetts Dartmouth, North Dartmouth, MA 02747, USA
[134] Niels Bohr Institute, Copenhagen University, 2100 København, Denmark
[135] Universidad de Guadalajara, 44430, Guadalajara, Jalisco, Mexico
[136] Istituto di Astrofisica e Planetologia Spaziali di Roma, 00133 Roma, Italy
[137] Colorado State University, Fort Collins, CO 80523, USA
[138] Departamento de Astronomía y Astrofísica, Universitat de València, E-46100 Burjassot, València, Spain
[139] Observatori Astronòmic, Universitat de València, E-46980 Paterna, València, Spain
[140] Niels Bohr Institute, University of Copenhagen, 2100 Kóbenhavn, Denmark
[141] National Central University, Taoyuan City 320317, Taiwan
[142] National Tsing Hua University, Hsinchu City 30013, Taiwan
[143] OzGrav, Charles Sturt University, Wagga Wagga, New South Wales 2678, Australia
[144] Vanderbilt University, Nashville, TN 37235, USA
[145] University of the Chinese Academy of Sciences/International Centre for Theoretical Physics Asia-Pacific, Beijing 100049, People's Republic of China
[146] Department of Electrophysics, National Yang Ming Chiao Tung University, 101 Univ. Street, Hsinchu, Taiwan
[147] Kamioka Branch, National Astronomical Observatory of Japan, 238 Higashi-Mozumi, Kamioka-cho, Hida City, Gifu 506-1205, Japan
[148] CaRT, California Institute of Technology, Pasadena, CA 91125, USA
[149] Northeastern University, Boston, MA 02115, USA
[150] Dipartimento di Ingegneria Industriale (DIIN), Università di Salerno, I-84084 Fisciano, Salerno, Italy
[151] Faculty of Science, University of Toyama, 3190 Gofuku, Toyama City, Toyama 930-8555, Japan
[152] Carleton College, Northfield, MN 55057, USA
[153] University of Szeged, Dóm tér 9, Szeged 6720, Hungary
[154] OzGrav, Swinburne University of Technology, Hawthorn VIC 3122, Australia
[155] INFN Sezione di Cagliari, Italy, and Physics Department, Università degli Studi di Cagliari, Cagliari 09042, Italy
[156] Università degli Studi di Cagliari, Via Università 40, 09124 Cagliari, Italy
[157] Université Libre de Bruxelles, Brussels 1050, Belgium
[158] INAF, Osservatorio Astronomico di Brera sede di Merate, I-23807 Merate, Lecco, Italy
[159] Departamento de Matemáticas, Universitat de València, E-46100 Burjassot, València, Spain
[160] Montana State University, Bozeman, MT 59717, USA
[161] The University of Utah, Salt Lake City, UT 84112, USA
[162] Johns Hopkins University, Baltimore, MD 21218, USA
[163] University of Rhode Island, Kingston, RI 02881, USA
[164] The University of Texas Rio Grande Valley, Brownsville, TX 78520, USA
[165] Université de Liège, B-4000 Liège, Belgium
[166] DIFA- Alma Mater Studiorum Università di Bologna, Via Zamboni, 33 - 40126 Bologna, Italy
[167] Istituto Nazionale di Fisica Nucleare Sezione di Bologna, viale Carlo Berti Pichat 6/2 - 40127 Bologna, Italy
[168] University of Manitoba, Winnipeg, MB R3T 2N2, Canada
[169] INFN-CNAF, Viale Carlo Berti Pichat, 6/2, 40127, Bologna BO, Italy
[170] Università degli Studi di Sassari, I-07100 Sassari, Italy
[171] INFN Laboratori Nazionali del Sud, I-95125 Catania, Italy
[172] Université de Normandie, ENSICAEN, UNICAEN, CNRS/IN2P3, LPC Caen, F-14000 Caen, France
[173] Laboratoire de Physique Corpusculaire Caen, 6 boulevard du Maréchal Juin, F-14050 Caen, France
[174] The University of Sheffield, Sheffield S10 2TN, UK
[175] Université Claude Bernard Lyon 1, CNRS, Laboratoire des Matériaux Avancés (LMA), IP2I Lyon / IN2P3, UMR 5822, F-69622 Villeurbanne, France
[176] Università di Firenze, Sesto Fiorentino I-50019, Italy
[177] IGFAE, Universidade de Santiago de Compostela, E-15782 Santiago de Compostela, Spain
[178] Dipartimento di Scienze Matematiche, Fisiche e Informatiche, Università di Parma, I-43124 Parma, Italy
[179] INFN Sezione di Milano-Bicocca, Gruppo Collegato di Parma, I-43124 Parma, Italy
[180] California State University, Los Angeles, Los Angeles, CA 90032, USA
[181] Marquette University, Milwaukee, WI 53233, USA
[182] Perimeter Institute, Waterloo, ON N2L 2Y5, Canada
[183] Corps des Mines, Mines Paris, Université PSL, 60 Bd Saint-Michel, 75272 Paris, France
[184] Dipartimento di Fisica, Università di Trieste, I-34127 Trieste, Italy
[185] Université Côte d'Azur, Observatoire de la Côte d'Azur, CNRS, Lagrange, F-06304 Nice, France
[186] National Center for Nuclear Research, 05-400 Świerk-Otwock, Poland
[187] Vrije Universiteit Brussel, 1050 Brussel, Belgium
[188] University of Zurich, Winterthurerstrasse 190, 8057 Zurich, Switzerland
[189] Universidad de Antioquia, Medellín, Colombia
[190] Canadian Institute for Theoretical Astrophysics, University of Toronto, Toronto, ON M5S 3H8, Canada
[191] Stony Brook University, Stony Brook, NY 11794, USA
[192] Center for Computational Astrophysics, Flatiron Institute, New York, NY 10010, USA
[193] Montclair State University, Montclair, NJ 07043, USA
[194] HUN-REN Institute for Nuclear Research, H-4026 Debrecen, Hungary
[195] Indian Institute of Technology Bombay, Powai, Mumbai 400 076, India

[196] Centro de Física, das Universidades do Minho e do Porto, Universidade do Minho, PT-4710-057 Braga, Portugal
[197] Aix Marseille Université, CNRS/IN2P3, CPPM, Marseille, France
[198] CNR-SPIN, I-84084 Fisciano, Salerno, Italy
[199] Scuola di Ingegneria, Università della Basilicata, I-85100 Potenza, Italy
[200] Western Washington University, Bellingham, WA 98225, USA
[201] SUPA, University of the West of Scotland, Paisley PA1 2BE, UK
[202] Barry University, Miami Shores, FL 33168, USA
[203] Eötvös University, Budapest 1117, Hungary
[204] Institute for Cosmic Ray Research, KAGRA Observatory, The University of Tokyo, 5-1-5 Kashiwa-no-Ha, Kashiwa City, Chiba 277-8582, Japan
[205] Department of Physics, Graduate School of Science, Osaka Metropolitan University, 3-3-138 Sugimoto-cho, Sumiyoshi-ku, Osaka City, Osaka 558-8585, Japan
[206] University of Sannio at Benevento, I-82100 Benevento, Italy and INFN Sezione di Napoli, I-80100 Napoli, Italy
[207] Instituto de Fisica Teorica UAM-CSIC, Universidad Autonoma de Madrid, 28049 Madrid, Spain
[208] Laboratoire d'Acoustique de l'Université du Mans, UMR CNRS 6613, F-72085 Le Mans, France
[209] University of Southampton, Southampton SO17 1BJ, UK
[210] University of California, Riverside, Riverside, CA 92521, USA
[211] Dipartimento di Ingegneria Industriale, Elettronica e Meccanica, Università degli Studi Roma Tre, I-00146 Roma, Italy
[212] University of Nevada, Las Vegas, Las Vegas, NV 89154, USA
[213] University of Nottingham, Nottingham NG7 2RD, UK
[214] Ariel University, Ramat HaGolan St 65, Ari'el, Israel
[215] The University of Mississippi, University, MS 38677, USA
[216] Graduate School of Science, Institute of Science Tokyo, 2-12-1 Ookayama, Meguro-ku, Tokyo 152-8551, Japan
[217] Institute of Physics, Academia Sinica, 128 Sec. 2, Academia Road, Nankang, Taipei 11529, Taiwan
[218] The Chinese University of Hong Kong, Shatin, NT, Hong Kong
[219] American University, Washington, DC 20016, USA
[220] Dipartimento di Fisica, Università degli studi di Milano, Via Celoria 16, I-20133, Milano, Italy
[221] INFN Sezione di Milano, Via Celoria 16, I-20133, Milano, Italy
[222] Department of Applied Physics, Fukuoka University, 8-19-1 Nanakuma, Jonan, Fukuoka City, Fukuoka 814-0180, Japan
[223] University of Cambridge, Cambridge CB2 1TN, UK
[224] University of Lancaster, Lancaster LA1 4YW, UK
[225] College of Industrial Technology, Nihon University, 1-2-1 Izumi, Narashino City, Chiba 275-8575, Japan
[226] Faculty of Engineering, Niigata University, 8050 Ikarashi-2-no-cho, Nishi-ku, Niigata City, Niigata 950-2181, Japan
[227] Department of Physics, Tamkang University, No. 151, Yingzhuan Road, Danshui Dist., New Taipei City 25137, Taiwan
[228] Rutherford Appleton Laboratory, Didcot OX11 0DE, UK
[229] Department of Physical Sciences, Aoyama Gakuin University, 5-10-1 Fuchinobe, Sagamihara City, Kanagawa 252-5258, Japan
[230] Helmut Schmidt University, D-22043 Hamburg, Germany
[231] Nambu Yoichiro Institute of Theoretical and Experimental Physics (NITEP), Osaka Metropolitan University, 3-3-138 Sugimoto-cho, Sumiyoshi-ku, Osaka City, Osaka 558-8585, Japan
[232] Directorate of Construction Services and Estate Management, Mumbai 400094, India
[233] Observatoire Astronomique de Strasbourg, 11 Rue de l'Université, 67000 Strasbourg, France
[234] Faculty of Physics, University of Białystok, 15-245 Białystok, Poland
[235] National Astronomical Observatories, Chinese Academic of Sciences, 20A Datun Road, Chaoyang District, Beijing, People's Republic of China
[236] School of Astronomy and Space Science, University of Chinese Academy of Sciences, 20A Datun Road, Chaoyang District, Beijing, People's Republic of China
[237] Sungkyunkwan University, Seoul 03063, Republic of Korea
[238] Department of Physics, Ulsan National Institute of Science and Technology (UNIST), 50 UNIST-gil, Ulju-gun, Ulsan 44919, Republic of Korea
[239] Institute for Cosmic Ray Research, The University of Tokyo, 5-1-5 Kashiwa-no-Ha, Kashiwa City, Chiba 277-8582, Japan
[240] Chung-Ang University, Seoul 06974, Republic of Korea
[241] University of Washington Bothell, Bothell, WA 98011, USA
[242] Laboratoire de Physique et de Chimie de l'Environnement, Université Joseph KI-ZERBO, 9GH2+3V5, Ouagadougou, Burkina Faso
[243] Ewha Womans University, Seoul 03760, Republic of Korea
[244] National Institute for Mathematical Sciences, Daejeon 34047, Republic of Korea
[245] Korea Astronomy and Space Science Institute, Daejeon 34055, Republic of Korea
[246] Department of Astronomy and Space Science, Chungnam National University, 9 Daehak-ro, Yuseong-gu, Daejeon 34134, Republic of Korea
[247] Institute of Particle and Nuclear Studies (IPNS), High Energy Accelerator Research Organization (KEK), 1-1 Oho, Tsukuba City, Ibaraki 305-0801, Japan
[248] Division of Science, National Astronomical Observatory of Japan, 2-21-1 Osawa, Mitaka City, Tokyo 181-8588, Japan
[249] Nagoya University, Nagoya, 464-8601, Japan
[250] Department of Physics, Aristotle University of Thessaloniki, 54124 Thessaloniki, Greece
[251] Bard College, Annandale-On-Hudson, NY 12504, USA
[252] Technical University of Braunschweig, D-38106 Braunschweig, Germany
[253] Institute of Mathematics, Polish Academy of Sciences, 00656 Warsaw, Poland
[254] Astronomical Observatory, Jagiellonian University, 31-007 Cracow, Poland
[255] Department of Physics and Astronomy, University of Padova, Via Marzolo, 8-35151 Padova, Italy
[256] Istituto Nazionale di Fisica Nucleare (INFN) Sezione di Padova, Via Marzolo, 8-35131 Padova, Italy
[257] Department of Physics, Nagoya University, ES building, Furocho, Chikusa-ku, Nagoya, Aichi 464-8602, Japan
[258] Université de Montréal/Polytechnique, Montreal, Quebec H3T 1J4, Canada
[259] Indian Institute of Science Education and Research, Kolkata, Mohanpur, West Bengal 741252, India
[260] Seoul National University, Seoul 08826, Republic of Korea
[261] Department of Computer Simulation, Inje University, 197 Inje-ro, Gimhae, Gyeongsangnam-do 50834, Republic of Korea
[262] NAVIER, École des Ponts, Université Gustave Eiffel, CNRS, Marne-la-Vallée, France
[263] Gravitational Wave Science Project, National Astronomical Observatory of Japan (NAOJ), Mitaka City, Tokyo 181-8588, Japan
[264] Department of Physics, National Cheng Kung University, No.1, University Road, Tainan City 701, Taiwan
[265] St. Thomas University, Miami Gardens, FL 33054, USA
[266] Scuola Normale Superiore, I-56126 Pisa, Italy
[267] Institució Catalana de Recerca i Estudis Avançats, E-08010 Barcelona, Spain
[268] Institut de Física d'Altes Energies, E-08193 Barcelona, Spain
[269] Institut fuer Theoretische Astrophysik, Zentrum fuer Astronomie Heidelberg, Universitaet Heidelberg, Albert Ueberle Str. 2, 69120 Heidelberg, Germany
[270] Institucio Catalana de Recerca i Estudis Avançats (ICREA), Passeig de Lluís Companys, 23, 08010 Barcelona, Spain

[271] Research Center for Space Science, Advanced Research Laboratories, Tokyo City University, 3-3-1 Ushikubo-Nishi, Tsuzuki-Ku, Yokohama, Kanagawa 224-8551, Japan
[272] Tsinghua University, Beijing 100084, People's Republic of China
[273] Institut des Hautes Etudes Scientifiques, F-91440 Bures-sur-Yvette, France
[274] Faculty of Law, Ryukoku University, 67 Fukakusa Tsukamoto-cho, Fushimi-ku, Kyoto City, Kyoto 612-8577, Japan
[275] Phenikaa Institute for Advanced Study (PIAS), Phenikaa University, Yen Nghia, Ha Dong, Hanoi, Vietnam
[276] University of Stavanger, 4021 Stavanger, Norway
[277] Physics Program, Graduate School of Advanced Science and Engineering, Hiroshima University, 1-3-1 Kagamiyama, Higashihiroshima City, Hiroshima 739-8526, Japan
[278] GRAPPA, Anton Pannekoek Institute for Astronomy and Institute for High-Energy Physics, University of Amsterdam, 1098 XH Amsterdam, The Netherlands
[279] University College London, London WC1E 6BT, UK
[280] Observatoire de Paris, 75014 Paris, France
[281] Laboratoire Univers et Théories, Observatoire de Paris, 92190 Meudon, France
[282] Graduate School of Science and Technology, Niigata University, 8050 Ikarashi-2-no-cho, Nishi-ku, Niigata City, Niigata 950-2181, Japan
[283] University of Maryland, Baltimore County, Baltimore, MD 21250, USA
[284] CSIR-Central Glass and Ceramic Research Institute, Kolkata, West Bengal 700032, India
[285] Consiglio Nazionale delle Ricerche - Istituto dei Sistemi Complessi, I-00185 Roma, Italy
[286] Department of Astronomy, Yonsei University, 50 Yonsei-Ro, Seodaemun-Gu, Seoul 03722, Republic of Korea
[287] Department of Physics, University of Guadalajara, Av. Revolucion 1500, Colonia Olimpica C.P. 44430, Guadalajara, Jalisco, Mexico
[288] Hobart and William Smith Colleges, Geneva, NY 14456, USA
[289] INAF-Osservatorio Astronomico di Padova, I-35122 Padova, Italy
[290] Dipartimento di Ingegneria, Università del Sannio, I-82100 Benevento, Italy
[291] Museo Storico della Fisica e Centro Studi e Ricerche "Enrico Fermi," I-00184 Roma, Italy
[292] Kennesaw State University, Kennesaw, GA 30144, USA
[293] Subatech, CNRS/IN2P3 - IMT Atlantique - Nantes Université, 4 rue Alfred Kastler BP 20722 44307 Nantes CÉDEX 03, France
[294] Departamento de Física - ETSIDI, Universidad Politécnica de Madrid, 28012 Madrid, Spain
[295] Department of Electronic Control Engineering, National Institute of Technology, Nagaoka College, 888 Nishikatakai, Nagaoka City, Niigata 940-8532, Japan
[296] Trinity College, Hartford, CT 06106, USA
[297] Dipartimento di Fisica e Scienze della Terra, Università degli Studi di Ferrara, Via Saragat, 1, 44121, Ferrara FE, Italy
[298] Faculty of Science, Toho University, 2-2-1 Miyama, Funabashi City, Chiba 274-8510, Japan
[299] Indian Institute of Technology, Palaj, Gandhinagar, Gujarat 382355, India
[300] Kavli Institute for Astronomy and Astrophysics, Peking University, Yiheyuan Road 5, Haidian District, Beijing 100871, People's Republic of China
[301] Laboratoire MSME, Cité Descartes, 5 Boulevard Descartes, Champs-sur-Marne, 77454, Marne-la-Vallée Cedex 2, France
[302] Faculty of Information Science and Technology, Osaka Institute of Technology, 1-79-1 Kitayama, Hirakata City, Osaka 573-0196, Japan
[303] NASA Goddard Space Flight Center, Greenbelt, MD 20771, USA
[304] Faculty of Science and Technology, Kochi University, 2-5-1 Akebono-cho, Kochi-shi, Kochi 780-8520, Japan
[305] Laboratoire de Physique de l'École Normale Supérieure, ENS, (CNRS, Université PSL, Sorbonne Université, Université Paris Cité), F-75005 Paris, France
[306] Faculty of Physics, University of Warsaw, Ludwika Pasteura 5, 02-093 Warszawa, Poland
[307] Laser Interferometry and Gravitational Wave Astronomy, Max Planck Institute for Gravitational Physics, Callinstrasse 38, 30167 Hannover, Germany
[308] The Hakubi Center for Advanced Research, Kyoto University, Yoshida-honmachi, Sakyou-ku, Kyoto City, Kyoto 606-8501, Japan
[309] Department of Physics, Kyoto University, Kita-Shirakawa Oiwake-cho, Sakyou-ku, Kyoto City, Kyoto 606-8502, Japan
[310] Yukawa Institute for Theoretical Physics (YITP), Kyoto University, Kita-Shirakawa Oiwake-cho, Sakyou-ku, Kyoto City, Kyoto 606-8502, Japan
[311] University of Catania, Department of Physics and Astronomy, Via S. Sofia, 64, 95123, Catania CT, Italy
[312] National Institute of Technology, Fukui College, Geshi-cho, Sabae-shi, Fukui 916-8507, Japan
[313] Department of Communications Engineering, National Defense Academy of Japan, 1-10-20 Hashirimizu, Yokosuka City, Kanagawa 239-8686, Japan
[314] Eindhoven University of Technology, 5600 MB Eindhoven, The Netherlands
[315] Kavli Institute for the Physics and Mathematics of the Universe (Kavli IPMU), WPI, The University of Tokyo, 5-1-5 Kashiwa-no-Ha, Kashiwa City, Chiba 277-8583, Japan
[316] Department of Astronomy, Beijing Normal University, Xinjiekouwai Street 19, Haidian District, Beijing 100875, People's Republic of China
[317] School of Physics and Technology, Wuhan University, Bayi Road 299, Wuchang District, Wuhan, Hubei, 430072, People's Republic of China
[318] University of California, Berkeley, Berkeley, CA 94720, USA

## Abstract

LIGO, Virgo, and KAGRA form a network of gravitational-wave observatories. Data and analysis results from this network are made publicly available through the Gravitational Wave Open Science Center. This paper describes open data from this network, including the addition of data from the first part of the fourth observing run and selected periods from the preceding engineering run, collected from 2023 May to 2024 January. The public dataset includes calibrated strain time series for each instrument, data from additional channels used for noise subtraction and detector characterization, and analysis data products from version 4.0 of the Gravitational-Wave Transient Catalog.

## 1. Introduction

The Laser Interferometer Gravitational-Wave Observatory (LIGO; J. Aasi et al. 2015), Virgo (F. Acernese et al. 2015), KAGRA (T. Akutsu et al. 2021a), and GEO 600 (H. Luck et al. 2010; C. Affeldt et al. 2014; K. L. Dooley et al. 2016) are observing the gravitational-wave (GW) Universe with unprecedented sensitivity (A. G. Abac et al. 2025c). Data from these observatories are shared and jointly analyzed as described in a Memorandum of Agreement (P. Brady et al. 2019), so that the instruments form a global network. The instruments operate in a series of observing runs, with breaks between observing periods for instrument upgrades and detector commissioning (B. P. Abbott et al. 2020a; E. Capote et al. 2025). Following a proprietary period for work by the LIGO-Virgo-KAGRA Collaboration (LVK), data from each observing run are publicly released via the Gravitational Wave Open Science Center (GWOSC) for use by the entire scientific community.[319] The schedule of data releases for LIGO is maintained in the LIGO Data Management Plan (S. Anderson et al. 2025).

Data from previous observing runs have been released with a history going back more than a decade, starting with data from initial LIGO released in 2014 (M. Vallisneri et al. 2015). Data from the advanced-detector era are organized into an ongoing series of observing runs, beginning with data from the first observing run (O1) and second observing run (O2; R. Abbott et al. 2021a), and continuing with data released from the third observing run (O3; R. Abbott et al. 2023a).

The fourth observing run is divided into several segments, with the first segment (O4a) spanning time from 2023 May 24 to 2024 January 16. This paper describes the publicly available data from the LVK with an emphasis on the new O4a public data release. At the time of this writing, as described in Section 2.1, this data release only includes data from LIGO.

Public data include time series strain data from the instruments, additional time series channels describing the instrument state, and segment information describing the data quality of each instrument. In addition, the full set of events known as the Gravitational-Wave Transient Catalog (GWTC) identified by the LVK is publicly available. The O4a data release also includes a number of analysis results from the fourth version of the catalog, GWTC-4.0 (LIGO Scientific Collaboration et al. 2025d), including an electronic version of the catalog that can be browsed online and queried through a Representational State Transfer Application Programming Interface (REST API). This paper is included in a suite of papers describing GWTC-4.0 (A. G. Abac et al. 2025c), and is best understood in the context of the related papers.

In Section 2, we discuss the primary data products from GW observatories and the epochs of data taking known as observing runs. In Section 3, we explain the basics of calibration of interferometer strain data. Section 4 outlines some nonastrophysical noise sources that impact GW data analysis and noise mitigation. In Section 5, we describe the details for the strain data available on GWOSC and alternative services. Section 6 discusses additional instrumental monitoring channels used for noise subtraction and production of data quality flags that are now also distributed by GWOSC. Finally, the GWOSC Event Portal that provides electronic catalogs of detected GW transients is described in Section 7.

[319] GWOSC Home Page, https://gwosc.org.

## 2. Dataset Overview

### 2.1. Observing Time

The instruments operated in a series of observing runs described in A. G. Abac et al. (2025c), with the following dates:

- O1: 2015 September 12 0:00 UTC (GPS 1126051217)to 2016 January 19 16:00 UTC (GPS 1137254417)
- O2: 2016 November 30 16:00 UTC (GPS 1164556817) to 2017 August 25 22:00 UTC (GPS 1187733618)
- O3: 2019 April 1 15:00 UTC (GPS 1238166018) to 2020 March 27 17:00 UTC (GPS 1269363618)
- O4a: 2023 May 24 15:00 UTC (GPS 1368975618) to 2024 January 16 16:00 UTC (GPS 1389456018)

The O4a data release also includes some additional data from the preceding engineering run (see Section 5.3). Some observational data are also available from other times studied in targeted LVK publications, including data from the GEO 600 (K. L. Dooley et al. 2016) and KAGRA (T. Akutsu et al. 2021a) detectors, as described in LIGO Scientific Collaboration & Virgo Collaboration (2017), and LIGO Scientific Collaboration et al. (2022), (2024).

The observatories do not record astrophysical data at all times. Many types of phenomena can interrupt data taking, including instrument maintenance and detector improvements, seismic activity, power outages, and instrument lock-loss for other reasons. A comparison of the total amount of observing time for each run is shown in Table 1. The amount of calendar time for each observing run is shown at the top of each table section (e.g., O4a spanned 237 calendar days). The O3 run includes observing time from both O3a and O3b, as well as a commissioning break with no observing time.

During O4a, Virgo was not observing due to commissioning activities. KAGRA and GEO 600 data were less sensitive than the other observatories and so were not used for most analyses. Therefore, only data from LIGO are included in this data release. The total coincident time with both LIGO Hanford Observatory (LHO) and LIGO Livingston Observatory (LLO) observing simultaneously during O4a was 126.5 days. The table shows similar statistics for each of O1, O2, and O3, with additional information on past data releases available in R. Abbott et al. (2021a) and R. Abbott et al. (2023a).

### 2.2. Time Series Data

The primary data product from a GW observatory is the calibrated strain data, often called $h(t)$. The strain data are recorded as a time series, so that each time sample records one measurement of the fractional difference in lengths in the interferometer arms, $\Delta L/L$. The LIGO, Virgo, and KAGRA observatories typically use a sampling rate of 16384 samples per second, corresponding to a data rate of ~4 TB per year per instrument for the strain data. This is a small fraction of the total data rate of several PB per year, which includes hundreds of thousands of diagnostic channels to monitor the state of the instrument and the local environment.

**Table 1**
The Amount of Observing Time in Days and the Corresponding Fraction of the Total Runtime for Each Detector (IFO) Combination[a]

| IFO Combination | | Observing Time | Fraction |
|---|---|---|---|
| | O1: 129.7 days | | |
| LHO | | 27.7 days | 21.4% |
| LLO | 16.8 days | 13.0% | |
| LHO, LLO | | 49.0 days | 37.8% |
| | O2: 268.3 days | | |
| LHO | | 37.8 days | 14.1% |
| LLO | | 33.5 days | 12.5% |
| Virgo | | 1.7 days | 0.6% |
| LHO, LLO | | 103.0 days | 38.4% |
| LHO, Virgo | | 1.7 days | 0.6% |
| LLO, Virgo | | 2.2 days | 0.8% |
| LHO, LLO, Virgo | | 15.3 days | 5.7% |
| | O3a: 183.0 days | | |
| LHO | | 5.6 days | 3.1% |
| LLO | | 6.3 days | 3.4% |
| Virgo | | 15.7 days | 8.6% |
| LHO, LLO | | 25.9 days | 14.2% |
| LHO, Virgo | | 17.5 days | 9.5% |
| LLO, Virgo | | 25.3 days | 13.8% |
| LHO, LLO, Virgo | | 80.8 days | 44.2% |
| | O3b: 147.1 days | | |
| LHO | | 4.5 days | 3.1% |
| LLO | | 3.4 days | 2.3% |
| Virgo | | 9.3 days | 6.3% |
| LHO, LLO | | 22.9 days | 15.6% |
| LHO, Virgo | | 14.5 days | 9.8% |
| LLO, Virgo | | 13.8 days | 9.4% |
| LHO, LLO, Virgo | | 73.4 days | 49.9% |
| | O3: 361.1 days | | |
| LHO | | 10.1 days | 2.8% |
| LLO | | 9.7 days | 2.7% |
| Virgo | | 25.0 days | 6.9% |
| LHO, LLO | | 48.9 days | 13.5% |
| LHO, Virgo | | 31.9 days | 8.8% |
| LLO, Virgo | | 39.1 days | 10.8% |
| LHO, LLO, Virgo | | 154.3 days | 42.7% |
| | O3GK: 13.7 days | | |
| GEO600 | | 4.5 days | 32.9% |
| KAGRA | | 0.9 days | 6.5% |
| GEO600, KAGRA | | 6.4 days | 46.7% |
| | O4a: 237.0 days | | |
| LHO | | 33.3 days | 14.1% |
| LLO | | 37.0 days | 15.6% |
| LHO, LLO | | 126.5 days | 53.4% |

**Note.**
[a] The amount of calendar time for each observing run is shown at the top of each table section.

For current instruments, the strain time series consists primarily of instrumental noise at most times (B. P. Abbott et al. 2020b). The noise levels in the instruments fluctuate over time, so that the sensitivities of the instruments vary from moment to moment. One long-used figure of merit for the instrument sensitivity is the binary neutron star (BNS) inspiral range (L. S. Finn & D. F. Chernoff 1993; H.-Y. Chen et al. 2021), which is briefly summarized in A. G. Abac et al. (2025c), and gives an estimate of the distance to which a detector could detect a BNS merger with a signal-to-noise ratio (SNR) of 8 after averaging over sky location and orientation. The variability in the BNS inspiral range over the course of O4a is shown in Figure 1. The plot shows that the observatories frequently operated near their peak sensitivity of around 160 Mpc in O4a. However, there are also times when elevated noise levels cause the sensitivity to drop, as well as times when the detectors are not operating at all or operating for periods too short to make a reliable inspiral range estimate, and in this case, the inspiral range is set to zero. This variability in instrument sensitivity is an important feature of the strain data and needs to be properly accounted for in any analysis. Over the course of O1 through O4, the typical BNS inspiral range of the instruments generally improved for each run due to occasional hardware upgrades and commissioning work at the observatories (A. F. Brooks et al. 2021; J. C. Driggers et al. 2019; M. Tse et al. 2019; A. Buikema et al. 2020; S. Soni et al. 2021; W. Jia et al. 2024; S. Soni et al. 2024; E. Capote et al. 2025). The evolution of this range is described in A. G. Abac et al. (2025c) and B. P. Abbott et al. (2020a).

In addition to the strain time series, the observatories also record a large number of auxiliary channels to measure the state of each instrument and the local environment (P. Nguyen et al. 2021; R. Huxford et al. 2024). These channels record quantities like laser-power levels, temperatures, seismic motion, angular alignments of mirrors, and many other types of information that can be used to assess instrument performance (J. G. Rollins 2016; S. Soni et al. 2025). A subset of these auxiliary channels is included in the O3 and O4a data releases, and is described in Section 6.

## 3. Calibration

Ground-based interferometers make use of a feedback control loop in order to hold the differential arm degree of freedom, $\Delta L(t) = \Delta L_x(t) - \Delta L_y(t)$, in resonance. Calibration refers to the process of reconstructing the external differential arm motion $\Delta L(t)$ through a model for the interferometer response function $R$ and the measured error signal from the differential arm feedback loop, $d_{\mathrm{err}}$ (B. P. Abbott et al. 2017; A. Viets et al. 2018). In the time domain, the strain can be reconstructed through a convolution of the differential arm error signal and the interferometer response function,

$$\Delta L(t) = R(t) \circledast d_{\mathrm{err}}(t), \tag{1}$$

where $\circledast$ represents a convolution, implemented using digital finite impulse response filters for the response function (A. Viets et al. 2018). The strain $h(t)$ is the ratio of the differential arm motion to the unperturbed interferometer arm length $L$,

$$h(t) = \frac{\Delta L(t)}{L}, \tag{2}$$

where $L$ is the length of an interferometer arm and $L \simeq 3995$ m for LIGO.

Imperfect calibration results in small errors in the measured strain values. For LIGO in O4a, the broadband uncertainty on the calibrated strain data is determined on an hourly cadence using a modified version of the methods described in L. Sun et al. (2020) that incorporates additional continuous measurements of the calibrated strain systematic error at discrete frequencies (M. Wade et al. 2025). Calibration uncertainty values for all runs are available in the LIGO Document

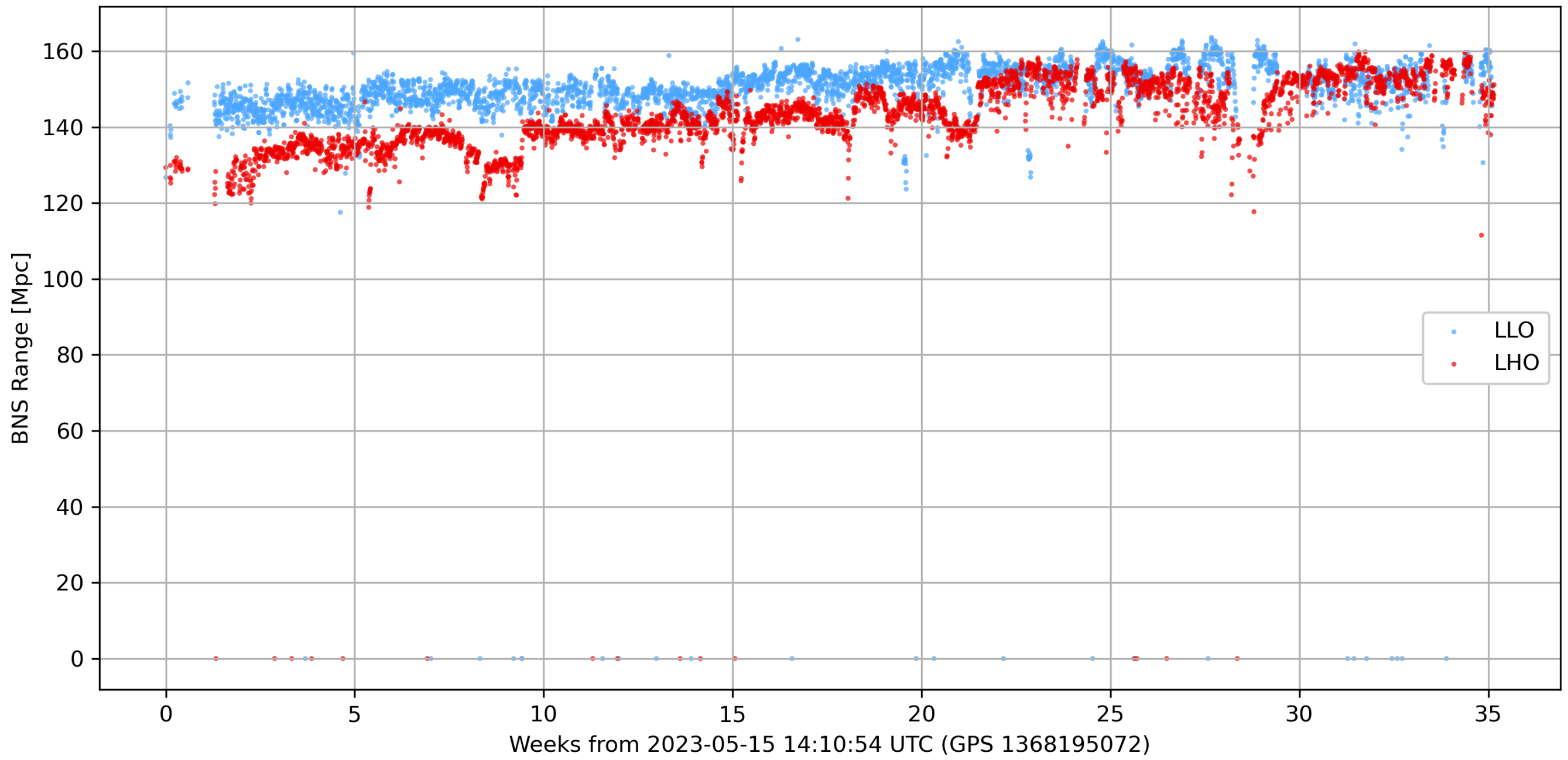

**Figure 1.** BNS inspiral range over time for O4a. Each point corresponds to a time interval of 4096 s, corresponding to the time range of one strain data file. The fast Fourier transform window length used is 8 s with a 4 s overlap. These inspiral range values are stored as metadata for each file in the GWOSC database and are accessible via the GWOSC website.

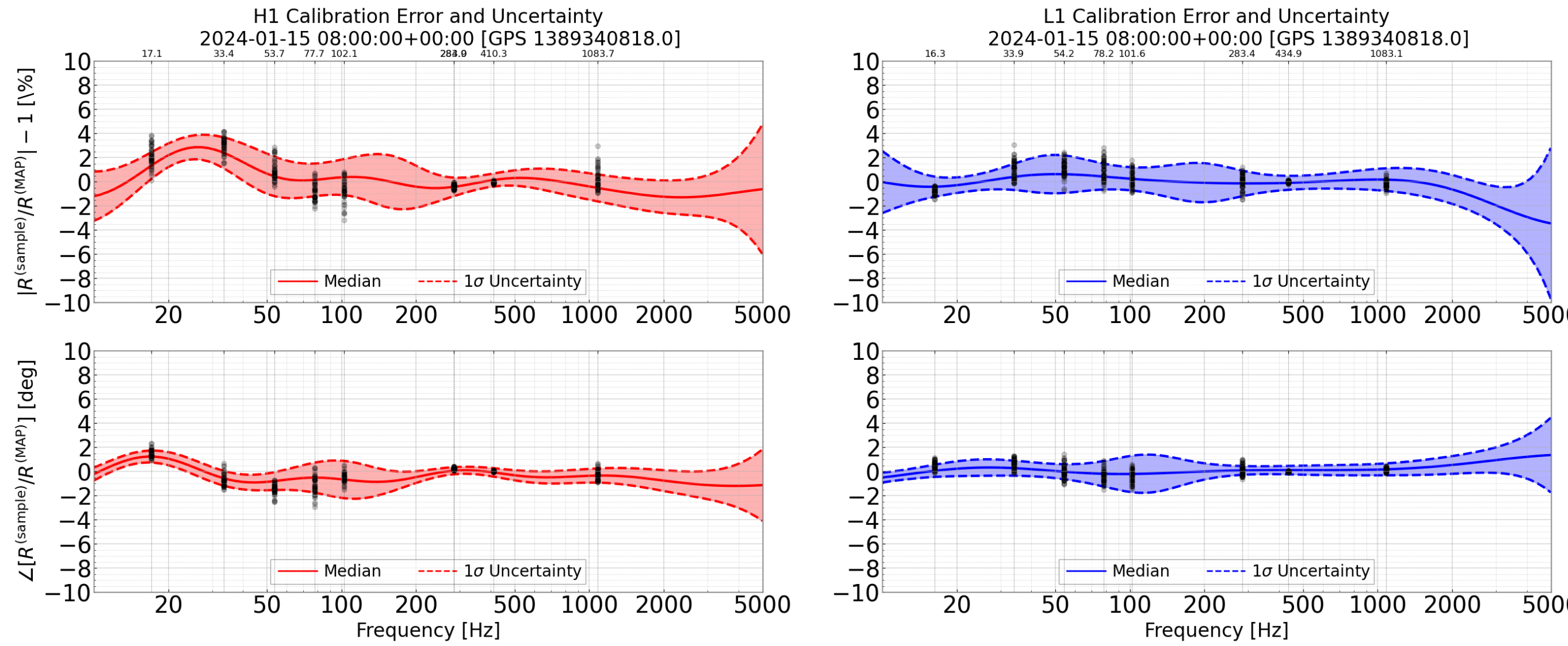

**Figure 2.** LIGO Hanford (left) and Livingston (right) frequency-dependent calibration error for a 1 hr time period in O4a starting at 2024 January 15 08:00:00 UTC. The magnitude of the calibration error is shown in the top plots, and the phase of the calibration error (in units of degrees) is shown in the bottom plots. The median systematic error of the calibration is given by the solid line. The $1\sigma$ uncertainty on this systematic error is given by the dotted lines. The black dots overlaid on the figure at specific frequencies are direct measurements of the calibration error at a specific frequency during this 1 hr time period. These measurements are made using sinusoidal injections with the photon-calibrator system. These uncertainties are representative of typical LIGO calibration uncertainties throughout O4a.

Control Center (LIGO-Virgo-KAGRA Collaboration 2021, 2025a).

Calibration methods for past observing runs are described in B. P. Abbott et al. (2017), C. Cahillane et al. (2017), L. Sun et al. (2020, 2021), D. Chen et al. (2025), and F. Acernese et al. (2018, 2022). During O4a, the final calibrated strain data for both LIGO observatories were computed in near real time. The photon-calibrator system was used in O4a to inject continuous, sinusoidal excitations at eight frequencies throughout the run (S. Karki et al. 2016; D. Bhattacharjee et al. 2024). The systematic error in the calibration at these frequencies was then inferred through the transfer function of the photon-calibrator excitation strain and the reconstructed strain (M. Wade et al. 2025). The measured systematic error at each of these frequencies was included in the modeled systematic error. The O4a run marked the first time calibration uncertainty estimates were provided for the near-real-time calibrated strain data in LIGO, making this data usable as the final calibrated strain data product. Example uncertainty envelopes for the LHO and LLO calibrated strain data are shown in Figure 2. These envelopes represent the median systematic calibration error as well as the 1$\sigma$ uncertainty on the systematic error for a 1 hr time period during O4a. The black dots overlaid on the envelopes are direct measurements of the calibration systematic error over the 1 hr time period using sinusoidal photon-calibrator injections at specific frequencies. The calibrated strain data for LIGO in O4a should be considered valid for analyses within the provided hourly uncertainty envelopes in the range of 10–5000 Hz.

## 4. Data Quality

Noise in the strain data is sometimes approximated as stationary and Gaussian. However, real strain data are also affected by disturbances originating from instrumental and environmental sources (B. P. Abbott et al. 2020b; S. Soni et al. 2025). These artifacts can manifest as broadband, non-Gaussian, short-duration features known as *glitches* (L. K. Nuttall 2018; J. Glanzer et al. 2023), or as narrowband spectral features referred to as *spectral lines* (P. B. Covas et al. 2018). In addition, GW observatories can experience periods of degraded performance due to issues in control systems, calibration procedures, or environmental disturbances. These episodes can compromise data quality over extended durations, reducing the sensitivity of the detectors and potentially biasing astrophysical parameter estimation (C. Pankow et al. 2018; J. Y. L. Kwok et al. 2022; R. Macas et al. 2022; S. Ghonge et al. 2024). To support robust analyses, the LVK employs a suite of data quality metrics designed to identify and allow mitigation of the impact of instrumental and environmental noise. An overview of LIGO detector characterization efforts is available in S. Soni et al. (2025). Detector characterization methods for past observing runs are described in D. Davis et al. (2021), F. Acernese et al. (2023a), L. Nuttall et al. (2015), B. P. Abbott et al. (2016, 2018), P. B. Covas et al. (2018), P. Nguyen et al. (2021), F. Acernese et al. (2023b), and T. Akutsu et al. (2021b).

Searches for different types of signals exhibit distinct susceptibility to different noise sources, but detector characterization is needed in all cases to remove signals of terrestrial origin and strengthen the confidence of detections. Data quality products are employed across four broad categories of searches, each with customized detector characterization methods (S. Caudill et al. 2021). Compact binary coalescence (CBC) searches (LIGO Scientific Collaboration et al. 2025c) target GW signals from the coalescence of neutron stars and/or black holes by applying matched-filtering techniques with modeled waveform templates. GW burst (BURST) analyses (A. G. Abac et al. 2025d) aim to identify short-duration transients without strong assumptions on the signal morphology, by detecting excess power in the time–frequency representation of the strain data. Continuous wave (CW) searches (R. Abbott et al. 2022a) focus on long-duration, nearly monochromatic signals such as GWs emitted by nonaxisymmetric rotating neutron stars. Stochastic (STOCH) searches (R. Abbott et al. 2021b) aim to detect a diffuse GW background resulting from the superposition of numerous unresolved sources.

### 4.1. Hardware Injections

Simulated GW signals, known as *hardware injections* (C. Biwer et al. 2017), are introduced by physically displacing the interferometer test masses to mimic true astrophysical events and characterize the performance of the detectors and analyses. Hardware injections with signals that mimic expected astrophysical sources are labeled by the simulation type, corresponding to the different classes of potential sources (CBC, BURST, STOCH, or CW). Detector characterization *safety injections* are labeled as DETCHAR injections. These are used to test couplings between channels used to monitor instrumental and environmental noise sources.

We identify auxiliary channels that can be used to indicate a detector problem, for example, when they display excess noise levels. However, some channels are *unsafe* for identifying noise because they respond to astrophysical signals, and therefore, a response in the auxiliary channel could indicate a real signal. An auxiliary channel is said to be safe to use for identifying noise transients only if it does not respond to astrophysical signals. Safety injections are used to identify unsafe channels. They are performed by actuating the end test mass mirror using a photon calibrator, thereby generating a response in the primary strain channel (S. Karki et al. 2016). Channels that exhibit a response to these injections are considered unsafe to use for vetoes, and are identified following the method in R. Essick et al. (2021).

Hardware injection times are available on the GWOSC website via a tool called Timelines.[320] Additional notes about hardware injections are also provided in the documentation for each data release. In O4a, there were no injections labeled CBC, BURST, DETCHAR, or STOCH during observing mode. The only hardware injections present in the released data are of the CW type and consist of simulations of 18 spinning neutron stars whose parameters are given on the GWOSC website,[321] with GW frequencies spanning 12–3000 Hz. Injecting simulated CW signals allows, within the uncertainties of the hardware injection system, a direct end-to-end assessment of degradation in CW signal detection due to imperfect calibration or due to noise subtraction. These injections are almost always present (in O4a, only 1.7% of the released LHO data and 2.6% of LLO data do not have CW injections), but because they are faint and nearly monochromatic, they have minimal impact on transient signal searches.

### 4.2. Data Quality Flags

Data quality flags are used to identify periods during which the strain data are affected by data quality issues. These flags are used by analysis pipelines to exclude or down-weight compromised data segments. Flags are classified in order of descending severity, such that the most severe problems are labeled Category 1 (`CAT1`). These flags mark intervals of time affected by well-understood and severe data quality issues (hardware faults, control system failures, or known periods of malfunction) during which the strain data are not considered suitable for astrophysical analyses; these times are systematically excluded from all GW searches and parameter estimation studies. Some searches also apply the less severe Category 2 (`CAT2`), which are typically shorter in duration than `CAT1` (LIGO Scientific Collaboration et al. 2025c). Category 3 (`CAT3`) flags are used to identify possible issues of unknown origin that are found through statistical studies.

[320] Timelines App, https://gwosc.org/timeline/.
[321] O4a Injections, https://gwosc.org/O4/o4_inj.

**Table 2**
The iDQ Channel Names in the Alternate Strain Release Described in Section 5.5, Sampled at 128 Hz

| | Channel Name |
|---|---|
| OK flag | *Ifo*:IDQ-OK_OVL_10_2048_AR |
| Rank | *Ifo*:IDQ-RANK_OVL_10_2048_AR |
| FAP | *Ifo*:IDQ-FAP_OVL_10_2048_AR |
| Efficiency | *Ifo*:IDQ-EFF_OVL_10_2048_AR |
| Log-likelihood ratio | *Ifo*:IDQ-LOGLIKE_OVL_10_2048_AR |

**Note.** *Ifo* corresponds to H1 for Hanford or L1 for Livingston, and AR stands for analysis ready.

During O4a, some searches used only the `CAT1` flag set. `CAT2` flags were applied only to BURST searches. The methodology for identifying and applying these flags during O4a is described in detail in S. Soni et al. (2025).

For CBC analyses, an additional data quality product used during O4a is the set of statistical flags derived from the `iDQ` supervised-learning framework (R. Essick et al. 2020). This pipeline evaluates the likelihood that transient noise is present in the strain data based on activity in auxiliary channels that are not expected to be sensitive to astrophysical sources. The `iDQ` pipeline produces several statistical outputs sampled at 128 Hz. The low-latency versions of the `iDQ` outputs are available in the alternate strain release described in Section 5.5; the corresponding channel names are listed in Table 2. The offline `iDQ` outputs are available in Zenodo as the LIGO-Virgo-KAGRA Collaboration (2025b). The `OK` flag indicates whether the `iDQ` outputs are reliable at a given time, taking the value 1 when valid and 0 otherwise. Other outputs include a normalized `RANK` score that reflects the raw classifier output, an estimated detection efficiency (`EFF`), the false alarm probability (`FAP`) of misclassifying clean times as glitchy, and the log likelihood-ratio (`LOGLIKE`), where positive values suggest the presence of a glitch and negative values suggest clean data. For CBC searches, the `iDQ` flags typically used are derived by thresholding the `LOGLIKE` channel; segments with `LOGLIKE` $>$ 5 are flagged as likely to be glitch-contaminated. Unlike `CAT1` and `CAT2` flags, `iDQ` flags are not applied universally. Searches algorithms may choose to use them either as vetoes or to re-rank the significance of candidate events.

### *4.3. Spectral Line Catalog*

Spectral lines are narrowband, often persistent features that appear in the amplitude spectral density of GW strain data. These features arise from a variety of instrumental and environmental sources, including mechanical resonances (e.g., violin-mode resonances from the suspension system), injected lines used for calibration, digital dither signals, and harmonics of the electrical mains. Although their impact on CBC and BURST searches is minimal, spectral lines are a significant concern for persistent searches. Spectral artifacts that overlap with the target signal frequency can severely impact CW searches, which rely on coherence over long timescales. Similarly, STOCH searches are sensitive to lines, especially those of common origin or at harmonics of shared noise sources.

To support the identification and mitigation of these features, the LVK maintains a curated catalog of instrumental spectral lines (P. B. Covas et al. 2018). The catalog includes both previously known artifacts and newly identified lines, and provides relevant metadata such as frequency, amplitude, detector, and, where possible, a suspected source. This information is used by various search pipelines to identify contaminated frequency bins, apply vetoes, or model noise contamination. Spectral line catalogs are publicly available on the GWOSC website as part of the documentation of each run's data release.[322]

### *4.4. Glitch Subtraction*

Glitches represent a significant source of contamination in strain data, particularly affecting transient searches while having a limited impact on persistent searches. In a small number of cases, a glitch occurs in close temporal and spectral proximity to a candidate GW signal, requiring targeted mitigation to reliably recover the astrophysical signal. Subtraction techniques are employed to remove localized excess power in the time-frequency domain (C. Pankow et al. 2018). One such method is `BayesWave` (N. J. Cornish et al. 2021; S. Hourihane et al. 2022), which models and subtracts transient noise features from the data without requiring a specific model for the astrophysical signal. A complementary method is linear noise subtraction (D. Davis et al. 2019), which removes known, repeatable noise contributions using information from auxiliary sensors that monitor the detector environment and instrumental systems.

During O4a, 16 candidate events with a false alarm rate (FAR) below 1 per year required such glitch mitigation in at least one interferometer (LIGO Scientific Collaboration et al. 2025d). These events are listed in Table 3, along with their GPS time, affected interferometer(s), and the time-frequency region used to constrain the subtraction and assess residuals. All glitch-mitigated data listed in Table 3 were produced with the `BayesWave` pipeline.

## 5. Strain Data

Sections 5.1 to 5.4 describe the *default strain data release*, which can be downloaded directly from the GWOSC website or accessed via the `fetch_open_data` method of the `gwpy` package (D. M. Macleod et al. 2021). Alternatively, especially for downloading large datasets, such as an entire observation run, users can download data from the network data server NDS2 (Z. Zweizig et al. 2021) or the Open Science Data Federation (OSDF) service that acts as a storage resource broker and optimizes data access through a worldwide network of data centers.[323] While much of the content in this section is similar across observing runs, Section 5.4 describes differences between runs.

Data in the LVK archives are stored as GW frame files (`gwf`; LIGO Scientific Collaboration & Virgo Collaboration 2022), a custom binary format developed within the GW community in which data are uniquely identified by a channel name and a frame type. Files for the default strain data release are created by repackaging the original LVK data, whose identifiers are listed in Table 4 for all observing runs. The best calibration version available is used in the default strain release. In addition to the default strain data release, several *alternate strain* channels have also been released starting from O3 (see Section 5.5).

[322] GWOSC data releases, https://gwosc.org/data.
[323] Large Scale Data and Computing Resources, https://gwosc.org/osdf.

**Table 3**
List of O4a Events from LIGO Scientific Collaboration et al. (2025d) with FAR < 1 per yr, for Which Glitch Subtraction Was Performed Using `BayesWave` (N. J. Cornish et al. 2021)

| Event | GPS Time | IFO | Time (s) | Frequency (Hz) |
|---|---|---|---|---|
| GW230606_004305 | 1370047403.79 | LHO | [1.11, 1.31] | [8.0, 512] |
| GW230707_124047 | 1372768865.35 | LHO | [−0.15, 0.1] | [15.0, 30.0] |
| GW230708_053705 | 1372829843.12 | LHO | [−2.02, 0.98] | [10.0, 50.0] |
| GW230806_204041 | 1375389659.94 | LHO | [1.8, 2.0] | [25.0, 65.0] |
| GW230819_171910 | 1376500768.45 | LLO | [−3.2,–2.8] | [8.0, 512] |
| GW231113_122623 | 1383913601.88 | LLO | [0.01, 0.22] | [70.0, 120.0] |
| GW231114_043211 | 1383971549.25 | LHO | [−0.95, −0.6] | [10.0, 30.0] |
| GW231118_090602 | 1384333580.01 | LHO | [−4.81, −4.51] | [15.0, 50.0] |
| GW231123_135430 | 1384782888.63 | LHO | [−1.7, −1.1] | [15.0, 30.0] |
| GW231129_081745 | 1385281083.64 | LLO | [1.4, 1.8] | [10.0, 170.0] |
| GW231221_135041 | 1387201859.32 | LHO | [0.3, 0.4] | [200.0, 450.0] |
| GW231223_032836 | 1387337334.05 | LHO | [−0.55, −0.25] | [10.0, 25.0] |

**Note.** For each event, we provide the GPS time, the interferometer (IFO) where subtraction was applied, and the time and frequency windows used for subtraction. The time window indicates the approximate time of the glitch and is referenced from the indicated GPS time, with negative values before the merger and positive values after the merger. A similar table is included in LIGO Scientific Collaboration et al. (2025d).

**Table 4**
Channel Names and Frame Types That Allow Tracing the Provenance of Strain Data Released on GWOSC

| Run | IFO | Channel Name | Frame Type |
|---|---|---|---|
| O1 | LHO | `H1:DCS-CALIB_STRAIN_C02` | `H1_HOFT_C02` |
| O1 | LLO | `L1:DCS-CALIB_STRAIN_C02` | `L1_HOFT_C02` |
| O2 | LHO | `H1:DCH-CLEAN_STRAIN_C02` | `H1_CLEANED_HOFT_C02` |
| O2 | LLO | `L1:DCH-CLEAN_STRAIN_C02` | `L1_CLEANED_HOFT_C02` |
| O2 | Virgo | `V1:Hrec_hoft_V1O2Repro2A_16384Hz` | `V1O2Repro2A` |
| O3a | LHO | `H1:DCS-CALIB_STRAIN_CLEAN_SUB60HZ_C01` | `H1_HOFT_CLEAN_SUB60HZ_C01` |
| O3a | LLO | `L1:DCS-CALIB_STRAIN_CLEAN_SUB60HZ_C01` | `L1_HOFT_CLEAN_SUB60HZ_C01` |
| O3a | Virgo | `V1:Hrec_hoft_16384Hz` | `V1Online` |
| O3a (last 2 weeks) | Virgo | `V1:Hrec_hoft_V1O3ARepro1A_16384Hz` | `V1O3Repro1A` |
| O3b | LHO | `H1:DCS-CALIB_STRAIN_CLEAN_SUB60HZ_C01` | `H1_HOFT_CLEAN_SUB60HZ_C01` |
| O3b | LLO | `L1:DCS-CALIB_STRAIN_CLEAN_SUB60HZ_C01` | `L1_HOFT_CLEAN_SUB60HZ_C01` |
| O3b | Virgo | `V1:Hrec_hoft_16384Hz` | `V1Online` |
| O3GK | GEO 600 | `G1:DER_DATA_HD_CLEAN` | `G1_RDS_C02_L3` |
| O3GK | KAGRA | `K1:DAC-STRAIN_C20` | `K1_HOFT_C20` |
| O4a | LHO | `H1:GDS-CALIB_STRAIN_CLEAN_AR` | `H1_HOFT_C00_AR` |
| O4a | LLO | `L1:GDS-CALIB_STRAIN_CLEAN_AR` | `L1_HOFT_C00_AR` |

**Note.** H1, L1, V1, G1, and K1 are short versions of the interferometer names. The attribute `CLEAN` in H1 and L1 indicates that noise subtraction (G. Vajente et al. 2020; D. Davis et al. 2019; A. Viets & M. Wade 2021) was used. `AR` stands for analysis ready and indicates that these channels contain only data for times ready to be analyzed. The other parts of the names refer in general to the calibration version.

The released calibrated strain data of the observing runs are divided into files containing 4096 s of data each. When data are missing or do not meet the necessary quality standards for analysis, strain values are recorded as `NaN`s. The strain data are available at both the original sampling rate of 16384 Hz and a reduced rate of 4096 Hz, referred to as 16 and 4 kHz in the remainder of this paper. The downsampling process is carried out using the standard decimation method `scipy.signal.decimate`, from the Python library `SciPy` (P. Virtanen et al. 2020). Before downsampling, an eighth-order Chebyshev type I filter (S. W. Smith 1999) was used as an anti-aliasing filter.

The maximum resolvable frequency in a given dataset is constrained by the Nyquist–Shannon sampling theorem (H. Nyquist 1924) and corresponds to half of the sampling rate. Due to the roll-off effect of the anti-aliasing filters applied during resampling, the valid frequency range is slightly lower than the Nyquist frequency. Consequently, for data sampled at 4 kHz, frequencies above approximately 1700 Hz are increasingly attenuated by the anti-aliasing filter, which can suppress but not entirely eliminate usable signal content. For data sampled at 16 kHz, however, the limiting factor is given by the calibration, which restricts the valid data to 5 kHz, as described in Section 3.

These considerations on the maximum frequency are a crucial factor to consider when selecting a dataset to download. On the other hand, data sampled at higher rates requires increased storage space and longer download times. Users should, therefore, select the dataset that best aligns with their specific requirements.

### 5.1. Structure of the Default Strain Data Files

The default strain files include not only the strain data but also information on data quality and injections. The main file formats used for the GWOSC open data are the Hierarchical Data Format (`hdf`; HDF Group & Q. Koziol 2018), a data format designed for storing and organizing large datasets that is easily readable by many programming languages, and the `gwf` format, which is the file format used in the LVK archives, as already mentioned.

For all available file formats, file names in the default strain release contain the following information:

1. *Obs*: The observatory, i.e., the site, that is indicated by one letter and can have values L for LIGO Livingston, H for LIGO Hanford, V for Virgo, G for GEO 600, or K for KAGRA.
2. *Ifo*: The interferometer label, created by adding a number to the letter that indicates the observatory. This choice allows the identification of multiple detectors installed in the same site, as was the case for Initial LIGO (B. P. Abbott et al. 2009). For the LVK observation runs, it can have values H1, L1, V1, G1, or K1.
3. *Run*: The observing run name, i.e., O4a.
4. *s*`KHZ` or *s*: The sampling rate in kHz, *s* can have values 4 or 16 (4096 or 16,384 Hz).
5. `R`*n* or `V`*n*: Release number or version number of the file. The letter `V` was used only for O1, then it was avoided because it could be confused with the Virgo interferometer name.
6. *GPSstart*: The starting time of the data contained in the file, as a 10-digit GPS value (in seconds).
7. *Dur*: The duration in seconds of the file, typically 4096 s.
8. *Ext*: The extension corresponding to the file format (`gwf`, `hdf` or `txt`).

The structure of the file names is

*Obs*—*FrameType*—*GPSstart*—*Dur*.*Ext*,

where the *FrameType* is *Ifo*_`GWOSC`_*Run*_*s*`KHZ`_`R`*n*.

The `hdf` files contain three folders (or groups).

1. The folder *meta* hosts the metadata of the file: *Description*, *DescriptionURL*, *Detector* (e.g., L1), *Observatory* (e.g., L), *Duration*, *GPSstart*, *UTCstart* (duration and starting time, using GPS and UTC standards, respectively, of the segment of data contained in a file), *StrainChannel,* and *FrameType* used in the LVK archives as listed in Table 4.
2. The folder *strain* contains the array *Strain* of $h(t)$ values with useful attributes such as *Xstart* and *Xspacing* that define the GPS start time of the data contained in the array and the temporal distance between points in the array.
3. The folder *quality* contains two subfolders: one for data quality and another for injections. Each subfolder includes a bitmask that indicates, for each second, the status of data quality or injections, along with a description of each bit in the mask (see Section 5.2 for details).

The GWOSC `gwf` files include three channels (for strain data, data quality, and injections), whose names are listed in Table 5.

**Table 5**
Channel Names in the GWOSC `gwf` Files.

| | Channel Name |
|---|---|
| Strain | *Ifo*:`GWOSC-`*s*`KHZ_R`*n*`_STRAIN` |
| Data quality mask | *Ifo*:`GWOSC-`*s*`KHZ_R`*n*`_DQMASK` |
| Injections mask | *Ifo*:`GWOSC-`*s*`KHZ_R`*n*`_INJMASK` |

**Note.** The nomenclature is similar to that described for the file name conventions: *Ifo* is the interferometer, *s* is the sampling rate, and *n* is the version or release (until now, only 1 has been used).

**Table 6**
Meaning of the Bits in the Data Quality Bitmask of the GWOSC Files

| Bit | Short Name | Description |
|---|---|---|
| 0 | `DATA` | Data present |
| 1 | `CBC_CAT1` | Pass the `CAT1` test for the CBC search |
| 2 | `CBC_CAT2` | Pass the `CAT1` and `CAT2` tests for the CBC search |
| 3 | `CBC_CAT3` | Pass the `CAT1`, `CAT2`, and `CAT3` tests for the CBC search |
| 4 | `BURST_CAT1` | Pass the `CAT1` test for the BURST search |
| 5 | `BURST_CAT2` | Pass the `CAT1` and `CAT2` tests for the BURST search |
| 6 | `BURST_CAT3` | Pass the `CAT1`, `CAT2`, and `CAT3` tests for the BURST search |
| 7 | `STOCH_CAT1` | Pass the `CAT1` test for the STOCH search |
| 8 | `CW_CAT1` | Pass the `CAT1` test for the CW search |

**Note.** Each bit has a value of 1 if the data quality passes the corresponding data quality requirement, the value is zero. Bits 7 and 8 were introduced in O4 and are not present in previous runs (R. Abbott et al. 2021a, 2023a).

### 5.2. Data Quality and Injections in GWOSC Files

The data quality information is encoded in the default strain release as a bitmask following the structure described in Table 6, with a 1 Hz sampling rate.[324] The meaning of the data quality flags is explained in Section 4. For template-based binary coalescence searches and minimally modeled searches, labeled respectively as CBC and BURST in Table 6, the bitmask allows 3 categories of data quality (labeled with `CAT1`, `CAT2`, and `CAT3`). We keep this structure even for runs in which not all the categories have been used. This choice avoids confusion that could arise if the same bit would have different meaning in different runs. A bit value of 1 indicates the data has passed both the indicated level and lower-level checks; for example, a 1 in the `CBC_CAT2` bit indicates the data passes both `CBC_CAT1` and `CBC_CAT2`.

The data quality bit labeled as `DATA` is obtained, requiring that both `CBC_CAT1` and `BURST_CAT1` are satisfied. This is the criterion that defines what data are released in the *default strain data release*.

In O4a, two additional bits have been added to accommodate Category 1 data quality for STOCH searches

[324] See tutorial 3 of https://github.com/gwosc-tutorial/introduction_gwosc_data to learn how to get data quality information from GWOSC files.

**Table 7**
Meaning of the Bits in the Injections Bitmask of the GWOSC Files

| Bit | Short Name | Description |
|---|---|---|
| 0 | NO_CBC_HW_INJ | No CBC injections |
| 1 | NO_BURST_HW_INJ | No BURST injections |
| 2 | NO_DETCHAR_HW_INJ | No DETCHAR injections |
| 3 | NO_CW_HW_INJ | No CW injections |
| 4 | NO_STOCH_HW_INJ | No STOCH injections |

**Note.** The mask indicates when injections are not present, so the bit value is set to 1 to indicate no injection and 0 when there is an injection. HW_INJ stands for hardware injection.

(STOCH_CAT1) and for CW searches (CW_CAT1). Moreover, Categories 2 and 3 have not been used for the CBC data quality flag, so the corresponding segments coincide with those marked as CBC_CAT1 (and also DATA in this case). BURST searches in O4a did not apply any Category 3 flags. The CW data quality flag CW_CAT1 is identical to CBC_CAT1.

As discussed in Section 4, a crucial way to test the response of the detector is the use of *hardware injections*. In the GWOSC files, the information about injections is provided as 1 Hz time series containing a bitmask detailed in Table 7.[325] Five bits are used to distinguish injections relevant for astrophysical searches or detector characterization studies.

### 5.3. Additional Segments in the O4a Strain Data Release

The O4a run started on 2023 May 24 at 15:00 UTC. However, a few segments of data from the engineering run just prior to the start of the run were used in the search for GW emitted from SN 2023ixf (A. G. Abac et al. 2025a) and for the neutron star–black hole binary candidate GW230518_125908 (LIGO Scientific Collaboration et al. 2025d), and they have been added to this release. These additional data segments span between times 2023 May 15 14:13:22 UTC (GPS 1368195220) and 2023 May 19 17:31:33 UTC (GPS 1368552711) when both LHO and LLO were in observing mode. These segments have a total duration of 0.8 days. Including this addition, the total amount of observing time in O4a is 160.8 days for LHO and 164.3 days for LLO (to compare with Table 1, 0.8 days have to be added to the coincident times). Using this total amount of time as a reference, the percentage of the observing time for which the data quality passes each category described in Table 6 in O4a can be calculated. Since these percentages are close to 100%, Table 8 summarizes the percentages of failing times.

### 5.4. Previous Runs' Data Releases

The structure described here for the default strain data is largely valid for all data released to date (R. Abbott et al. 2021a, 2023a), with few differences. These concern details in the naming conventions of the files and the channels within them. For example, the *FrameType* mentioned in Section 5.1 is *Ifo*_LOSC_*s*_V*n* in O1, the *StrainChannel* and *FrameType* have been present in files since O2, while the channel names in the gwf files used only in O1 are shown in Table 9.

[325] See tutorials in https://gwosc.org/tutorials/ to learn how to get hardware injection information from GWOSC files.

**Table 8**
Percentage of the Observing Time in O4a, Including the Extra 0.8 days of Engineering Run Data, for Which Data Quality Fails the Categories Described in Table 6.

| Short Name | Percentage in LHO | Percentage in LLO |
|---|---|---|
| DATA | 0.0971% | 0.0059% |
| CBC_CAT1 | 0.0971% | 0.0059% |
| CBC_CAT2 | 0.0971% | 0.0059% |
| CBC_CAT3 | 0.0971% | 0.0059% |
| BURST_CAT1 | 0.0971% | 0.0059% |
| BURST_CAT2 | 0.2242% | 0.1188% |
| BURST_CAT3 | 0.2242% | 0.1188% |
| STOCH_CAT1 | 0.0979% | 0.0065% |
| CW_CAT1 | 0.0971% | 0.0059% |

**Table 9**
Channel Names in the GWOSC gwf Files, Used Only in O1 for the 4 kHz Release (*Ifo* is the Interferometer Name)

| | Channel Name |
|---|---|
| Strain | *Ifo*:LOSC-STRAIN |
| Data quality mask | *Ifo*:LOSC-DQMASK |
| Injections mask | *Ifo*:LOSC-INJMASK |

The main difference that the user will notice is that up to O3, for each published GW detection, data snippets centered on the event detection time are also released via the GWOSC Event Portal, described in Section 7. These data snippets are provided as plain text files in addition to the hdf and gwf files described above. The text files contain strain values in a single column. As of O4, data snippets for most events are no longer published, but instead the associated files from the default strain data release are linked for each event in the Event Portal, if available at the time of posting the event.

### 5.5. Alternate Strain Release

The default strain files described in Sections 5.1–5.4 are prepared after each observing run in order to provide a user-friendly file format that includes the preferred strain channel and the final data quality and injection segments. For O3 and O4a, in addition to these user-friendly files, each data release also includes an alternate strain release, which includes several versions of the strain channel and more closely matches the data available to LVK members before the data release. Detailed documentation is available on the GWOSC website (LIGO Scientific Collaboration & Virgo Collaboration 2021a, 2021b; LIGO-Virgo-KAGRA Collaboration 2025a).

Alternate strain releases from O3 and O4a are available for all times the detectors passed the conditions for ANALYSIS_READY, meaning that the detectors were nominally in a good working state. This set of times includes a small amount of time that fails CAT1 data quality, and so there are some times in the alternate strain releases that are not available in the default strain files. Files and strain channels are marked with the tag AR, to indicate that only ANALYSIS_READY times are included. For O3, this release was prepared after the run had ended, with three channels of varying noise subtraction levels for LIGO, and two channels for Virgo: a main and a short-duration one covering a period when additional noise subtraction was needed (R. Abbott et al. 2023a).

**Table 10**
Alternate Strain Channels in the O4a Release and the Layers of Noise Subtraction Applied to Them

| Channel name | Narrowband Noise Subtraction | Broadband Noise Subtraction | Glitch Removal |
|---|---|---|---|
| *Ifo*:GDS-CALIB_STRAIN_AR | ⋯ | ⋯ | ⋯ |
| *Ifo*:GDS-CALIB_STRAIN_NOLINES_AR | ✓ | ⋯ | ⋯ |
| *Ifo*:GDS-CALIB_STRAIN_CLEAN_AR | ✓ | ✓ | ⋯ |
| *Ifo*:GDS-GATED_STRAIN_AR | ✓ | ⋯ | ✓ |

**Note.** NOLINES means that narrowband noise features have been subtracted, while CLEAN means that broadband noise features have also been subtracted, and GATED means the removal of loud glitches.

Starting in O4a, the AR frames are produced during the run. A calibration pipeline runs at all times and writes out strain data in 1 s long frame files in order to produce low-latency strain data, including the preferred channel GDS-CALIB_STRAIN_CLEAN. Low-latency analyses, including pipelines that produce public transient alerts, use these low-latency frame files. To produce the AR frames, a frame aggregator collects the 1 s files into ANALYSIS_READY frame files of up to 4096 s long, excludes any times that are not marked as ready for analysis, and adds the tag AR to the channel names. The AR frame files are then used for most offline LVK analyses, so that the released product exactly matches the data used inside the collaboration.

The alternate strain releases are not currently available for direct download from the GWOSC website; instead, they are available via OSDF and NDS2, as discussed earlier in Section 5.

The O4a alternate strain release includes four versions of the strain channel for each LIGO instrument, which are summarized in Table 10 and described below. In addition to the strain channels, the O4a ANALYSIS_READY frames also include a number of auxiliary channels that record data quality information and calibration parameters. The O4a strain channels are (*Ifo* corresponds to H1 for Hanford or L1 for Livingston):

1. *Ifo*:GDS-CALIB_STRAIN_AR This is the calibrated strain data with no noise subtraction applied and corresponds to the channel name DCS-CALIB_STRAIN_C01_AR in O3 (R. Abbott et al. 2023a).
2. *Ifo*:GDS-CALIB_STRAIN_NOLINES_AR This is the calibrated strain data after removing narrowband noise. This channel removes the calibration lines, 60 Hz power-mains line, and harmonics of the power-mains line as described in A. Viets & M. Wade (2021). It corresponds to the channel name DCS-CALIB_STRAIN_CLEAN_C01_AR in O3 (R. Abbott et al. 2023a).
3. *Ifo*:GDS-CALIB_STRAIN_CLEAN_AR This is the calibrated strain data, after subtracting both narrowband and broadband noise. This applies additional broadband noise subtraction onto the *Ifo*: GDS-CALIB_-STRAIN_NOLINES_AR channel, using the method described in G. Vajente et al. (2020). This is the recommended channel for transient searches, and it corresponds to the channel name DCS-CALIB_STRAIN_CLEAN_SUB60HZ_C01_AR in O3 (R. Abbott et al. 2023a).
4. *Ifo*:GDS-GATED_STRAIN_AR This is the *Ifo*: GDS-CALIB_STRAIN_NOLINES_AR channel where times with loud glitches are replaced with zeros by applying a window function around short segments of elevated noise (S. A. Usman et al. 2016; J. Zweizig & K. Riles 2020). This channel was not used for analyses by the LVK.

During some intervals of O4a, the broadband noise subtraction was turned off, so that the channels *Ifo:*GDS-CALIB_STRAIN_NOLINES_AR and *Ifo*:GDS-CALIB_-STRAIN_CLEAN_AR contained identical data. At LHO, broadband noise subtraction was turned off from 2023 June 21 16:00 UTC until 2023 November 11 01:00:00 UTC. At LLO, this period was from 2023 May 24 15:00:00 UTC until 2023 November 10 23:00:00 UTC.

## 6. Auxiliary Channels

In addition to the strain time series, LIGO records around 200,000 *auxiliary channels* for each instrument. Auxiliary channels are stored as time series data and record the state of the instrument, the local environment, and a number of digital-filter settings. Auxiliary channels include readings from seismometers, accelerometers, photodiodes, magnetometers, wind monitors, and microphones (P. Nguyen et al. 2021; R. Huxford et al. 2024).

The O3 and O4a data releases include the set of LIGO auxiliary channels used for noise subtraction and to produce data quality flags. These are the ways that auxiliary channels directly contribute to the GWTC-4.0 analysis. The auxiliary data release consists of about 40 channels for each LIGO detector for each run. The online documentation includes the name of each available channel along with the associated sampling rate and a short description of the channel content.[326]

Auxiliary channels should be used with caution, as the data are not necessarily calibrated and may include times when the data values are missing or corrupt. For example, sensors sometimes break or become disconnected, leading to periods of corrupted data. Auxiliary data are available at all times when the detector is in ANALYSIS_READY, with the exact times marked in segment lists that are available with the data release documentation.

In addition to channels used for noise subtraction and data quality flags, some auxiliary data have been released on request and are now available on the GWOSC website. For O3, these releases include a dataset to support a study of machine learning applied to detector characterization (R. Gurav et al. 2024) and a study of noise subtraction methods (B. Zackay et al. 2023).

Auxiliary channel data cannot currently be downloaded directly through the GWOSC website or the API. Instead,

[326] See https://gwosc.org/O3/auxiliary/ for O3 and https://gwosc.org/O4/auxiliary/ for O4a.

**Table 11**
List of Releases Currently Available via the GWOSC Event Portal[a]

| Release Name | Notes |
|---|---|
| `GWTC` | Cumulative set of GW transients maintained by the LVK (A. G. Abac et al. 2025c) |
| `Initial_LIGO_Virgo` | Event release from initial LIGO and Virgo, 2005–2010 (J. Abadie et al. 2012) |
| `O1_O2-Preliminary` | Notable events in O1 and O2 published prior to the GWTC-1 release (B. P. Abbott et al. 2019) |
| `GWTC-1-confident` | Confident detections from the O1 and O2 runs (B. P. Abbott et al. 2019) |
| `GWTC-1-marginal` | Marginal candidates from the O1 and O2 runs (B. P. Abbott et al. 2019) |
| `GWTC-2` | Confident detections from the O3a run (R. Abbott et al. 2021c) |
| `GWTC-2.1-confident` | Confident detections from the O3a run based on a revised analysis (R. Abbott et al. 2024) |
| `GWTC-2.1-marginal` | Marginal candidates from O3a run based on a revised analysis (R. Abbott et al. 2024) |
| `GWTC-2.1-auxiliary` | Candidates from GWTC-2 that were demoted in GWTC-2.1 (R. Abbott et al. 2021c) |
| `GWTC-3-confident` | Confident detection from the O3b run (R. Abbott et al. 2023b) |
| `GWTC-3-marginal` | Marginal candidates from the O3b run (R. Abbott et al. 2023b) |
| `IAS-O3a` | Events from O3a as described in a community catalog (S. Olsen et al. 2022) |
| `O3_Discovery_Papers` | Notable events in O3 run published independently (B. P. Abbott et al. 2020c; R. Abbott et al. 2020a, 2020c, 2020b, 2021d) |
| `O3_IMBH_marginal` | Marginal intermediate-mass black hole candidates from O3 run (R. Abbott et al. 2022b) |
| `O4_Discovery_Papers` | Notable events in O4 run published independently (A. G. Abac et al. 2024, 2025b) |
| `GWTC-4.0` | All events from the O4a data product release (LIGO Scientific Collaboration et al. 2025d) |

**Note.**
[a] GWOSC Event Portal, https://gwosc.org/eventapi/html.

auxiliary channel data are available through the NDS2 and OSDF interfaces described in Section 5.

## 7. Event Portal

The GWOSC Event Portal provides an online interface to a database of published GW transient events.[327] These events are short-duration GW signals identified in the strain data by search analyses. They are typically attributed to a signal consistent with a CBC source. Not all events in the Event Portal will be from CBCs; some may be noise with an instrumental origin.

Users of the Event Portal can discover an array of data products associated with each event, including strain data, segment lists, detection confidence, astrophysical source parameters, and documentation. In most cases, the database also includes links to additional analysis products, such as posterior samples describing source properties and source localizations.

The Event Portal may be accessed as browsable HTML web pages or via a REST API. A Python client, named *gwosc*, can be used to query the API and download content.[328] Documentation of the Event Portal and REST API are available on the GWOSC website.

### *7.1. User Interfaces to the Data*

A collection of GW transients in the Event Portal is labeled a *release*. A release might be a set of events published in a single paper, such as the new CBC candidates added for GWTC-4.0 (A. G. Abac et al. 2025c), or a collection of events that are related in some other way, such as all the GW transients published in stand-alone discovery papers, such as the O3 observing run. As a result, there are a number of ways to interact with the data through the Event Portal.

The *Release List View* in the Event Portal displays a list of all available releases, summarized in Table 11. Several releases are labeled marginal, i.e., `GWTC-1-marginal`, `GWTC-2.1-marginal`, `GWTC-3-marginal`, `O3_IMBH_marginal`, which include candidate triggers that have a plausible instrumental origin, falling short of the criteria required for the event to be labeled as confident.

In addition, there is one release, `GWTC-2.1-auxilary`, in which revised analyses in `GWTC-2.1`(R. Abbott et al. 2024) resulted in the demotion of events that were previously identified in `GWTC-2`(R. Abbott et al. 2021c) as statistically significant. They are included separately for completeness.

With the exception of `GWTC-4.0`, events available in the individual releases are documented in the tables found in the publications that first reported the events. In the case of `GWTC-4.0` the events found in the Event Portal satisfy

$$p_{\rm astro} > 0.5 \quad \text{or} \quad \text{FAR} < 1\ \text{yr}^{-1}, \tag{3}$$

where $p_{\rm astro}$ is the probability that a GW candidate is of astrophysical origin and FAR is the FAR (the rate of noise triggers with a detection statistic at least as high as the candidate).

Detailed documentation and references to publications are available for each release on the GWOSC website.

By selecting one of the releases, an *Event List* table is generated showing all events in the release, along with source parameters and credible intervals where available. The Event List view includes estimates of a number of physical parameters, such as mass, distance, and spin, as well as search-pipeline outputs such as SNR and FAR. If more than one parameter estimation analysis is available for any event, then values from the analysis marked as `Default PE` for the specific event are displayed in the Event List to indicate which of the multiple results are captured in the table. Additionally, the lowest FAR and highest probability of astrophysical origin $p_{\rm astro}$ from all available search-analysis results are included in the Event List table.

A special case of the Event List view is the electronic version of the GWTC, a collection of GW candidates identified by the LVK (A. G. Abac et al. 2025c). Selecting the GWTC option returns this set of events as an Event List. It includes confidently detected events from multiple data releases.

[327] GWOSC Event Portal, https://gwosc.org/eventapi.
[328] GWOSC Client API, https://pypi.org/project/gwosc/.

Event Lists are also available to browse specific events using customized queries. Event Lists may include more than one version of an event, as some events have appeared in multiple publications or have been reanalyzed with newer code or models. The *Query Page* presents the user with a form that allows selection of events based on name, membership in particular releases, various ranges of masses, distance, redshift, effective inspiral spin ($\chi_{\rm eff}$), $p_{\rm astro}$, SNR, FAR, UTC, and GPS times. The query also allows specification of the output format to return, with support for HTML, JSON, CSV, and ASCII.

The *Single Event View* shows key parameters for an individual event in the Event Portal. The view can include sets of parameters from different pipelines, including search-pipeline outputs such as SNR and FAR, and inferred source properties such as masses and spins. The Single Event View also includes images showing a spectrogram (S. K. Chatterji 2005) of the strain data for each detector at the time of the event, and links to download the associated strain time series in multiple file formats, as described in Section 5. The Single Event View also provides links to documentation, low-latency information in the Gravitational-Wave Candidate Event Database (GraceDB),[329] General Coordinates Network Circulars and Notices, and links to segment lists for times around the event (see Section 5.2 for details). A new feature found on these pages is the Event Viewer, which leads to a web app that generates plots of waveforms, source parameters, and sky localization for each GW event.[330]

### 7.2. Parameter Estimation Results

Most events in the Event Portal include one or more sets of credible intervals drawn from the publication associated with the event. The naming of parameters is based on the LVK standard names convention.[331] Many entries include multiple sets of values attributed to different priors, waveforms, or analysis changes (LIGO Scientific Collaboration et al. 2025c). In some cases, a publication will present credible intervals describing samples combined from several analyses using different waveform models. The results for any individual parameter are expressed as the median value with its 90% symmetric credible interval. Details regarding how the posterior samples were constructed can be found in the publication from which the event is released or in the supplemental releases linked from the Single Event View.

### 7.3. Supplemental Data Releases

For most events, the Single Event View includes links to supplemental data. These data vary between releases, but in general will include links to the parameter estimation posterior samples, source localization, and additional resources. Supplemental data releases are published either in the LIGO Document Control Center[332] or under the LVK Zenodo Community.[333] In addition, GWOSC maintains a snapshot of the Event Portal's version history as a dataset available on Zenodo.[334] Each snapshot consists of JSON files downloaded from the Event Portal with data for each event.

### 7.4. Community Catalogs

The Event Portal has recently been extended to provide GW events from catalog authors outside of the LVK, which are collectively organized as *Community Catalogs*. The presentation of results for Community Catalogs is identical to that available with any LVK release of events, adhering to the website layout and search capabilities discussed previously. This is achieved by standardizing on a `JSON` schema file format for GWOSC staff to use to ingest search and parameter estimation values. The details for this standard JSON are publicly available in GitHub,[335] along with useful test scripts, notebooks, and documentation. The criteria for authors outside the LVK to have their catalogs included in GWOSC are documented in the *GWOSC Community Catalogs Guidelines*.[336]

## 8. Summary

Data from LIGO recorded during O4a are now publicly available through the GWOSC website at gwosc.org. This paper can serve as a practical guide for users interested in analyzing this set of open data. It provides important details about calibration and data quality, explains the structure of the datasets available online, and offers instructions for using the Event Portal, which gives access to the list of astrophysical sources identified by the LVK.

Making these resources available maximizes the scientific potential of the dataset. The O4a data contain a wealth of compact object mergers, many of which are listed in version 4.0 of the GWTC (LIGO Scientific Collaboration et al. 2025d). By releasing the data publicly, the community is empowered to further explore and discover new insights.

The present release covers only the first part of the O4 observing run. The second and third parts are planned for release in 2026 May and December, respectively (S. Anderson et al. 2025). With these forthcoming datasets, we anticipate even more opportunities for discovery.

## Acknowledgments

This material is based upon work supported by NSF's LIGO Laboratory, which is a major facility fully funded by the National Science Foundation. The authors also gratefully acknowledge the support of the Science and Technology Facilities Council (STFC) of the United Kingdom, the Max Planck Society (MPS), and the State of Niedersachsen/Germany for support of the construction of Advanced LIGO and the construction and operation of the GEO 600 detector. Additional support for Advanced LIGO was provided by the Australian Research Council. The authors gratefully acknowledge the Italian Istituto Nazionale di Fisica Nucleare (INFN), the French Centre National de la Recherche Scientifique (CNRS), and the Netherlands Organization for Scientific

[329] GraceDB website, https://gracedb.ligo.org/.
[330] GW Event Viewer, https://peviewer.igwn.org.
[331] Parameter Estimation Name Standard, https://lscsoft.docs.ligo.org/pesummary/stable/gw/parameters.html.
[332] Public LIGO Document Control Center, https://dcc.ligo.org/cgi-bin/DocDB/DocumentDatabase.
[333] LVK Zenodo Data Releases, https://zenodo.org/communities/ligo-virgo-kagra/.
[334] GWOSC Zenodo Event Portal Snapshots, https://zenodo.org/records/10071492.
[335] GWOSC GitHub Community Catalog Schema, https://github.com/gwosc-tutorial/gwosc-catalog.
[336] GWOSC Community Catalogs Guidelines, https://dcc.ligo.org/LIGO-M2500012/public.

Research (NWO) for the construction and operation of the Virgo detector and the creation and support of the EGO consortium. The authors also gratefully acknowledge research support from these agencies as well as by the Council of Scientific and Industrial Research of India, the Department of Science and Technology, India, the Science & Engineering Research Board (SERB), India, the Ministry of Human Resource Development, India, the Spanish Agencia Estatal de Investigación (AEI), the Spanish Ministerio de Ciencia, Innovación y Universidades, the European Union NextGenerationEU/PRTR (PRTR-C17.I1), the ICSC—CentroNazionale di Ricerca in High Performance Computing, Big Data and Quantum Computing, funded by the European Union NextGenerationEU, the Comunitat Autonòma de les Illes Balears through the Conselleria d'Educació i Universitats, the Conselleria d'Innovació, Universitats, Ciència i Societat Digital de la Generalitat Valenciana and the CERCA Programme Generalitat de Catalunya, Spain, the Polish National Agency for Academic Exchange, the National Science Centre of Poland and the European Union—European Regional Development Fund; the Foundation for Polish Science (FNP), the Polish Ministry of Science and Higher Education, the Swiss National Science Foundation (SNSF), the Russian Science Foundation, the European Commission, the European Social Funds (ESF), the European Regional Development Funds (ERDF), the Royal Society, the Scottish Funding Council, the Scottish Universities Physics Alliance, the Hungarian Scientific Research Fund (OTKA), the French Lyon Institute of Origins (LIO), the Belgian Fonds de la Recherche Scientifique (FRS-FNRS), Actions de Recherche Concertées (ARC) and Fonds Wetenschappelijk Onderzoek—Vlaanderen (FWO), Belgium, the Paris Île-de-France Region, the National Research, Development and Innovation Office of Hungary (NKFIH), the National Research Foundation of Korea, the Natural Sciences and Engineering Research Council of Canada (NSERC), the Canadian Foundation for Innovation (CFI), the Brazilian Ministry of Science, Technology, and Innovations, the International Center for Theoretical Physics South American Institute for Fundamental Research (ICTP-SAIFR), the Research Grants Council of Hong Kong, the National Natural Science Foundation of China (NSFC), the Israel Science Foundation (ISF), the US-Israel Binational Science Fund (BSF), the Leverhulme Trust, the Research Corporation, the National Science and Technology Council (NSTC), Taiwan, the United States Department of Energy, and the Kavli Foundation. The authors gratefully acknowledge the support of the NSF, STFC, INFN, and CNRS for the provision of computational resources.

This work was supported by MEXT, the JSPS Leading-edge Research Infrastructure Program, JSPS Grant-in-Aid for Specially Promoted Research 26000005, JSPS Grant-in-Aid for Scientific Research on Innovative Areas 2402: 24103006, 24103005, and 2905: JP17H06358, JP17H06361, and JP17H06364, JSPS Core-to-Core Program A. Advanced Research Networks, JSPS Grants-in-Aid for Scientific Research (S) 17H06133 and 20H05639, JSPS Grant-in-Aid for Transformative Research Areas (A) 20A203: JP20H05854, the joint research program of the Institute for Cosmic Ray Research, University of Tokyo, the National Research Foundation (NRF), the Computing Infrastructure Project of the Global Science experimental Data hub Center (GSDC) at KISTI, the Korea Astronomy and Space Science Institute (KASI), the Ministry of Science and ICT (MSIT) in Korea, Academia Sinica (AS), the AS Grid Center (ASGC) and the National Science and Technology Council (NSTC) in Taiwan under grants including the Science Vanguard Research Program, the Advanced Technology Center (ATC) of NAOJ, and the Mechanical Engineering Center of KEK.

Additional acknowledgments for support of individual authors may be found in the following document: https://dcc.ligo.org/LIGO-M2300033/public. For the purpose of open access, the authors have applied a Creative Commons Attribution (CC BY) license to any Author Accepted Manuscript version arising. We request that citations to this article use “A. G. Abac et al. (LIGO-Virgo-KAGRA Collaboration), ...” or similar phrasing, depending on journal convention.

*Software:* This software infrastructure used to publish data products onto the GWOSC website made significant use of the IGWN software environment (LIGO-Virgo-KAGRA Collaboration et al. 2025e) curated by the LVK, to provide a reproducible computing environment built around tools needed for GW data analysis. Plots were prepared with `Matplotlib` (J. D. Hunter 2007).

## Data Availability

All data products described in this work are publicly available on the Gravitational Wave Open Science Center,[337] the Zenodo LVK community page,[338] and through the OSDF and NDS2 data repositories, as described in the previous sections of this paper. Data products are provided under a Creative Commons Attribution 4.0 International license.[339]

## ORCID iDs

A. G. Abac https://orcid.org/0000-0003-4786-2698
K. Ackley https://orcid.org/0000-0002-8648-0767
C. Adamcewicz https://orcid.org/0000-0001-5525-6255
S. Adhicary https://orcid.org/0009-0004-2101-5428
N. Adhikari https://orcid.org/0000-0002-4559-8427
R. X. Adhikari https://orcid.org/0000-0002-5731-5076
S. Afroz https://orcid.org/0009-0004-4459-2981
D. Agarwal https://orcid.org/0000-0002-8735-5554
M. Agathos https://orcid.org/0000-0002-9072-1121
O. D. Aguiar https://orcid.org/0000-0002-2139-4390
L. Aiello https://orcid.org/0000-0003-2771-8816
A. Ain https://orcid.org/0000-0003-4534-4619
P. Ajith https://orcid.org/0000-0001-7519-2439
T. Akutsu https://orcid.org/0000-0003-0733-7530
S. Albanesi https://orcid.org/0000-0001-7345-4415
A. Allocca https://orcid.org/0000-0002-5288-1351
P. A. Altin https://orcid.org/0000-0001-8193-5825
S. Alvarez-Lopez https://orcid.org/0009-0003-8040-4936
A. Amato https://orcid.org/0000-0001-9557-651X
F. Amicucci https://orcid.org/0009-0005-2139-4197
S. B. Anderson https://orcid.org/0000-0003-2219-9383
W. G. Anderson https://orcid.org/0000-0003-0482-5942
M. Andia https://orcid.org/0000-0003-3675-9126
M. Andrés-Carcasona https://orcid.org/0000-0002-8738-1672
T. Andrić https://orcid.org/0000-0002-9277-9773

[337] GWOSC Home Page: https://gwosc.org.
[338] Zenodo LVK Community Page: https://zenodo.org/communities/ligo-virgo-kagra/
[339] CC BY 4.0 license: https://creativecommons.org/licenses/by/4.0/

S. Ansoldi https://orcid.org/0000-0002-5613-7693
J. M. Antelis https://orcid.org/0000-0003-3377-0813
S. Antier https://orcid.org/0000-0002-7686-3334
S. K. Apple https://orcid.org/0009-0007-4490-5804
K. Arai https://orcid.org/0000-0001-8916-8915
A. Araya https://orcid.org/0000-0002-6884-2875
M. C. Araya https://orcid.org/0000-0002-6018-6447
M. Arca Sedda https://orcid.org/0000-0002-3987-0519
J. S. Areeda https://orcid.org/0000-0003-0266-7936
F. Armato https://orcid.org/0000-0002-8856-8877
S. Armstrong https://orcid.org/6512-3515-4685-5112
N. Arnaud https://orcid.org/0000-0001-6589-8673
M. Arogeti https://orcid.org/0000-0001-5124-3350
S. M. Aronson https://orcid.org/0000-0001-7080-8177
G. Ashton https://orcid.org/0000-0001-7288-2231
Y. Aso https://orcid.org/0000-0002-1902-6695
P. Astone https://orcid.org/0000-0003-4981-4120
F. Attadio https://orcid.org/0009-0008-8916-1658
F. Aubin https://orcid.org/0000-0003-1613-3142
K. AultONeal https://orcid.org/0000-0002-6645-4473
G. Avallone https://orcid.org/0000-0001-5482-0299
E. A. Avila https://orcid.org/0009-0008-9329-4525
S. Babak https://orcid.org/0000-0001-7469-4250
S. Bae https://orcid.org/0000-0003-2429-3357
S. Bagnasco https://orcid.org/0000-0001-6062-6505
L. Baiotti https://orcid.org/0000-0003-0458-4288
R. Bajpai https://orcid.org/0000-0003-0495-5720
T. Baker https://orcid.org/0000-0001-5470-7616
G. Baldi https://orcid.org/0000-0001-8963-3362
N. Baldicchi https://orcid.org/0009-0009-8888-291X
S. Banagiri https://orcid.org/0000-0001-7852-7484
B. Banerjee https://orcid.org/0000-0002-8008-2485
D. Bankar https://orcid.org/0000-0002-6068-2993
P. Baral https://orcid.org/0000-0001-6308-211X
M. Baratti https://orcid.org/0009-0003-5744-8025
P. Barneo https://orcid.org/0000-0002-8883-7280
F. Barone https://orcid.org/0000-0002-8069-8490
B. Barr https://orcid.org/0000-0002-5232-2736
L. Barsotti https://orcid.org/0000-0001-9819-2562
M. Barsuglia https://orcid.org/0000-0002-1180-4050
D. Barta https://orcid.org/0000-0001-6841-550X
M. A. Barton https://orcid.org/0000-0002-9948-306X
A. Basalaev https://orcid.org/0000-0001-5623-2853
R. Bassiri https://orcid.org/0000-0001-8171-6833
A. Basti https://orcid.org/0000-0003-2895-9638
M. Bawaj https://orcid.org/0000-0003-3611-3042
J. C. Bayley https://orcid.org/0000-0003-2306-4106
A. C. Baylor https://orcid.org/0000-0003-0918-0864
F. Beirnaert https://orcid.org/0000-0002-4003-7233
M. Bejger https://orcid.org/0000-0002-4991-8213
D. Belardinelli https://orcid.org/0000-0001-9332-5733
A. S. Bell https://orcid.org/0000-0003-1523-0821
L. Bellizzi https://orcid.org/0000-0002-2071-0400
W. Benoit https://orcid.org/0000-0003-4750-9413
I. Bentara https://orcid.org/0009-0000-5074-839X
J. D. Bentley https://orcid.org/0000-0002-4736-7403
S. Bera https://orcid.org/0000-0003-0907-6098
F. Bergamin https://orcid.org/0000-0002-1113-9644
B. K. Berger https://orcid.org/0000-0002-4845-8737
S. Bernuzzi https://orcid.org/0000-0002-2334-0935
M. Beroiz https://orcid.org/0000-0001-6486-9897
C. P. L. Berry https://orcid.org/0000-0003-3870-7215
D. Bersanetti https://orcid.org/0000-0002-7377-415X
J. Betzwieser https://orcid.org/0000-0003-1533-9229
D. Beveridge https://orcid.org/0000-0002-1481-1993
G. Bevilacqua https://orcid.org/0000-0002-7298-6185
N. Bevins https://orcid.org/0000-0002-4312-4287
D. Bhattacharjee https://orcid.org/0000-0001-6623-9506
S. Bhaumik https://orcid.org/0000-0001-8492-2202
V. Biancalana https://orcid.org/0000-0002-1642-5391
G. Billingsley https://orcid.org/0000-0002-4141-2744
A. Binetti https://orcid.org/0000-0001-6449-5493
S. Bini https://orcid.org/0000-0002-0267-3562
O. Birnholtz https://orcid.org/0000-0002-7562-9263
S. Biscoveanu https://orcid.org/0000-0001-7616-7366
M. Bitossi https://orcid.org/0000-0002-9862-4668
M.-A. Bizouard https://orcid.org/0000-0002-4618-1674
J. K. Blackburn https://orcid.org/0000-0002-3838-2986
N. Bode https://orcid.org/0000-0002-7101-9396
G. Boileau https://orcid.org/0000-0002-3576-6968
M. Boldrini https://orcid.org/0000-0001-9861-821X
G. N. Bolingbroke https://orcid.org/0000-0002-7350-5291
L. D. Bonavena https://orcid.org/0000-0002-2630-6724
R. Bondarescu https://orcid.org/0000-0003-0330-2736
F. Bondu https://orcid.org/0000-0001-6487-5197
E. Bonilla https://orcid.org/0000-0002-6284-9769
M. S. Bonilla https://orcid.org/0000-0003-4502-528X
R. Bonnand https://orcid.org/0000-0001-5013-5913
V. Boschi https://orcid.org/0000-0001-8665-2293
Y. Bothra https://orcid.org/0000-0002-9380-6390
P. R. Brady https://orcid.org/0000-0002-4611-9387
M. Branchesi https://orcid.org/0000-0003-1643-0526
T. Briant https://orcid.org/0000-0002-6013-1729
E. Brockmueller https://orcid.org/0000-0002-1489-942X
A. F. Brooks https://orcid.org/0000-0003-4295-792X
M. L. Brozzetti https://orcid.org/0000-0002-5260-4979
R. Bruntz https://orcid.org/0000-0002-0840-8567
F. Bucci https://orcid.org/0000-0003-1726-3838
O. Bulashenko https://orcid.org/0000-0003-1720-4061
A. Buonanno https://orcid.org/0000-0002-5433-1409
R. Buscicchio https://orcid.org/0000-0002-7387-6754
C. Buy https://orcid.org/0000-0003-2872-8186
G. S. Cabourn Davies https://orcid.org/0000-0002-4289-3439
R. Cabrita https://orcid.org/0000-0003-0133-1306
V. Cáceres-Barbosa https://orcid.org/0000-0001-9834-4781
L. Cadonati https://orcid.org/0000-0002-9846-166X
G. Cagnoli https://orcid.org/0000-0002-7086-6550
C. Cahillane https://orcid.org/0000-0002-3888-314X
S. R. Callos https://orcid.org/0000-0003-0639-9342
G. Caneva Santoro https://orcid.org/0000-0002-2935-1600
K. C. Cannon https://orcid.org/0000-0003-4068-6572
E. Capocasa https://orcid.org/0000-0003-3762-6958
E. Capote https://orcid.org/0009-0007-0246-713X
G. Capurri https://orcid.org/0000-0003-0889-1015
J. B. Carlin https://orcid.org/0000-0001-5694-0809
M. Carpinelli https://orcid.org/0000-0002-8205-930X
J. J. Carter https://orcid.org/0000-0001-8845-0900
G. Carullo https://orcid.org/0000-0001-9090-1862
J. Casanueva Diaz https://orcid.org/0000-0002-2948-5238
C. Casentini https://orcid.org/0000-0001-8100-0579
M. Cavaglià https://orcid.org/0000-0002-3835-6729
R. Cavalieri https://orcid.org/0000-0001-6064-0569

G. Cella https://orcid.org/0000-0002-0752-0338
P. Cerdá-Durán https://orcid.org/0000-0003-4293-340X
E. Cesarini https://orcid.org/0000-0001-9127-3167
A. Chakraborty https://orcid.org/0009-0004-4937-4633
P. Chakraborty https://orcid.org/0000-0002-0994-7394
S. Chalathadka Subrahmanya https://orcid.org/0000-0002-9207-4669
J. C. L. Chan https://orcid.org/0000-0002-3377-4737
S. Chao https://orcid.org/0000-0003-3853-3593
P. Charlton https://orcid.org/0000-0002-4263-2706
E. Chassande-Mottin https://orcid.org/0000-0003-3768-9908
C. Chatterjee https://orcid.org/0000-0001-8700-3455
Debarati Chatterjee https://orcid.org/0000-0002-0995-2329
Deep Chatterjee https://orcid.org/0000-0003-0038-5468
S. Chaty https://orcid.org/0000-0002-5769-8601
K. Chatziioannou https://orcid.org/0000-0002-5833-413X
A. Chen https://orcid.org/0000-0001-9174-7780
D. Chen https://orcid.org/0000-0003-1433-0716
H. Y. Chen https://orcid.org/0000-0001-5403-3762
Yitian Chen https://orcid.org/0000-0002-8664-9702
P. Chessa https://orcid.org/0000-0001-9092-3965
H. T. Cheung https://orcid.org/0000-0003-3905-0665
F. Chiadini https://orcid.org/0000-0002-9339-8622
A. Chincarini https://orcid.org/0000-0003-4094-9942
M. L. Chiofalo https://orcid.org/0000-0002-6992-5963
A. Chiummo https://orcid.org/0000-0003-2165-2967
S. Choudhary https://orcid.org/0000-0003-0949-7298
N. Christensen https://orcid.org/0000-0002-6870-4202
S. S. Y. Chua https://orcid.org/0000-0001-8026-7597
G. Ciani https://orcid.org/0000-0003-4258-9338
P. Ciecielag https://orcid.org/0000-0002-5871-4730
M. Cieślar https://orcid.org/0000-0001-8912-5587
M. Cifaldi https://orcid.org/0009-0007-1566-7093
J. A. Clark https://orcid.org/0000-0003-3243-1393
T. A. Clarke https://orcid.org/0000-0002-6714-5429
E. Codazzo https://orcid.org/0000-0001-7170-8733
P.-F. Cohadon https://orcid.org/0000-0003-3452-9415
S. Colace https://orcid.org/0009-0007-9429-1847
M. Colleoni https://orcid.org/0000-0002-7214-9088
S. Colloms https://orcid.org/0009-0009-9828-3646
A. Colombo https://orcid.org/0000-0002-7439-4773
L. Conti https://orcid.org/0000-0003-2731-2656
T. R. Corbitt https://orcid.org/0000-0002-5520-8541
I. Cordero-Carrión https://orcid.org/0000-0002-1985-1361
S. Corezzi https://orcid.org/0000-0002-3437-5949
N. J. Cornish https://orcid.org/0000-0002-7435-0869
A. Corsi https://orcid.org/0000-0001-8104-3536
M. W. Coughlin https://orcid.org/0000-0002-8262-2924
P. Couvares https://orcid.org/0000-0002-2823-3127
R. Coyne https://orcid.org/0000-0002-5243-5917
J. D. E. Creighton https://orcid.org/0000-0003-3600-2406
P. Cremonese https://orcid.org/0000-0001-6472-8509
J. R. Cudell https://orcid.org/0000-0002-2003-4238
T. J. Cullen https://orcid.org/0000-0001-8075-4088
A. Cumming https://orcid.org/0000-0003-4096-7542
E. Cuoco https://orcid.org/0000-0002-6528-3449
M. Cusinato https://orcid.org/0000-0003-4075-4539
T. Dal Canton https://orcid.org/0000-0001-5078-9044
S. Dal Pra https://orcid.org/0000-0002-1057-2307
G. Dálya https://orcid.org/0000-0003-3258-5763
B. D'Angelo https://orcid.org/0000-0001-9143-8427
S. Danilishin https://orcid.org/0000-0001-7758-7493
S. D'Antonio https://orcid.org/0000-0003-0898-6030
L. P. Dartez https://orcid.org/0000-0002-2216-0465
V. Dattilo https://orcid.org/0000-0002-8816-8566
D. Davis https://orcid.org/0000-0001-5620-6751
M. C. Davis https://orcid.org/0000-0001-7663-0808
P. Davis https://orcid.org/0009-0004-5008-5660
E. J. Daw https://orcid.org/0000-0002-3780-5430
M. Dax https://orcid.org/0000-0001-8798-0627
J. De Bolle https://orcid.org/0000-0002-5179-1725
J. Degallaix https://orcid.org/0000-0002-1019-6911
M. De Laurentis https://orcid.org/0000-0002-3815-4078
F. De Lillo https://orcid.org/0000-0003-4977-0789
S. Della Torre https://orcid.org/0000-0002-7669-0859
W. Del Pozzo https://orcid.org/0000-0003-3978-2030
F. De Marco https://orcid.org/0000-0002-5411-9424
F. De Matteis https://orcid.org/0000-0001-7860-9754
T. Dent https://orcid.org/0000-0003-1354-7809
A. Depasse https://orcid.org/0000-0003-1014-8394
R. De Pietri https://orcid.org/0000-0003-1556-8304
R. De Rosa https://orcid.org/0000-0002-4004-947X
C. De Rossi https://orcid.org/0000-0002-5825-472X
M. Desai https://orcid.org/0009-0003-4448-3681
R. DeSalvo https://orcid.org/0000-0002-4818-0296
A. Dhani https://orcid.org/0000-0001-9930-9101
M. C. Díaz https://orcid.org/0000-0002-7555-8856
M. Di Cesare https://orcid.org/0009-0003-0411-6043
T. Dietrich https://orcid.org/0000-0003-2374-307X
C. Di Fronzo https://orcid.org/0000-0002-2693-6769
M. Di Giovanni https://orcid.org/0000-0003-4049-8336
T. Di Girolamo https://orcid.org/0000-0003-2339-4471
J. Ding https://orcid.org/0000-0003-1693-3828
S. Di Pace https://orcid.org/0000-0001-6759-5676
I. Di Palma https://orcid.org/0000-0003-1544-8943
F. Di Renzo https://orcid.org/0000-0002-5447-3810
Divyajyoti https://orcid.org/0000-0002-2787-1012
A. Dmitriev https://orcid.org/0000-0002-0314-956X
Z. Doctor https://orcid.org/0000-0002-2077-4914
L. D'Onofrio https://orcid.org/0000-0001-9546-5959
K. L. Dooley https://orcid.org/0000-0002-1636-0233
S. Doravari https://orcid.org/0000-0001-8750-8330
M. Drago https://orcid.org/0000-0002-3738-2431
J. C. Driggers https://orcid.org/0000-0002-6134-7628
M. Dubois https://orcid.org/0000-0003-1490-7271
L. Dunn https://orcid.org/0000-0002-1769-6097
P.-A. Duverne https://orcid.org/0000-0002-3906-0997
D. D'Urso https://orcid.org/0000-0002-8215-4542
H. Duval https://orcid.org/0000-0002-2475-1728
M. Ebersold https://orcid.org/0000-0003-4631-1771
T. Eckhardt https://orcid.org/0000-0002-1224-4681
G. Eddolls https://orcid.org/0000-0002-5895-4523
A. Effler https://orcid.org/0000-0001-8242-3944
J. Eichholz https://orcid.org/0000-0002-2643-163X
M. Emma https://orcid.org/0000-0001-7943-0262
R. Enficiaud https://orcid.org/0000-0003-3908-1912
L. Errico https://orcid.org/0000-0003-2112-0653
M. Esposito https://orcid.org/0009-0009-8482-9417
R. C. Essick https://orcid.org/0000-0001-8196-9267
H. Estellés https://orcid.org/0000-0001-6143-5532
M. Evans https://orcid.org/0000-0001-8459-4499
J. M. Ezquiaga https://orcid.org/0000-0002-7213-3211
F. Fabrizi https://orcid.org/0000-0002-3809-065X

V. Fafone https://orcid.org/0000-0003-1314-1622
S. Fairhurst https://orcid.org/0000-0001-8480-1961
A. M. Farah https://orcid.org/0000-0002-6121-0285
B. Farr https://orcid.org/0000-0002-2916-9200
W. M. Farr https://orcid.org/0000-0003-1540-8562
G. Favaro https://orcid.org/0000-0002-0351-6833
M. Favata https://orcid.org/0000-0001-8270-9512
M. Fays https://orcid.org/0000-0002-4390-9746
M. Fazio https://orcid.org/0000-0002-9057-9663
R. Felicetti https://orcid.org/0009-0005-6263-5604
E. Fenyvesi https://orcid.org/0000-0003-2777-3719
T. Fernandes https://orcid.org/0009-0006-6820-2065
S. Ferraiuolo https://orcid.org/0009-0005-5582-2989
F. Fidecaro https://orcid.org/0000-0002-6189-3311
P. Figura https://orcid.org/0000-0002-8925-0393
A. Fiori https://orcid.org/0000-0003-3174-0688
I. Fiori https://orcid.org/0000-0002-0210-516X
M. Fishbach https://orcid.org/0000-0002-1980-5293
R. Fittipaldi https://orcid.org/0000-0003-2096-7983
V. Fiumara https://orcid.org/0000-0003-3644-217X
S. M. Fleischer https://orcid.org/0000-0001-7884-9993
J. A. Font https://orcid.org/0000-0001-6650-2634
B. Fornal https://orcid.org/0000-0003-3271-2080
F. Frasconi https://orcid.org/0000-0003-4204-6587
Z. Frei https://orcid.org/0000-0002-0181-8491
A. Freise https://orcid.org/0000-0001-6586-9901
O. Freitas https://orcid.org/0000-0002-2898-1256
R. Frey https://orcid.org/0000-0003-0341-2636
G. G. Fronzé https://orcid.org/0000-0003-0966-4279
M. Fuentes-Garcia https://orcid.org/0000-0003-3390-8712
B. Gadre https://orcid.org/0000-0002-1534-9761
J. R. Gair https://orcid.org/0000-0002-1671-3668
S. Galaudage https://orcid.org/0000-0002-1819-0215
A. Gamboa https://orcid.org/0000-0001-8391-5596
D. Ganapathy https://orcid.org/0000-0003-3028-4174
A. Ganguly https://orcid.org/0000-0001-7394-0755
B. Garaventa https://orcid.org/0000-0003-2490-404X
J. García-Bellido https://orcid.org/0000-0002-9370-8360
C. García-Quirós https://orcid.org/0000-0002-8059-2477
J. W. Gardner https://orcid.org/0000-0002-8592-1452
J. Gargiulo https://orcid.org/0000-0002-3507-6924
X. Garrido https://orcid.org/0000-0002-7088-5831
A. Garron https://orcid.org/0000-0002-1601-797X
F. Garufi https://orcid.org/0000-0003-1391-6168
C. Gasbarra https://orcid.org/0000-0001-8335-9614
F. Gautier https://orcid.org/0000-0001-8006-9590
V. Gayathri https://orcid.org/0000-0002-7167-9888
G. Gemme https://orcid.org/0000-0002-1127-7406
A. Gennai https://orcid.org/0000-0003-0149-2089
V. Gennari https://orcid.org/0000-0002-0190-9262
R. George https://orcid.org/0000-0002-7797-7683
O. Gerberding https://orcid.org/0000-0001-7740-2698
L. Gergely https://orcid.org/0000-0003-3146-6201
A. Ghosh https://orcid.org/0000-0003-0423-3533
Shaon Ghosh https://orcid.org/0000-0001-9901-6253
Suprovo Ghosh https://orcid.org/0000-0002-1656-9870
T. Ghosh https://orcid.org/0000-0001-9848-9905
J. A. Giaime https://orcid.org/0000-0002-3531-817X
C. Gier https://orcid.org/0000-0003-0897-7943
S. Gkaitatzis https://orcid.org/0000-0001-9420-7499
J. Glanzer https://orcid.org/0009-0000-0808-0795
F. Glotin https://orcid.org/0000-0003-2637-1187
P. Godwin https://orcid.org/0000-0002-7489-4751
A. S. Goettel https://orcid.org/0000-0002-6215-4641
E. Goetz https://orcid.org/0000-0003-2666-721X
S. Gomez Lopez https://orcid.org/0000-0002-9557-4706
B. Goncharov https://orcid.org/0000-0003-3189-5807
G. González https://orcid.org/0000-0003-0199-3158
P. Goodarzi https://orcid.org/0009-0008-1093-6706
A. W. Goodwin-Jones https://orcid.org/0000-0002-0395-0680
R. Gouaty https://orcid.org/0000-0001-5372-7084
A. Grado https://orcid.org/0000-0002-0501-8256
V. Graham https://orcid.org/0000-0003-3633-0135
A. E. Granados https://orcid.org/0000-0003-2099-9096
M. Granata https://orcid.org/0000-0003-3275-1186
V. Granata https://orcid.org/0000-0003-2246-6963
R. Gray https://orcid.org/0000-0002-5556-9873
A. C. Green https://orcid.org/0000-0002-6287-8746
S. R. Green https://orcid.org/0000-0002-6987-6313
H. L. Griggs https://orcid.org/0000-0001-5018-7908
C. Grimaud https://orcid.org/0000-0001-7736-7730
H. Grote https://orcid.org/0000-0002-0797-3943
S. Grunewald https://orcid.org/0000-0003-4641-2791
D. Guerra https://orcid.org/0000-0003-0029-5390
D. Guetta https://orcid.org/0000-0002-7349-1109
G. M. Guidi https://orcid.org/0000-0002-3061-9870
F. Gulminelli https://orcid.org/0000-0003-4354-2849
H. Guo https://orcid.org/0000-0002-3777-3117
W. Guo https://orcid.org/0000-0002-4320-4420
Y. Guo https://orcid.org/0000-0002-6959-9870
A. Gupta https://orcid.org/0000-0002-5441-9013
I. Gupta https://orcid.org/0000-0001-6932-8715
V. Gupta https://orcid.org/0000-0002-7672-0480
F. Guzman https://orcid.org/0000-0001-9136-929X
M. Haberland https://orcid.org/0000-0001-9816-5660
E. D. Hall https://orcid.org/0000-0001-9018-666X
E. Z. Hamilton https://orcid.org/0000-0003-0098-9114
G. Hammond https://orcid.org/0000-0002-1414-3622
C. Hanna https://orcid.org/0000-0002-0965-7493
O. A. Hannuksela https://orcid.org/0000-0002-3887-7137
A. G. Hanselman https://orcid.org/0000-0002-8304-0109
S. Harikumar https://orcid.org/0000-0002-2653-7282
T. Harmark https://orcid.org/0000-0002-2795-7035
J. Harms https://orcid.org/0000-0002-7332-9806
G. M. Harry https://orcid.org/0000-0002-8905-7622
I. W. Harry https://orcid.org/0000-0002-5304-9372
C. J. Haster https://orcid.org/0000-0001-8040-9807
K. Haughian https://orcid.org/0000-0002-1223-7342
J. Heinze https://orcid.org/0000-0001-8692-2724
H. Heitmann https://orcid.org/0000-0003-0625-5461
F. Hellman https://orcid.org/0000-0002-9135-6330
A. F. Helmling-Cornell https://orcid.org/0000-0002-7709-8638
G. Hemming https://orcid.org/0000-0001-5268-4465
O. Henderson-Sapir https://orcid.org/0000-0002-1613-9985
M. Hendry https://orcid.org/0000-0001-8322-5405
M. H. Hennig https://orcid.org/0000-0003-1531-8460
C. Henshaw https://orcid.org/0000-0002-4206-3128
M. Heurs https://orcid.org/0000-0002-5577-2273
A. L. Hewitt https://orcid.org/0000-0002-1255-3492
Y. Himemoto https://orcid.org/0000-0002-6856-3809
I. J. Hollows https://orcid.org/0000-0002-3404-6459

D. E. Holz https://orcid.org/0000-0002-0175-5064
J. Hough https://orcid.org/0000-0003-3242-3123
S. Hourihane https://orcid.org/0000-0002-9152-0719
E. J. Howell https://orcid.org/0000-0001-7891-2817
C. G. Hoy https://orcid.org/0000-0002-8843-6719
H.-F. Hsieh https://orcid.org/0000-0002-8947-723X
W.-F. Hsu https://orcid.org/0000-0001-5234-3804
Q. Hu https://orcid.org/0000-0002-3033-6491
H. Y. Huang https://orcid.org/0000-0002-1665-2383
Y. Huang https://orcid.org/0000-0002-2952-8429
V. Hui https://orcid.org/0000-0002-0233-2346
S. Husa https://orcid.org/0000-0002-0445-1971
L. Iampieri https://orcid.org/0009-0004-1161-2990
G. A. Iandolo https://orcid.org/0000-0003-1155-4327
G. Iannone https://orcid.org/0000-0001-8347-7549
G. Iorio https://orcid.org/0000-0003-0293-503X
P. Iosif https://orcid.org/0000-0003-1621-7709
J. Irwin https://orcid.org/0000-0002-2364-2191
M. Isi https://orcid.org/0000-0001-8830-8672
K. S. Isleif https://orcid.org/0000-0001-7032-9440
Y. Itoh https://orcid.org/0000-0003-2694-8935
B. R. Iyer https://orcid.org/0000-0002-4141-5179
P.-E. Jacquet https://orcid.org/0000-0001-9552-0057
S. P. Jadhav https://orcid.org/0000-0003-0554-0084
A. L. James https://orcid.org/0000-0001-9165-0807
K. Jani https://orcid.org/0000-0003-1007-8912
J. Janquart https://orcid.org/0000-0003-2888-7152
S. Jaraba https://orcid.org/0000-0002-4759-143X
P. Jaranowski https://orcid.org/0000-0001-8085-3414
R. Jaume https://orcid.org/0000-0001-8691-3166
J. Jiang https://orcid.org/0000-0002-0154-3854
H.-B. Jin https://orcid.org/0000-0002-6217-2428
M. C. Johnston https://orcid.org/0000-0002-0663-9193
D. H. Jones https://orcid.org/0000-0003-3987-068X
P. Joshi https://orcid.org/0000-0002-4148-4932
L. Ju https://orcid.org/0000-0002-7951-4295
K. Jung https://orcid.org/0000-0003-4789-8893
J. Junker https://orcid.org/0000-0002-3051-4374
H. B. Kabagoz https://orcid.org/0000-0002-0900-8557
T. Kajita https://orcid.org/0000-0003-1207-6638
V. Kalogera https://orcid.org/0000-0001-9236-5469
M. Kalomenopoulos https://orcid.org/0000-0001-6677-949X
M. Kamiizumi https://orcid.org/0000-0001-7216-1784
N. Kanda https://orcid.org/0000-0001-6291-0227
S. Kandhasamy https://orcid.org/0000-0002-4825-6764
G. Kang https://orcid.org/0000-0002-6072-8189
S. J. Kapadia https://orcid.org/0000-0001-5318-1253
D. P. Kapasi https://orcid.org/0000-0001-8189-4920
M. Kasprzack https://orcid.org/0000-0003-4618-5939
R. Kaushik https://orcid.org/0000-0003-4888-5154
D. Keitel https://orcid.org/0000-0002-2824-626X
L. J. Kemperman https://orcid.org/0009-0009-5254-8397
J. Kennington https://orcid.org/0000-0002-6899-3833
R. Kesharwani https://orcid.org/0009-0002-2528-5738
J. S. Key https://orcid.org/0000-0003-0123-7600
F. Y. Khalili https://orcid.org/0000-0001-7068-2332
F. Khan https://orcid.org/0000-0001-6176-853X
W. Kiendrebeogo https://orcid.org/0000-0002-9108-5059
N. Kijbunchoo https://orcid.org/0000-0002-2874-1228
K. Kim https://orcid.org/0000-0003-1653-3795
M. H. Kim https://orcid.org/0009-0009-9894-3640
S. Kim https://orcid.org/0000-0003-1437-4647
Y.-M. Kim https://orcid.org/0000-0001-8720-6113
C. Kimball https://orcid.org/0000-0001-9879-6884
J. S. Kissel https://orcid.org/0000-0002-1702-9577
A. M. Knee https://orcid.org/0000-0003-0703-947X
N. Knust https://orcid.org/0000-0002-5984-5353
S. M. Koehlenbeck https://orcid.org/0000-0002-3842-9051
K. Kohri https://orcid.org/0000-0003-3764-8612
K. Kokeyama https://orcid.org/0000-0002-2896-1992
S. Koley https://orcid.org/0000-0002-5793-6665
P. Kolitsidou https://orcid.org/0000-0002-6719-8686
A. E. Koloniari https://orcid.org/0000-0002-0546-5638
K. Komori https://orcid.org/0000-0002-4092-9602
A. K. H. Kong https://orcid.org/0000-0002-5105-344X
A. Kontos https://orcid.org/0000-0002-1347-0680
M. Korobko https://orcid.org/0000-0002-3839-3909
A. Koushik https://orcid.org/0000-0002-7638-4544
N. Kouvatsos https://orcid.org/0000-0002-5497-3401
N. V. Krishnendu https://orcid.org/0000-0002-3483-7517
A. Królak https://orcid.org/0000-0003-4514-7690
J. Kubisz https://orcid.org/0000-0001-7258-8673
S. Kulkarni https://orcid.org/0000-0001-8057-0203
A. Kulur Ramamohan https://orcid.org/0000-0003-3681-1887
Praveen Kumar https://orcid.org/0000-0002-2288-4252
Prayush Kumar https://orcid.org/0000-0001-5523-4603
J. Kume https://orcid.org/0000-0003-3126-5100
K. Kuns https://orcid.org/0000-0003-0630-3902
S. Kuroyanagi https://orcid.org/0000-0001-6538-1447
S. Kuwahara https://orcid.org/0009-0009-2249-8798
K. Kwak https://orcid.org/0000-0002-2304-7798
S. Kwon https://orcid.org/0009-0006-3770-7044
D. Laghi https://orcid.org/0000-0001-7462-3794
M. Lalleman https://orcid.org/0000-0002-2254-010X
R. N. Lang https://orcid.org/0000-0002-4804-5537
R. Langgin https://orcid.org/0000-0002-5116-6217
B. Lantz https://orcid.org/0000-0002-7404-4845
I. L. Rosa https://orcid.org/0000-0003-0107-1540
A. Lartaux-Vollard https://orcid.org/0000-0003-1714-365X
P. D. Lasky https://orcid.org/0000-0003-3763-1386
J. Lawrence https://orcid.org/0000-0003-1222-0433
M. Laxen https://orcid.org/0000-0001-7515-9639
C. Lazarte https://orcid.org/0000-0002-6964-9321
A. Lazzarini https://orcid.org/0000-0002-5993-8808
P. Leaci https://orcid.org/0000-0002-3997-5046
Y. K. Lecoeuche https://orcid.org/0000-0002-9186-7034
H. M. Lee https://orcid.org/0000-0003-4412-7161
H. W. Lee https://orcid.org/0000-0002-1998-3209
K. Lee https://orcid.org/0000-0003-0470-3718
R.-K. Lee https://orcid.org/0000-0002-7171-7274
Sungho Lee https://orcid.org/0000-0001-6034-2238
M. Le Jean https://orcid.org/0009-0003-8047-3958
A. Lemaître https://orcid.org/0000-0002-6865-9245
M. Lenti https://orcid.org/0000-0002-2765-3955
M. Leonardi https://orcid.org/0000-0002-7641-0060
N. Leroy https://orcid.org/0000-0002-2321-1017
M. Lethuillier https://orcid.org/0000-0001-6185-2045
K. L. Li https://orcid.org/0000-0001-8229-2024
X. Li https://orcid.org/0000-0002-3780-7735
E. T. Lin https://orcid.org/0000-0002-0030-8051
L. C.-C. Lin https://orcid.org/0000-0003-4083-9567

Y.-C. Lin https://orcid.org/0000-0003-4939-1404
A. Liu https://orcid.org/0000-0003-1081-8722
G. C. Liu https://orcid.org/0000-0001-5663-3016
J. Liu https://orcid.org/0000-0001-6726-3268
J. Llobera-Querol https://orcid.org/0000-0003-3322-6850
R. K. L. Lo https://orcid.org/0000-0003-1561-6716
A. Longo https://orcid.org/0000-0003-4254-8579
D. Lopez https://orcid.org/0000-0003-3342-9906
A. Lorenzo-Medina https://orcid.org/0009-0006-0860-5700
G. Losurdo https://orcid.org/0000-0003-0452-746X
T. P. Lott IV https://orcid.org/0009-0002-2864-162X
J. D. Lough https://orcid.org/0000-0002-5160-0239
C. O. Lousto https://orcid.org/0000-0002-6400-9640
N. Lu https://orcid.org/0000-0002-8861-9902
L. Lucchesi https://orcid.org/0000-0002-5916-8014
D. Lumaca https://orcid.org/0000-0002-3628-1591
A. P. Lundgren https://orcid.org/0000-0002-0363-4469
A. W. Lussier https://orcid.org/0000-0002-4507-1123
R. Macas https://orcid.org/0000-0002-6096-8297
D. M. Macleod https://orcid.org/0000-0002-1395-8694
I. A. O. MacMillan https://orcid.org/0000-0002-6927-1031
A. Macquet https://orcid.org/0000-0001-5955-6415
S. Maenaut https://orcid.org/0000-0003-1464-2605
R. M. Magee https://orcid.org/0000-0001-9769-531X
E. Maggio https://orcid.org/0000-0002-1960-8185
M. Magnozzi https://orcid.org/0000-0003-4512-8430
D. Malakar https://orcid.org/0000-0003-4234-4023
U. Mali https://orcid.org/0009-0003-1285-2788
L. Mallick https://orcid.org/0000-0001-8624-9162
A.-K. Malz https://orcid.org/0009-0004-7196-4170
M. Mancarella https://orcid.org/0000-0002-0675-508X
V. Mandic https://orcid.org/0000-0001-6333-8621
V. Mangano https://orcid.org/0000-0001-7902-8505
G. L. Mansell https://orcid.org/0000-0003-4736-6678
M. Manske https://orcid.org/0000-0002-7778-1189
M. Mantovani https://orcid.org/0000-0002-4424-5726
M. Mapelli https://orcid.org/0000-0001-8799-2548
C. Marinelli https://orcid.org/0000-0002-3596-4307
F. Marion https://orcid.org/0000-0002-8184-1017
S. Marsat https://orcid.org/0000-0001-9449-1071
F. Martelli https://orcid.org/0000-0003-3761-8616
I. W. Martin https://orcid.org/0000-0001-7300-9151
R. M. Martin https://orcid.org/0000-0001-9664-2216
V. Martinez https://orcid.org/0000-0001-5852-2301
J. C. Martins https://orcid.org/0000-0002-6099-4831
M. Masso-Reid https://orcid.org/0000-0001-6177-8105
S. Mastrogiovanni https://orcid.org/0000-0003-1606-4183
T. Matcovich https://orcid.org/0009-0004-1209-008X
M. Matiushechkina https://orcid.org/0000-0002-9957-8720
N. Mavalvala https://orcid.org/0000-0003-0219-9706
D. E. McClelland https://orcid.org/0000-0001-6210-5842
L. McCuller https://orcid.org/0000-0003-0851-0593
G. I. McGhee https://orcid.org/0000-0001-5038-2658
J. McIver https://orcid.org/0000-0003-0316-1355
A. McLeod https://orcid.org/0000-0001-5424-8368
I. McMahon https://orcid.org/0000-0002-4529-1505
D. Meacher https://orcid.org/0000-0001-5882-0368
C. S. Menoni https://orcid.org/0000-0001-9185-2572
R. A. Mercer https://orcid.org/0000-0001-8372-3914
J. R. Mérou https://orcid.org/0000-0002-5776-6643
C. Messick https://orcid.org/0000-0002-8230-3309
M. Meyer-Conde https://orcid.org/0000-0003-2230-6310
F. Meylahn https://orcid.org/0000-0002-9556-142X
A. Miani https://orcid.org/0000-0001-7737-3129
C. Michel https://orcid.org/0000-0003-0606-725X
Y. Michimura https://orcid.org/0000-0002-2218-4002
H. Middleton https://orcid.org/0000-0001-5532-3622
D. P. Mihaylov https://orcid.org/0000-0002-8820-407X
S. J. Miller https://orcid.org/0000-0001-5670-7046
M. Millhouse https://orcid.org/0000-0002-8659-5898
E. Milotti https://orcid.org/0000-0001-7348-9765
V. Milotti https://orcid.org/0000-0003-4732-1226
Ll. M. Mir https://orcid.org/0000-0002-4276-715X
L. Mirasola https://orcid.org/0009-0004-0174-1377
M. Miravet-Tenés https://orcid.org/0000-0002-8766-1156
C.-A. Miritescu https://orcid.org/0000-0002-7716-0569
C. Mishra https://orcid.org/0000-0002-8115-8728
T. Mishra https://orcid.org/0000-0002-7881-1677
S. Mitra https://orcid.org/0000-0002-0800-4626
V. P. Mitrofanov https://orcid.org/0000-0002-6983-4981
O. Miyakawa https://orcid.org/0000-0002-9085-7600
S. Miyoki https://orcid.org/0000-0002-1213-8416
G. Mo https://orcid.org/0000-0001-6331-112X
L. Mobilia https://orcid.org/0009-0000-3022-2358
S. R. Mohite https://orcid.org/0000-0003-1356-7156
M. Molina-Ruiz https://orcid.org/0000-0003-4892-3042
A. More https://orcid.org/0000-0001-7714-7076
S. More https://orcid.org/0000-0002-2986-2371
C. Moreno https://orcid.org/0000-0002-0496-032X
E. A. Moreno https://orcid.org/0000-0001-5666-3637
S. Morisaki https://orcid.org/0000-0002-8445-6747
Y. Moriwaki https://orcid.org/0000-0002-4497-6908
G. Morras https://orcid.org/0000-0002-9977-8546
A. Moscatello https://orcid.org/0000-0001-5480-7406
M. Mould https://orcid.org/0000-0001-5460-2910
B. Mours https://orcid.org/0000-0002-6444-6402
C. M. Mow-Lowry https://orcid.org/0000-0002-0351-4555
L. Muccillo https://orcid.org/0009-0000-6237-0590
F. Muciaccia https://orcid.org/0000-0003-0850-2649
D. Mukherjee https://orcid.org/0000-0001-7335-9418
Suvodip Mukherjee https://orcid.org/0000-0002-3373-5236
N. Mukund https://orcid.org/0000-0002-8666-9156
P. G. Murray https://orcid.org/0000-0002-8218-2404
D. Nabari https://orcid.org/0009-0006-8500-7624
N. Nagarajan https://orcid.org/0000-0003-3695-0078
K. Nakamura https://orcid.org/0000-0001-6148-4289
H. Nakano https://orcid.org/0000-0001-7665-0796
D. Nanadoumgar-Lacroze https://orcid.org/0009-0009-7255-8111
P. Narayan https://orcid.org/0009-0009-0599-532X
I. Nardecchia https://orcid.org/0000-0001-5558-2595
L. Naticchioni https://orcid.org/0000-0003-2918-0730
R. K. Nayak https://orcid.org/0000-0002-6814-7792
A. Nelson https://orcid.org/0000-0002-5909-4692
A. Neunzert https://orcid.org/0000-0003-0323-0111
L. Nguyen Quynh https://orcid.org/0000-0002-1828-3702
A. B. Nielsen https://orcid.org/0000-0001-8694-4026
A. Nishizawa https://orcid.org/0000-0003-3562-0990
W. Niu https://orcid.org/0000-0003-1470-532X
J. Novak https://orcid.org/0000-0002-6029-4712
R. Nowicki https://orcid.org/0009-0008-6626-0725
J. F. Nuño Siles https://orcid.org/0000-0001-8304-8066

L. K. Nuttall https://orcid.org/0000-0002-8599-8791
J. Oberling https://orcid.org/0009-0001-4174-3973
E. Oelker https://orcid.org/0000-0002-3916-1595
M. Oertel https://orcid.org/0000-0002-1884-8654
M. Ohashi https://orcid.org/0000-0001-8072-0304
F. Ohme https://orcid.org/0000-0003-0493-5607
R. Oliveri https://orcid.org/0000-0002-7497-871X
K. Oohara https://orcid.org/0000-0002-7518-6677
B. O'Reilly https://orcid.org/0000-0002-3874-8335
M. Orselli https://orcid.org/0000-0003-3563-8576
R. O'Shaughnessy https://orcid.org/0000-0001-5832-8517
S. Oshino https://orcid.org/0000-0002-2794-6029
I. Ota https://orcid.org/0000-0001-5045-2484
D. J. Ottaway https://orcid.org/0000-0001-6794-1591
B. J. Owen https://orcid.org/0000-0003-3919-0780
A. E. Pace https://orcid.org/0009-0003-4044-0334
R. Pagano https://orcid.org/0000-0001-8362-0130
M. A. Page https://orcid.org/0000-0002-5298-7914
A. Pai https://orcid.org/0000-0003-3476-4589
S. Pal https://orcid.org/0000-0003-2172-8589
M. A. Palaia https://orcid.org/0009-0007-3296-8648
C. Palomba https://orcid.org/0000-0002-4450-9883
P. Palud https://orcid.org/0000-0002-5850-6325
K. C. Pan https://orcid.org/0000-0002-1473-9880
F. Pannarale https://orcid.org/0000-0002-7537-3210
F. Paoletti https://orcid.org/0000-0001-8898-1963
A. Paolone https://orcid.org/0000-0002-4839-7815
A. Papadopoulos https://orcid.org/0009-0006-1882-996X
L. Papalini https://orcid.org/0000-0002-5219-0454
G. Papigkiotis https://orcid.org/0009-0008-2205-7426
A. Parisi https://orcid.org/0000-0003-0251-8914
J. Park https://orcid.org/0000-0002-7510-0079
W. Parker https://orcid.org/0000-0002-7711-4423
D. Pascucci https://orcid.org/0000-0003-1907-0175
A. Pasqualetti https://orcid.org/0000-0003-0620-5990
R. Passaquieti https://orcid.org/0000-0003-4753-9428
O. Patane https://orcid.org/0000-0002-4850-2355
A. V. Patel https://orcid.org/0000-0001-6872-9197
B. Patricelli https://orcid.org/0000-0001-6709-0969
K. Paul https://orcid.org/0000-0002-8406-6503
S. Paul https://orcid.org/0000-0002-4449-1732
E. Payne https://orcid.org/0000-0003-4507-8373
A. Pele https://orcid.org/0000-0002-1873-3769
F. E. Peña Arellano https://orcid.org/0000-0002-8516-5159
S. Penn https://orcid.org/0000-0003-4956-0853
A. Perego https://orcid.org/0000-0002-0936-8237
C. Périgois https://orcid.org/0000-0002-9779-2838
G. Perna https://orcid.org/0000-0002-7364-1904
A. Perreca https://orcid.org/0000-0002-6269-2490
J. Perret https://orcid.org/0009-0006-4975-1536
S. Perriès https://orcid.org/0000-0003-2213-3579
H. P. Pfeiffer https://orcid.org/0000-0001-9288-519X
K. A. Pham https://orcid.org/0000-0002-7650-1034
K. S. Phukon https://orcid.org/0000-0003-1561-0760
L. Piccari https://orcid.org/0009-0000-0247-4339
O. J. Piccinni https://orcid.org/0000-0001-5478-3950
M. Pichot https://orcid.org/0000-0002-4439-8968
M. Piendibene https://orcid.org/0000-0003-2434-488X
F. Piergiovanni https://orcid.org/0000-0001-8063-828X
L. Pierini https://orcid.org/0000-0003-0945-2196
G. Pierra https://orcid.org/0000-0003-3970-7970
V. Pierro https://orcid.org/0000-0002-6020-5521
M. Pillas https://orcid.org/0000-0003-3224-2146
F. Pilo https://orcid.org/0000-0003-4967-7090
L. Pinard https://orcid.org/0000-0002-8842-1867
I. M. Pinto https://orcid.org/0000-0002-2679-4457
M. Pinto https://orcid.org/0009-0003-4339-9971
B. J. Piotrzkowski https://orcid.org/0000-0001-8919-0899
M. D. Pitkin https://orcid.org/0000-0003-4548-526X
A. Placidi https://orcid.org/0000-0001-8032-4416
E. Placidi https://orcid.org/0000-0002-3820-8451
M. L. Planas https://orcid.org/0000-0001-8278-7406
W. Plastino https://orcid.org/0000-0002-5737-6346
C. Plunkett https://orcid.org/0000-0002-1144-6708
R. Poggiani https://orcid.org/0000-0002-9968-2464
L. Pompili https://orcid.org/0000-0002-0710-6778
C. Posnansky https://orcid.org/0009-0009-7137-9795
R. Poulton https://orcid.org/0000-0003-2049-520X
J. Powell https://orcid.org/0000-0002-1357-4164
M. Pracchia https://orcid.org/0009-0001-8343-719X
B. K. Pradhan https://orcid.org/0000-0002-2526-1421
T. Pradier https://orcid.org/0000-0001-5501-0060
K. Prasai https://orcid.org/0000-0001-6552-097X
G. Pratten https://orcid.org/0000-0003-4984-0775
G. Principe https://orcid.org/0000-0003-0406-7387
G. A. Prodi https://orcid.org/0000-0001-5256-915X
A. Puecher https://orcid.org/0000-0003-1357-4348
J. Pullin https://orcid.org/0000-0001-8248-603X
M. Pürrer https://orcid.org/0000-0002-3329-9788
H. Qi https://orcid.org/0000-0001-6339-1537
J. Qin https://orcid.org/0000-0002-7120-9026
G. Quéméner https://orcid.org/0000-0001-6703-6655
B. Rajbhandari https://orcid.org/0000-0001-7568-1611
K. E. Ramirez https://orcid.org/0000-0003-2194-7669
F. A. Ramis Vidal https://orcid.org/0000-0001-6143-2104
M. Ramos Arevalo https://orcid.org/0009-0003-1528-8326
A. Ramos-Buades https://orcid.org/0000-0002-6874-7421
S. Ranjan https://orcid.org/0000-0001-7480-9329
P. Rapagnani https://orcid.org/0000-0002-1865-6126
A. Ray https://orcid.org/0000-0002-7322-4748
V. Raymond https://orcid.org/0000-0003-0066-0095
M. Razzano https://orcid.org/0000-0003-4825-1629
D. H. Reitze https://orcid.org/0000-0002-5756-1111
B. Revenu https://orcid.org/0000-0002-7629-4805
L. Ricca https://orcid.org/0009-0002-1638-0610
F. Ricci https://orcid.org/0000-0001-5475-4447
M. Ricci https://orcid.org/0009-0008-7421-4331
A. Ricciardone https://orcid.org/0000-0002-5688-455X
J. W. Richardson https://orcid.org/0000-0002-1472-4806
K. Riles https://orcid.org/0000-0002-6418-5812
S. Rinaldi https://orcid.org/0000-0001-5799-4155
A. Rocchi https://orcid.org/0000-0002-1382-9016
L. Rolland https://orcid.org/0000-0003-0589-9687
J. G. Rollins https://orcid.org/0000-0002-9388-2799
A. E. Romano https://orcid.org/0000-0002-0314-8698
R. Romano https://orcid.org/0000-0002-0485-6936
A. Romero https://orcid.org/0000-0003-2275-4164
S. Ronchini https://orcid.org/0000-0003-0020-687X
T. J. Roocke https://orcid.org/0000-0003-2640-9683
D. Rosińska https://orcid.org/0000-0002-3681-9304
M. P. Ross https://orcid.org/0000-0002-8955-5269
M. Rossello-Sastre https://orcid.org/0000-0002-3341-3480
S. Rowan https://orcid.org/0000-0002-0666-9907
S. K. Roy https://orcid.org/0000-0001-9295-5119

S. Roy https://orcid.org/0000-0003-2147-5411
D. Rozza https://orcid.org/0000-0002-7378-6353
E. Ruiz Morales https://orcid.org/0000-0002-0995-595X
S. Sachdev https://orcid.org/0000-0002-0525-2317
P. Saffarieh https://orcid.org/0009-0000-7504-3660
S. Safi-Harb https://orcid.org/0000-0001-6189-7665
M. R. Sah https://orcid.org/0009-0005-9881-1788
S. Saha https://orcid.org/0000-0002-3333-8070
T. Sainrat https://orcid.org/0009-0003-0169-266X
S. Sajith Menon https://orcid.org/0009-0008-4985-1320
Y. Sakai https://orcid.org/0000-0001-8810-4813
M. Sakellariadou https://orcid.org/0000-0002-2715-1517
S. Sakon https://orcid.org/0000-0002-5861-3024
O. S. Salafia https://orcid.org/0000-0003-4924-7322
F. Salces-Carcoba https://orcid.org/0000-0001-7049-4438
M. Saleem https://orcid.org/0000-0002-3836-7751
F. Salemi https://orcid.org/0000-0002-9511-3846
M. Sallé https://orcid.org/0000-0002-6620-6672
S. Salvador https://orcid.org/0000-0003-3444-7807
A. Samajdar https://orcid.org/0000-0002-0857-6018
N. Sanchis-Gual https://orcid.org/0000-0001-5375-7494
E. M. Sänger https://orcid.org/0009-0003-6642-8974
F. Santoliquido https://orcid.org/0000-0003-3752-1400
A. Sasli https://orcid.org/0000-0001-7357-0889
P. Sassi https://orcid.org/0000-0002-4920-2784
B. Sassolas https://orcid.org/0000-0002-3077-8951
B. S. Sathyaprakash https://orcid.org/0000-0003-3845-7586
O. Sauter https://orcid.org/0000-0003-2293-1554
R. L. Savage https://orcid.org/0000-0003-3317-1036
T. Sawada https://orcid.org/0000-0001-5726-7150
M. G. Schiworski https://orcid.org/0000-0001-9298-004X
P. Schmidt https://orcid.org/0000-0003-1542-1791
S. Schmidt https://orcid.org/0000-0002-8206-8089
R. Schnabel https://orcid.org/0000-0003-2896-4218
K. Schouteden https://orcid.org/0000-0002-5975-585X
E. Schwartz https://orcid.org/0000-0001-8922-7794
M. Scialpi https://orcid.org/0009-0007-6434-1460
J. Scott https://orcid.org/0000-0001-6701-6515
S. M. Scott https://orcid.org/0000-0002-9875-7700
R. M. Sedas https://orcid.org/0000-0001-8961-3855
M. Seglar-Arroyo https://orcid.org/0000-0001-8654-409X
Y. Sekiguchi https://orcid.org/0000-0002-2648-3835
A. S. Sengupta https://orcid.org/0000-0002-3212-0475
E. G. Seo https://orcid.org/0000-0002-8588-4794
J. W. Seo https://orcid.org/0000-0003-4937-0769
M. Serra https://orcid.org/0000-0002-6093-8063
U. S. Shah https://orcid.org/0000-0001-8249-7425
M. A. Shaikh https://orcid.org/0000-0003-0826-6164
L. Shao https://orcid.org/0000-0002-1334-8853
A. K. Sharma https://orcid.org/0000-0003-0067-346X
P. Shawhan https://orcid.org/0000-0002-8249-8070
N. S. Shcheblanov https://orcid.org/0000-0001-8696-2435
H. Shinkai https://orcid.org/0000-0003-1082-2844
D. H. Shoemaker https://orcid.org/0000-0002-4147-2560
D. M. Shoemaker https://orcid.org/0000-0002-9899-6357
H. Siegel https://orcid.org/0000-0001-5161-4617
D. Sigg https://orcid.org/0000-0003-4606-6526
L. Silenzi https://orcid.org/0000-0001-7316-3239
L. Silvestri https://orcid.org/0009-0008-5207-661X
L. P. Singer https://orcid.org/0000-0001-9898-5597
D. Singh https://orcid.org/0000-0001-9675-4584
N. Singh https://orcid.org/0000-0002-1135-3456
A. M. Sintes https://orcid.org/0000-0001-9050-7515
V. Skliris https://orcid.org/0000-0003-0902-9216
B. J. J. Slagmolen https://orcid.org/0000-0002-2471-3828
J. R. Smith https://orcid.org/0000-0003-0638-9670
L. Smith https://orcid.org/0000-0002-3035-0947
R. J. E. Smith https://orcid.org/0000-0001-8516-3324
W. J. Smith https://orcid.org/0009-0003-7949-4911
K. Somiya https://orcid.org/0000-0003-2601-2264
I. Song https://orcid.org/0000-0002-4301-8281
S. Soni https://orcid.org/0000-0003-3856-8534
V. Sordini https://orcid.org/0000-0003-0885-824X
H. Sotani https://orcid.org/0000-0002-3239-2921
F. Spada https://orcid.org/0000-0001-5664-1657
V. Spagnuolo https://orcid.org/0000-0002-0098-4260
A. P. Spencer https://orcid.org/0000-0003-4418-3366
P. Spinicelli https://orcid.org/0000-0001-8078-6047
F. Stachurski https://orcid.org/0000-0002-8658-5753
D. A. Steer https://orcid.org/0000-0002-8781-1273
N. Steinle https://orcid.org/0000-0003-0658-402X
S. Steinlechner https://orcid.org/0000-0003-4710-8548
N. Stergioulas https://orcid.org/0000-0002-5490-5302
S. Sudhagar https://orcid.org/0000-0001-8578-4665
L. Suleiman https://orcid.org/0000-0003-3783-7448
J. Sun https://orcid.org/0009-0008-8278-0077
L. Sun https://orcid.org/0000-0001-7959-892X
J. Suresh https://orcid.org/0000-0003-2389-6666
P. J. Sutton https://orcid.org/0000-0003-1614-3922
B. L. Swinkels https://orcid.org/0000-0002-3066-3601
A. Syx https://orcid.org/0009-0000-6424-6411
M. J. Szczepańczyk https://orcid.org/0000-0002-6167-6149
P. Szewczyk https://orcid.org/0000-0002-1339-9167
M. Tacca https://orcid.org/0000-0003-1353-0441
H. Tagoshi https://orcid.org/0000-0001-8530-9178
H. Takahashi https://orcid.org/0000-0003-0596-4397
R. Takahashi https://orcid.org/0000-0003-1367-5149
A. Takamori https://orcid.org/0000-0001-6032-1330
S. Takano https://orcid.org/0000-0002-1266-4555
H. Takeda https://orcid.org/0000-0001-9937-2557
N. Tamanini https://orcid.org/0000-0001-8760-5421
S. J. Tanaka https://orcid.org/0000-0002-8796-1992
S. Tanioka https://orcid.org/0000-0003-3321-1018
L. Tao https://orcid.org/0000-0003-4382-5507
E. N. Tapia San Martín https://orcid.org/0000-0002-4817-5606
A. Taruya https://orcid.org/0000-0002-4016-1955
J. D. Tasson https://orcid.org/0000-0002-4777-5087
J. G. Tau https://orcid.org/0009-0004-7428-762X
R. Tenorio https://orcid.org/0000-0002-3582-2587
A. Theodoropoulos https://orcid.org/0000-0003-4486-7135
L. M. Thomas https://orcid.org/0000-0003-3271-6436
J. E. Thompson https://orcid.org/0000-0002-0419-5517
E. Thrane https://orcid.org/0000-0002-4418-3895
J. Tissino https://orcid.org/0000-0003-2483-6710
S. Tiwari https://orcid.org/0000-0003-1611-6625
V. Tiwari https://orcid.org/0000-0002-1602-4176
A. M. Toivonen https://orcid.org/0009-0008-9546-2035
K. Toland https://orcid.org/0000-0001-9537-9698
A. E. Tolley https://orcid.org/0000-0001-9841-943X
T. Tomaru https://orcid.org/0000-0002-8927-9014
T. Tomura https://orcid.org/0000-0002-7504-8258

H. Tong https://orcid.org/0000-0002-4534-0485
A. Torres-Forné https://orcid.org/0000-0001-8709-5118
I. Tosta e Melo https://orcid.org/0000-0001-5833-4052
E. Tournefier https://orcid.org/0000-0002-5465-9607
A. Trapananti https://orcid.org/0000-0001-7763-5758
R. Travaglini https://orcid.org/0000-0002-5288-1407
F. Travasso https://orcid.org/0000-0002-4653-6156
M. C. Tringali https://orcid.org/0000-0001-5087-189X
A. Tripathee https://orcid.org/0000-0002-6976-5576
G. Troian https://orcid.org/0000-0001-6837-607X
A. Trovato https://orcid.org/0000-0002-9714-1904
T. Tsang https://orcid.org/0000-0003-3666-686X
S. Tsuchida https://orcid.org/0000-0001-8217-0764
L. Tsukada https://orcid.org/0000-0003-0596-5648
K. Turbang https://orcid.org/0000-0002-9296-8603
M. Turconi https://orcid.org/0000-0001-9999-2027
H. Ubach https://orcid.org/0000-0002-0679-9074
N. Uchikata https://orcid.org/0000-0003-0030-3653
T. Uchiyama https://orcid.org/0000-0003-2148-1694
R. P. Udall https://orcid.org/0000-0001-6877-3278
T. Uehara https://orcid.org/0000-0003-4375-098X
K. Ueno https://orcid.org/0000-0003-3227-6055
V. Undheim https://orcid.org/0000-0003-4028-0054
T. Ushiba https://orcid.org/0000-0002-5059-4033
M. Vacatello https://orcid.org/0009-0006-0934-1014
H. Vahlbruch https://orcid.org/0000-0003-2357-2338
N. Vaidya https://orcid.org/0000-0003-1843-7545
G. Vajente https://orcid.org/0000-0002-7656-6882
J. Valencia https://orcid.org/0000-0003-2648-9759
M. Valentini https://orcid.org/0000-0003-1215-4552
S. A. Vallejo Peña https://orcid.org/0000-0002-6827-9509
V. Valsan https://orcid.org/0000-0003-0315-4091
M. van Dael https://orcid.org/0000-0002-6061-8131
E. Van den Bossche https://orcid.org/0009-0009-2070-0964
J. F. J. van den Brand https://orcid.org/0000-0003-4434-5353
M. van der Sluys https://orcid.org/0000-0003-1231-0762
J. van Dongen https://orcid.org/0000-0003-0964-2483
H. van Haevermaet https://orcid.org/0000-0003-2386-957X
J. V. van Heijningen https://orcid.org/0000-0002-8391-7513
P. Van Hove https://orcid.org/0000-0002-2431-3381
N. van Remortel https://orcid.org/0000-0003-4180-8199
A. F. Vargas https://orcid.org/0000-0001-8396-5227
V. Varma https://orcid.org/0000-0002-9994-1761
A. Vecchio https://orcid.org/0000-0002-6254-1617
J. Veitch https://orcid.org/0000-0002-6508-0713
P. J. Veitch https://orcid.org/0000-0002-2597-435X
R. C. Venterea https://orcid.org/0000-0003-3299-3804
P. Verdier https://orcid.org/0000-0003-3090-2948
D. Verkindt https://orcid.org/0000-0003-4344-7227
Y. Verma https://orcid.org/0000-0003-4147-3173
S. M. Vermeulen https://orcid.org/0000-0003-4227-8214
A. Veutro https://orcid.org/0009-0002-9160-5808
A. Viceré https://orcid.org/0000-0003-0624-6231
A. D. Viets https://orcid.org/0000-0002-4241-1428
A. Vijaykumar https://orcid.org/0000-0002-4103-0666
V. Villa-Ortega https://orcid.org/0000-0001-7983-1963
E. T. Vincent https://orcid.org/0000-0002-0442-1916
S. Vitale https://orcid.org/0000-0003-2700-0767
H. Vocca https://orcid.org/0000-0002-1200-3917
D. Voigt https://orcid.org/0000-0001-9075-6503
L. Vujeva https://orcid.org/0000-0001-7697-8361
S. P. Vyatchanin https://orcid.org/0000-0002-6823-911X
M. Wade https://orcid.org/0000-0002-5703-4469
K. J. Wagner https://orcid.org/0000-0002-7255-4251
H. Wang https://orcid.org/0000-0002-6589-2738
Y. F. Wang https://orcid.org/0000-0002-2928-2916
G. Waratkar https://orcid.org/0000-0003-3630-9440
M. Was https://orcid.org/0000-0002-1890-1128
T. Washimi https://orcid.org/0000-0001-5792-4907
N. L. Weickhardt https://orcid.org/0000-0002-3923-5806
A. J. Weinstein https://orcid.org/0000-0002-0928-6784
L. Wen https://orcid.org/0000-0001-7987-295X
K. Wette https://orcid.org/0000-0002-4394-7179
J. T. Whelan https://orcid.org/0000-0001-5710-6576
B. F. Whiting https://orcid.org/0000-0002-8501-8669
C. Whittle https://orcid.org/0000-0002-8833-7438
D. Wilken https://orcid.org/0000-0002-7290-9411
D. Williams https://orcid.org/0000-0003-3772-198X
M. J. Williams https://orcid.org/0000-0003-2198-2974
N. S. Williams https://orcid.org/0000-0002-5656-8119
J. L. Willis https://orcid.org/0000-0002-9929-0225
B. Willke https://orcid.org/0000-0003-0524-2925
M. Wils https://orcid.org/0000-0002-1544-7193
G. Woan https://orcid.org/0000-0003-0381-0394
H. T. Wong https://orcid.org/0000-0003-4145-4394
I. C. F. Wong https://orcid.org/0000-0003-2166-0027
M. Wright https://orcid.org/0000-0003-1829-7482
C. Wu https://orcid.org/0000-0003-3191-8845
D. S. Wu https://orcid.org/0000-0003-2849-3751
H. Wu https://orcid.org/0000-0003-4813-3833
Z. Wu https://orcid.org/0000-0002-0032-5257
D. M. Wysocki https://orcid.org/0000-0001-9138-4078
V. A. Xu https://orcid.org/0000-0002-3020-3293
Y. Xu https://orcid.org/0000-0001-8697-3505
N. Yadav https://orcid.org/0009-0009-5010-1065
H. Yamamoto https://orcid.org/0000-0001-6919-9570
K. Yamamoto https://orcid.org/0000-0002-3033-2845
T. S. Yamamoto https://orcid.org/0000-0002-8181-924X
T. Yamamoto https://orcid.org/0000-0002-0808-4822
R. Yamazaki https://orcid.org/0000-0002-1251-7889
K. Z. Yang https://orcid.org/0000-0001-8083-4037
Y. Yang https://orcid.org/0000-0002-3780-1413
Z. Yarbrough https://orcid.org/0000-0002-9825-1136
A. B. Yelikar https://orcid.org/0000-0002-8065-1174
J. Yokoyama https://orcid.org/0000-0001-7127-4808
H. Yuzurihara https://orcid.org/0000-0002-3710-6613
M. Zeeshan https://orcid.org/0000-0002-6494-7303
M. Zeoli https://orcid.org/0009-0007-1898-4844
M. Zevin https://orcid.org/0000-0002-0147-0835
R. Zhang https://orcid.org/0000-0001-8095-483X
C. Zhao https://orcid.org/0000-0001-5825-2401
Z.-C. Zhao https://orcid.org/0000-0001-5180-4496
Y. Zheng https://orcid.org/0000-0002-5432-1331
H. Zhong https://orcid.org/0000-0001-8324-5158
Z.-H. Zhu https://orcid.org/0000-0002-3567-6743
A. B. Zimmerman https://orcid.org/0000-0002-7453-6372
M. E. Zucker https://orcid.org/0000-0002-2544-1596
J. Zweizig https://orcid.org/0000-0002-1521-3397